\documentclass[prd,aps,nofootinbib,amssymb,twocolumn,superscriptaddress,10pt,floatfix]{revtex4-1}
\usepackage{CJK}
\usepackage{amsfonts}
\usepackage{amsmath}
\usepackage{ dsfont }
\usepackage{hyperref}
\usepackage{amssymb}
\usepackage{xcolor}
\usepackage{xspace}
\usepackage{ragged2e}
\usepackage{relsize}
\usepackage[bottom]{footmisc}

\usepackage{tikz}
\usetikzlibrary{calc}
\newcommand{\nocontentsline}[3]{}
\newcommand{\tocless}[2]{\bgroup\let\addcontentsline=\nocontentsline#1{#2}\egroup}

\newcommand{\be}{\begin{equation}}
\newcommand{\ee}{\end{equation}}
\newcommand{\bea}{\begin{equation} \begin{aligned}}
\newcommand{\eea}{\end{aligned} \end{equation} }
\newcommand{\bi}{\begin{itemize}}
\newcommand{\ei}{\end{itemize}}

\newcommand{\la}{\lambda}
\renewcommand{\be}{\beta}
\newcommand{\al}{\alpha}
\newcommand{\bpm}{\begin{pmatrix}}
\newcommand{\epm}{\end{pmatrix}}
\newcommand{\eps}{\epsilon}

\renewcommand{\th}{\theta}

\newcommand{\lp}{\left(}
\newcommand{\rp}{\right)}

\newcommand{\Tr}{\text{Tr} \ }

\newcommand{\mbf}[1]{\mathbf{#1}}
\renewcommand{\d}{\downarrow}

\usepackage{color}
\usepackage{graphicx}
\usepackage{verbatim}
\usepackage{amsmath}
\usepackage{amssymb}
\usepackage{wasysym}
\usepackage{mathrsfs}
\usepackage[caption=false]{subfig}
\usepackage{url}
\usepackage{bbold}
\usepackage{slashed}
\usepackage{epstopdf}
\usepackage{braket}
\usepackage{float}
\usepackage[percent]{overpic}

\DeclareRobustCommand{\Sec}[1]{Sec.~\ref{#1}}

\DeclareRobustCommand{\App}[1]{App.~\ref{#1}}

\DeclareRobustCommand{\Tab}[1]{Table~\ref{#1}}

\DeclareRobustCommand{\Fig}[1]{Fig.~\ref{#1}}

\DeclareRobustCommand{\Eq}[1]{Eq.~(\ref{#1})}
\DeclareRobustCommand{\Eqs}[2]{Eqs.~(\ref{#1}) and (\ref{#2})}

\DeclareMathAlphabet\mathbfcal{OMS}{cmsy}{b}{n}

\RequirePackage[normalem]{ulem} 
\RequirePackage{color}\definecolor{RED}{rgb}{1,0,0}\definecolor{BLUE}{rgb}{0,0,1} 

\begin{document}

\title{Kekul\'e Spiral Order from Strained Topological Heavy Fermions}
\author{Jonah Herzog-Arbeitman}
\thanks{These authors contributed equally to this work.}
\affiliation{Department of Physics, Princeton University, Princeton, New Jersey 08544, USA}
\author{Dumitru C\u{a}lug\u{a}ru}
\thanks{These authors contributed equally to this work.}
\affiliation{Department of Physics, Princeton University, Princeton, New Jersey 08544, USA}
\affiliation{Rudolf Peierls Centre for Theoretical Physics, University of Oxford, Oxford OX1 3PU, United Kingdom}
\author{Haoyu Hu}
\affiliation{Donostia International Physics Center, P. Manuel de Lardizabal 4, 20018 Donostia-San Sebastian, Spain}
\author{Jiabin Yu}
\affiliation{Department of Physics, University of Florida, Gainesville, FL, USA}
\affiliation{Department of Physics, Princeton University, Princeton, New Jersey 08544, USA}
\author{Nicolas Regnault}
\affiliation{Center for Computational Quantum Physics, Flatiron Institute, 162 5th Avenue, New York, NY 10010, USA}
\affiliation{Department of Physics, Princeton University, Princeton, New Jersey 08544, USA}
\affiliation{Laboratoire de Physique de l'Ecole normale sup\'{e}rieure, ENS, Universit\'{e} PSL, CNRS, Sorbonne Universit\'{e}, Universit\'{e} Paris-Diderot, Sorbonne Paris Cit\'{e}, 75005 Paris, France}
\author{Jian Kang}
\affiliation{School of Physical Science and Technology, ShanghaiTech University, Shanghai 200031, China}
\author{B.~Andrei Bernevig}
\email{bernevig@princeton.edu}
\affiliation{Department of Physics, Princeton University, Princeton, New Jersey 08544, USA}
\affiliation{Donostia International Physics Center, P. Manuel de Lardizabal 4, 20018 Donostia-San Sebastian, Spain}
\affiliation{IKERBASQUE, Basque Foundation for Science, Bilbao, Spain}
\author{Oskar Vafek}
\email{vafek@magnet.fsu.edu}
\affiliation{Department of Physics, Florida State University, Tallahassee, Florida 32306, USA}
\affiliation{National High Magnetic Field Laboratory, Tallahassee, Florida, 32310, USA}

\date{\today}
\begin{abstract}
The topological heavy fermion (THF) model of twisted bilayer graphene is a framework for treating its strongly interacting topological flat bands. In this work, we employ the THF model with heterostrain and particle-hole symmetry breaking corrections to study its symmetry-broken ground states. We find that the heterostrain correction motivates a specific parent-state wavefunction which dictates the presence or absence of an incommensurate Kekul\'e spiral (IKS) at each integer filling by invoking Dirac node braiding and annihilation as a mechanism to achieve low energy gapped states. We then show that one-shot Hartree-Fock faithfully replicates the numerical results of fully self-consistent states and motivates an analytical approximation for the IKS wavevector. We can also account for the particle-hole asymmetry in the correlated insulator gaps. In particular, the THF model predicts stronger correlated states on the electron side rather than hole side in agreement with magic angle experiments, despite the electron side being more dispersive in the single-particle band structure. This work demonstrates that we can analytically explain even the more subtle symmetry breaking order properties observed in experiments where heterostrain, relaxation, and interactions together determine the ground state. 
\end{abstract}
\maketitle

\section{Introduction.} The correlated insulators in twisted bilayer graphene (TBG)\cite{GHO24,MER24,DIE23,TIA23,DI22a,LIA21c,YU23c,HUB22,GRO22,JAO22,PAU22,GHA22,WU21a,DAS21,PAR21c,LU21,SAI21a,CAO21,CHE20b,SAI21,POL19,CAO20,SER20,Stepanov2019SCCorrTBG,SAI20,YAN19,LU19,CAO18,ZHO23a,BEN21,NUC23,HES21,LIS21,CHO21,TSC21,CHO20,WON20,JIA19,KER19,CHO19,NUC20,XIE19} have been extensively studied\cite{MAC23,HON22,THO21,BRI22,KWA21b,ANG19,BLA22,CAL22d,LED21,RAD18,CHA21,PAD18,Zaletel2020SkyrmionTBG,HUA19,EUG20,KEN18,pizarro2019nature,CLA19,DA19,Liu2019May17NematicTSMMATBG,CHR20,SOE20,KAN20a,YUA18,DOD18,THO18,WU19,VEN18,XU18b,KOS18,LIU19,VAF20,KAN19,VAF23,CAL20,FAN19,CAR19a,KAN23b,CAR19,NAM17,KWA20,SHE21,DAV22,CAO21a,WAN21a,REN21,BER21,CAR20a,WIL20,HUA20a,KAN18,EFI18,FU20,JAI16,DAI16,WIJ15,KON20,UCH14,ZOU18,SUA10,LOP07,Po2019TBGFragile,BIS11,TAR19,davis2023kinetic,PhysRevB.107.045426,PhysRevB.107.235155,HEJ19a,Lian2020LLFragileTBG,HEJ19,AHN19,XIE20,SON19,khosravian2024moire} within the context of the Bistritzer-MacDonald (BM) continuum model\cite{Bistritzer2011BMModel} with  screened Coulomb interactions. In the strong coupling limit of the strain and relaxation-free interacting Hamiltonian, a manifold of exactly solvable ground states appears\cite{KAN19,SEO19,LIA21,BUL20a}, signaling close competition between various spin and valley orderings. Although the small kinetic energy of the 8 nearly flat BM bands splits this manifold, it is dwarfed by the effects of heterostrain, the layer-asymmetric component of the strain field, and moir\'e-scale lattice relaxation\cite{2024arXiv240314078L,Bi2019TBG,PhysRevB.96.075311,Koshino2018TBGFragile,2024arXiv241209654L,2023arXiv231107819L,PhysRevB.105.245408,KWA21,KWA21,2018PhRvL.120o6405H,PhysRevLett.127.126405,PhysRevB.107.125112,2021PhRvL.127l6405M,PAR21a,2023PNAS..12007151W,WAG22,WAN23b}. 

By and large transport experiments indicate the presence of interaction-driven insulators at $\nu = \pm2$ and semi-metals at $\nu = 0$ (although not without exception \cite{LU19,SHA19,SER20}), as well as the simple band insulators at $\pm 4$. Common but not as ubiquitous are insulators at $\nu = + 3$ and to a lesser extent $\nu = -3$, as well as metallic states $\nu = \pm 1$. It has been understood that most TBG devices are heterostrained\cite{KER19,CHO19,XIE19,kazmierczak2021strain}, which disrupts the strong coupling phase diagram characterized by generalized ferromagnets at integer filling \cite{XIE20b,WU20,XIE23a,KWA21a,VAF21,ZHA21,LIU19a,HOF22,KAN21,XIE21a,POT21,CHI20b,CHE21,DA21,XIE21,REP20,ZHA20,CEA20,BUL20b,OCH18,BUL20a,PO18a,RAD19,HEJ21}, with gaps possible at each integer filling. 

Local probes have provided further evidence for the importance of heterostrain. Scanning tunneling microscopy \cite{NUC23} has shown that, even at low and moderate heterostrain, the ground state at $\nu = -2$ (two filled bands below charge neutrality) displays a clear Kekul\'e pattern on the graphene scale that appears in and around the insulating phase. Similar behavior is found in twisted trilayer graphene \cite{2023Natur.623..942K,PhysRevB.109.L201119,YU22}. This Kekul\'e pattern can be visualized as a tripling of the graphene unit cell originating from the interference of $e^{\pm i \mbf{K} \cdot \mbf{r} }$ Bloch function oscillations which hybridize electrons in the graphene $K$ and $K'$ valleys. The leading candidate ground state of the BM model in the strong coupling limit, called the Kramers inter-valley coherent state or KIVC\cite{KAN19,LIA21,BUL20a}, is forbidden from showing a Kekul\'e pattern despite possessing inter-valley coherence \cite{CAL22d,HON22} and is therefore ruled out experimentally by Ref.\cite{NUC23}.

The experiment has also shown moir\'e scale variation of the Kekul\'e pattern in a device with $\sim 0.2\%$ heterostrain \cite{NUC23}. A state compatible with these observations was theoretically predicted in the presence of heterostrain in Ref.\cite{KWA21} and dubbed Incommensurate Kekul\'e Spiral, or IKS. While true incommensuration is difficult to prove in an experiment, the key feature of the IKS, namely modulation of the Kekul\'e pattern between moir\'e unit cells, has been observed. This is consistent with broken moir\'e lattice translation symmetry $T_\mbf{R}$ but the preservation of a valley-boosted moir\'e translation $e^{i \tau_3 \mbf{q} \cdot \mbf{R}/2} T_\mbf{R}$; here $\tau_3$ is the Pauli matrix acting on the two valleys and $\mbf{q}$ is the IKS boost vector. At very low heterostrain $\sim 0.05\%$, there is  evidence for a time-reversal invariant inter-valley coherent state \cite{KAN19,SEO19,LIA21,BUL20a,VAF21} without moir\'e translation breaking \cite{NUC23}. Such a state was argued to be stabilized by strong enough coupling of electrons to inter-valley phonons \cite{2020EPJP..135..630A,PhysRevX.9.041010,2024PhRvB.110h5160K,2024arXiv240211824S,2024arXiv240711116W}.

The heterostrained phase diagram \cite{KWA21,WAG22} obtained from Hartree-Fock calculations is in good agreement with experiment\cite{2024arXiv240801599H}. In particular, it explains the semi-metal at $\nu=0$ as a state without spontaneous symmetry breaking. Such a state must have Dirac nodes protected by $C_{2z}\mathcal{T}$ and $U(1)$ valley conservation\cite{PO18a,ZOU18,SON19,AHN19}. They cannot annihilate because the two Dirac nodes in the same valley have the same winding number\cite{PO18a,ZOU18,KAN20a}. What is more surprising is the appearance of gapped states at $\nu = \pm3$ with $C_{2z}\mathcal{T}$ still preserved\cite{KWA21}; such states offer a natural explanation for the experimentally observed absence of anomalous Hall conductivity, because electrical current and electric field have opposite parity under $C_{2z}\mathcal{T}$. In addition, experimental lack of spin polarization\cite{CAO18,YAN19} at $\nu = \pm 2$ excludes a $C_{2z}\mathcal{T}$-preserving spin-valley polarized state filling both flat bands. For comparison, we recall that the mechanism of gap opening in and around the chiral limit (at zero strain) is Chern band polarization \cite{KAN19,BUL20a,LIA21,VAF21}. In this limit, the 8 flat bands are split into decoupled Chern $\pm1$ bands that spontaneously break $C_{2z}\mathcal{T}$, which gaps their charge excitations. $U(4) \times U(4)$ rotations of such states can restore $C_{2z}\mathcal{T}$, but such states still inherit the gap. For instance, at $\nu = 0$ the valley polarized state occupies both $C = \pm 1$ bands and both spins in the $K$ valley and therefore globally preserves the intra-valley $C_{2z}\mathcal{T}$ symmetry, but its charge excitation spectrum block diagonalizes into Chern sectors, and remains gapped. Without this mechanism available in realistic devices, how can the system open a robust insulating gap, and how does this lead to IKS order?

The purpose of this paper is to offer new insights into the heterostrained and relaxed phase diagram. We show that the presence or absence of gaps, and their connection to IKS order, can be explained through a topological analysis of the Dirac nodes and in particular their non-abelian braiding properties\cite{Tomas2019NonAbelianNodalLine} in multi-band systems with $C_{2z}\mathcal{T}$ symmetry intact. To carry out these arguments analytically, we make extensive use of the topological heavy fermion (THF) representation of TBG \cite{Song20211110MATBGHF,CAL23,SHI22a}, which has been used to explain the cascade phase transitions \cite{KAN21,2023NatCo..14.5036D,ZON20,CAL22d,2023arXiv230908529R}, local moment entropy \cite{SAI21a,ROZ21}, thermoelectric effect \cite{2024arXiv240214057C,2024arXiv240211749L,2024arXiv240212296B}, and to motivate proposals of Kondo physics \cite{2023PhRvL.131b6502H,2024PhRvB.109d5419Z,2023arXiv230302670L,2023PhRvL.131b6501C,PhysRevB.108.125106,2024PhRvB.110d5123L}. We will provide a physical, local moment picture of the intricate IKS symmetry breaking in TBG.

In this work, we obtain analytical model wavefunctions for the candidate ground states at each integer filling which all preserve $C_{2z}\mathcal{T}$. From these ans\"atze, we can explain how $|\nu| = 3,2$ can open gaps by developing IKS order without breaking $C_{2z}\mathcal{T}$ while $|\nu| = 1,0$ remain gapless. We then derive analytical estimates of the critical heterostrain determining the IKS phase boundary and the inter-valley IKS boost vector $\mbf{q}$. Using symmetry analysis and topological arguments, we explain why the IKS boost $\mbf{q}$ is required to open a gap: it does so by driving a non-abelian Dirac node braiding process \cite{Tomas2019NonAbelianNodalLine} which we study in detail. 

Finally incorporating particle-hole asymmetry, which can arise from lattice relaxation \cite{Yoo_2019,ZHU20a,PhysRevB.95.075420,PhysRevB.98.224102} or the inclusion of momentum-dependence in the interlayer hopping\cite{SON21}, we are able to simply explain the particle-hole asymmetry in the correlated insulator gaps. Intriguingly, these gaps are larger on the more dispersive side of the single-particle band structure \cite{NUC20,2021NatPh..17.1210P,2022NatPh..18..825Y,2021NatPh..17..710D,LU19}. Our work shows the THF model can capture subtle experimental details of TBG which are beyond the minimal BM model. This work focuses on the Hartree-Fock description of the ground states at integer fillings and can serve as a starting point for more realistic dynamical mean-field theory calculations at higher temperatures and arbitrary fillings \cite{CAL20,DAT23,ZHO24,2023arXiv230908529R,Hu2023KondoMATBGStrain}.

\section{Single-Particle Heavy Fermion Model.}
\begin{figure}
\centering
\includegraphics[width=\columnwidth]{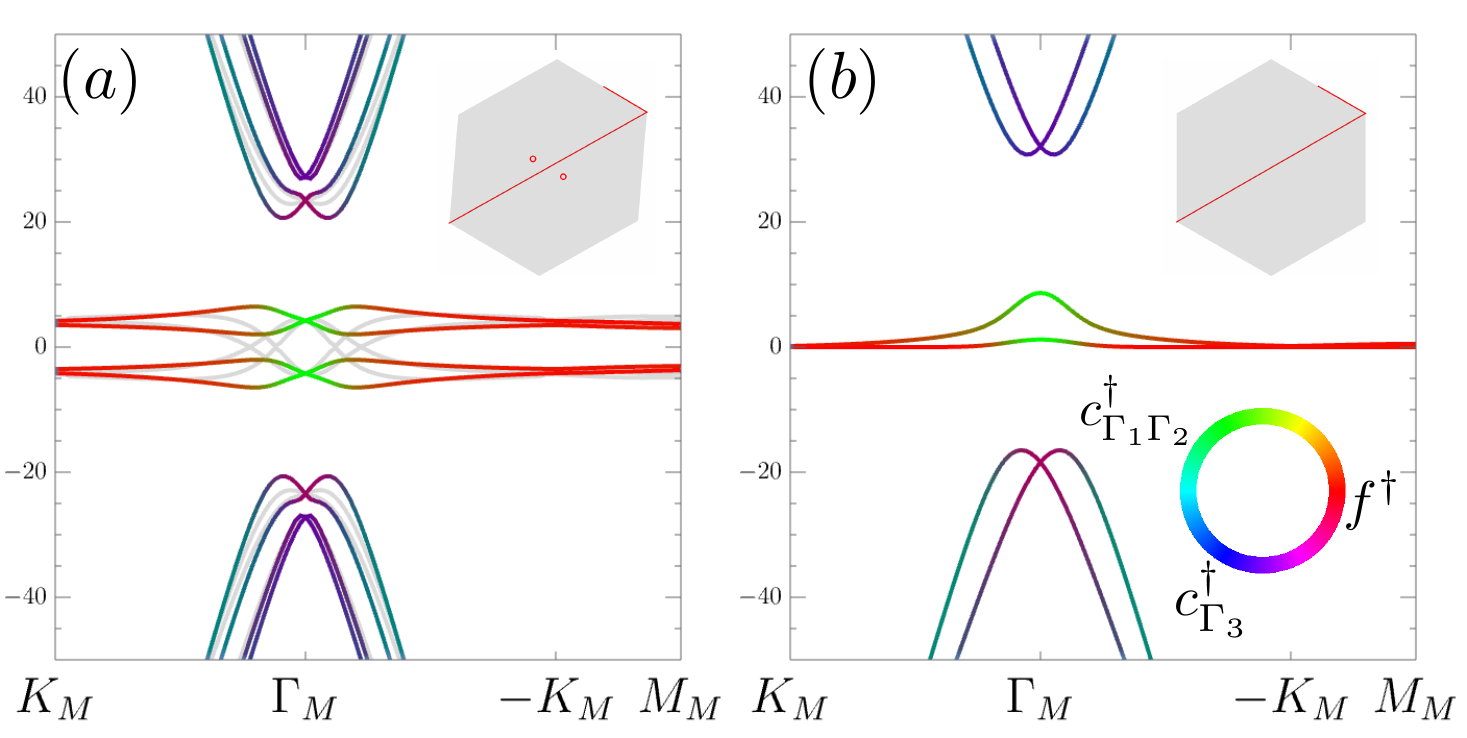}
\caption{Single-particle band structure with both valleys, colored by $f$/$c$ character according to the color wheel. $(a)$ shows the band structures in $\eps = .15\%$ heterostrain, the main effect of which is to split the density of states, reduce the gap to the dispersive bands, and unpin the Dirac nodes from the $K_M$ points (see red circles in the inset). The grey bands plot the band structure along a rotated path which goes through the $C_{2z}\mathcal{T}$-protected Dirac nodes. $(b)$ includes gradient tunneling terms that break particle-hole symmetry with $\eps = 0$.
}
\label{fig0}
\end{figure}

The band structure of TBG has been studied extensively. It has evolved to include key experimental perturbations beyond the minimal model, namely heterostrain and relaxation which reduce the symmetries of the Hamiltonian \cite{VAF23,CAL20,FAN19,CAR19a,KAN23b,CAR19,NAM17,KWA20,LED21,RAD19,SHE21,DAV22,CAO21a,HEJ21,WAN21a,REN21,BER21,CAR20a,WIL20,HUA20a,LIU19a,KOS18,KAN18,EFI18,FU20,JAI16,DAI16,WIJ15,UCH14,ZOU18,SUA10,LOP07,Po2019TBGFragile,BIS11,TAR19,davis2023kinetic,PhysRevB.107.045426,PhysRevB.107.235155}. In a companion paper \cite{HER24a}, we have shown that the THF theory of TBG can incorporate these effects merely by introducing symmetry-breaking \emph{single-particle} terms. The underlying fermion character of the low energy Hilbert space, as well as the interacting part of the periodic Anderson model, remain unchanged. 

We start with a brief summary of the heavy fermion theory of TBG. The flat bands of TBG are topological: there is no unitary transformation respecting valley $U(1)$ and space group symmetries which maps the Bloch states of the flat bands into exponentially decaying Wannier states. Thus, in any model of the flat bands alone, one is forced to have a Wannier basis where symmetries are not represented locally\cite{BER21a}\cite{VAF21} or where the Wannier functions decay as a power law\cite{2022arXiv221011573Z}, leading to long-range Coulomb matrix elements that prevent a purely local Hubbard-like description. Although mean field theory has been shown to be reliable at integer fillings\cite{KAN20a,SOE20,LIA21,XIE21,PAR21a,PhysRevB.108.235128}, it cannot, for example, reproduce the observed Hubbard bands above the ordering temperature \cite{WON20} or be trustworthy away from integer filling.  In order to study the effects of strong correlations believed to be important away from integer fillings, access to a real-space, microscopically-derived Anderson model is advantageous.

The topological obstruction to a Hubbard-like model is resolved by including the low energy states of the dispersive bands in the Hilbert space. The result\cite{Song20211110MATBGHF} is a mapping of the low energy Hilbert space of the BM model into localized and symmetric Wannier states $f^\dag_{\mbf{R}, \al \eta s}$ coupled to a set of topological conduction ``$c$"-electrons $c^\dag_{\mbf{k},a \eta s}$. Here $\eta = \pm , s =\uparrow,\downarrow$ are the valley and spin indices, and the flavor index $\al = 1,2$ corresponds to $p_x \pm i p_y$-like orbitals of the $f$-electron at the AA stacking site of the moir\'e lattice. The index $a$ corresponds to the two flavors of Dirac electrons carrying the 2D irrep $\Gamma_3$ ($a = 1,2$) and two 1D representations $\Gamma_1 \oplus \Gamma_2$ ($a = 3,4$). Interestingly, the $c$-electrons happen to carry the representations of the low-energy states of Bernal stacked bilayer graphene\cite{Song20211110MATBGHF}. In principle, this mapping is exact and unitary within the low energy subspace, but can be truncated\cite{Song20211110MATBGHF} to obtain a simple yet faithful model. The position space wavefunction of the BM flat bands has been derived from this model in Ref.\cite{Song20211110MATBGHF}. The wavefunction is delocalized in position space at the $\Gamma$ point, but elsewhere in the BZ the wavefunction is localized on the AA sites. Subsequent work in Ref.\cite{2024arXiv240816761L} derived a similar wavefunction.

The original BM model at the magic angle $1.05^\circ$ yields the THF Hamiltonian acting on a 6-component $c-f$ spinor,
\bea
\label{hkinetic}
h_K(\mbf{k}) &=\!\! \lp \begin{array}{c c|c}
  & v \mbf{k} \cdot (\sigma_0, i \sigma_3) &  \gamma \sigma_0 + v' \mbf{k} \cdot \pmb{\sigma} \\
v \mbf{k} \cdot (\sigma_0, -i \sigma_3)& M \sigma_1 & \\
\hline
 \gamma \sigma_0 + v' \mbf{k} \cdot \pmb{\sigma}  & & 0 \\
\end{array} \rp, 
\eea
written for convenience in the $K$ valley ($K'$ is obtained by spinless time-reversal). The vanishing dispersion in the $2\times 2$ $f$ block is due to the strong localization of the $f$-mode wavefunctions. Here we have only kept the $\mbf{G}=0$ shell of $c$-electrons, so $\mbf{k}$ is restricted to the first BZ. The low energy band structure is highly accurate and the inclusion of higher shells has a weak effect because of the large monolayer graphene Fermi velocity. Ref.\cite{HER24a} has obtained the corrections to this Hamiltonian from heterostrain, relaxation and other particle-hole symmetry breaking effects. The dominant terms are
\bea
\label{strainrelaxperturb}
\delta h_K &= 
\lp \begin{array}{c c|c}
c \, \pmb{\eps} \cdot \pmb{\sigma} + \mu_1 \sigma_0 & c' \pmb{\eps} \cdot \pmb{\sigma}^*  & i \gamma' \eps_+ \sigma_3 \\
 &  M' \eps_+  \sigma_2 + \mu_2 \sigma_0 & c'' \pmb{\eps} \cdot (\sigma_0, -i \sigma_3) \\
 \hline
h.c. &  & M_f \pmb{\eps} \cdot \pmb{\sigma} \\
\end{array} \rp
\eea
where $c,c',c'',M', \gamma'$ and $M_f$ describe the coupling to the $C_{3z}$-breaking heterostrain. The heterostrain is parameterized by $\pmb{\epsilon}=(\eps_{xy},\eps_-)$ and $\eps_+$, i.e. the shear $\eps_{xy}$, anisotropic $\eps_{-} = (\eps_{xx} - \eps_{yy})/2$, and isotropic  $\eps_{+} = (\eps_{xx} + \eps_{yy})/2$ components of the tensor. We take a Poisson ratio $\nu_G = 0.16$ \cite{Bi2019TBG}, so that compressive heterostrain $\eps>0$ along the $x$ axis results in $\eps_{xy} = 0$ and $\eps_\mp = - (1\pm \nu_G)\eps/2$. A derivation of this model may be found in Ref.\cite{HER24a}, and the magnitudes of various parameters can be found in \App{appreview} as calculated from symmetry-breaking terms in the BM model employed in Ref.\cite{KWA21}. We will refer to $\mu_1,\mu_2$ as the ``$\mu$ terms''; they arise from lattice relaxation and gradient inter-layer tunnelings.

The band structures in \Fig{fig0} depict the effects of heterostrain and $\mu$ terms separately. Heterostrain has the dramatic effect of splitting the flat bands in two at the edge of the BZ. This splitting is easy to understand from the THF model. At the edge of the BZ, the $f$-electrons are isolated due to the large kinetic energy $|v\mbf{k}|$ of the $c$-electrons. Then from \Eq{strainrelaxperturb}, we see that the $f$-modes are split by $\pm M_f |\pmb{\epsilon}|$, as their local irreducible representation on the triangular lattice is no longer 2-dimensional upon adding strain. However, the bands cannot be split entirely: $C_{2z}\mathcal{T}$ symmetry protects Dirac nodes that connect the bands as shown in \Fig{fig0}. 

Typical values of $|\pmb{\eps}| \sim 0.1\%$ yield $2M_f |\pmb{\epsilon}| \sim 10$meV. This splitting separates out four low energy flat bands (one per spin per valley), and completely breaks the $U(4) \times U(4)$ symmetry of the many-body THF model that emerges when $M = v' = 0$ (see \App{appreview}). This splitting and the presence of Dirac nodes is the key to the IKS as we will show. The effect of the $\mu$ terms is restricted to the vicinity of the $\Gamma$ point, where the energy of both states is shifted upwards; their effect is negligible at large $\mbf{k}$ where the kinetic energy dominates as can be seen from \Eq{hkinetic} and \Eq{strainrelaxperturb}.

While both the heterostrain and the $\mu$ terms in $\delta h_{K}$ are on the order of $10$meV, only heterostrain destabilizes the strong coupling KIVC ground state. This is easily seen from the symmetries of the THF model. In \App{app:relaxapp}, we prove that while the $\mu$ terms break (anti-commuting) particle-hole, they do not break the (commuting) $U(4) \times U(4)$ symmetry which appears in the $M=v'=0$ limit of the THF model (without $\delta h_K$). Thus the $\mu$ terms do not perturb the ground state away from the $U(4) \times U(4)$ manifold, although they do affect the Hartree-Fock charge gaps (in agreement with experimental trends) as we will show in \Sec{sec:mainPH}. In contrast, heterostrain provides the largest symmetry-breaking scale in the problem. For simplicity, we neglect $\mu_1,\mu_2$ until \Sec{sec:mainPH}.

\section{Interacting Hamiltonian and Parent States.}

Our analysis relies on the accuracy of heavy fermion ``parent states," which are product states of Slater determinants with symmetry-broken $f$-modes and the half-filled $c$-mode Fermi sea (with no $f$-$c$ hybridization). Considered as initial conditions for Hartree-Fock iterations, it has been shown that the parent states rapidly converge to self-consistency \cite{Song20211110MATBGHF}. Even after a single iteration, their density matrices and spectra are typically in quantitative agreement with the self-consistent states. We refer to the Hartree-Fock Hamiltonian after one iteration (i.e. the Hamiltonian obtained from a parent state) as the ``one-shot" Hamiltonian, whereas the self-consistent Hamiltonian is obtained after convergence. We limit ourselves to a Hartree-Fock treatment in this work, noting that under certain conditions the IKS has been demonstrated to exist beyond Hartree-Fock\cite{PhysRevB.108.235128}.

The full interacting Hamiltonian can be written as\cite{Song20211110MATBGHF}
\bea
H &= H_0 + H_U + H_W + H_J + H_V 
\eea
where $H_0 = H_f + H_{c} + H_{fc}$ is the single-particle Hamiltonian which is written in first quantization in \Eq{hkinetic} and \Eq{strainrelaxperturb}. Explicitly, $H_f$ reads
\bea
\label{eq:Hfindices}
H_f = \sum_{\mbf{R}, \al \eta s} M_f [\pmb{\eps} \cdot ( \sigma_1, \eta \sigma_2)]_{\al \al'} f^\dag_{\mbf{R},\al \eta s}f_{\mbf{R},\al' \eta s}
\eea
which defines the flat (spatially decoupled) $f$-mode levels. The largest interaction is
\bea
H_U &= \frac{U_1}{2} \sum_\mbf{R} (:f^\dag_\mbf{R} f_\mbf{R}: )^2 + \frac{U_2}{2} \sum_{\braket{\mbf{R},\mbf{R}'}} :f^\dag_\mbf{R} f_\mbf{R}: \, :f^\dag_{\mbf{R}'} f_{\mbf{R}'} :
\eea
which is an onsite Hubbard term  for the total $f$-mode density $f^\dag_\mbf{R} f_\mbf{R} = \sum_{\al \eta s} f^\dag_{\mbf{R} \al \eta s} f_{\mbf{R} \al \eta s}$ with $U_1 = 58$meV and a much smaller nearest-neighbor interaction $U_2 = 2.3$meV. Our analytical results will demonstrate that varying $U_1$ within the same order will not alter our conclusions. The normal ordering is with respect to charge neutrality where $4$ of the $8$ $f$-modes are filled. $H_V$ is the residual Coulomb interaction of the $c$-modes. The $f$-$c$ interactions yield a repulsion $H_W$ between the $f$ density and the $\Gamma_3$ ($\Gamma_1\oplus \Gamma_2$) electrons with strength $W_1 = 44$meV ($W_3 = 50.2$meV) and a ferromagnetic exchange interaction $H_J$ with strength  $J = 16.4$meV. The explicit expressions for $H_W$, $H_J$, and $H_V$ can also be found  in \App{appintreview}. The localization of the $f$-modes afforded by the THF mapping makes $U_1$ the dominant energy scale.  We can then write down parent states as many-body ground states of the decoupled terms $H_f + H_c + H_U$ at integer filling $\nu \in (-4,4)$ which take the form\cite{YU23a}
\bea
\label{eq:parentstates}
\ket{\nu} &= \prod_{\mbf{R},n=1}^{\nu+4} f^\dag_{\mbf{R},n} \ket{\text{FS}}, \quad f^\dag_{\mbf{R},n}=\sum_{\al \eta s} f^\dag_{\mbf{R} \al \eta s} [\zeta_n(\mbf{R})]_{\al \eta s} \ .
\eea
Here $[\zeta_n(\mbf{R})]_{\al \eta s}$ are normalized ordering vectors of the heavy $f$ modes and $\ket{\text{FS}}$ is the  Fermi sea filling the negative energy $c$-electron states of $H_c$ in the absence of heterostrain and $\mu$ terms, i.e. up to the quadratic (when $M\ne 0$) band touching point. Going forward, we will abbreviate $f^\dag_\mbf{R} \zeta_n(\mbf{R}) = \sum_{\al \eta s} f^\dag_{\mbf{R} \al \eta s} [\zeta_n(\mbf{R})]_{\al \eta s}$ for various orders to be introduced shortly. 

\subsection{Strong Coupling Hierarchy}

The variational parameters $\zeta_n(\mbf{R})$ in \Eq{eq:parentstates} define different trial states. To motivate our discussion in the heterostrained case, we first review the heavy fermion construction of the ferromagnetic states in absence of strain. 

Without heterostrain, all 8 $f$-modes are degenerate and translation-preserving states with different $U(4) \times U(4)$ polarizations are appropriate parent states. Their competition is primarily decided by the $U(4) \times U(4)$ breaking perturbation $M\sim 4$meV in \Eq{hkinetic}, which favors the KIVC over other $U(4)\times U(4)$ directions at $\nu = 0, \pm 2$ due to a generalized Hund's rule\cite{Song20211110MATBGHF}. At odd fillings, the parent states are products of polarized $p_x\pm i p_y$ $f$-modes and IVC parent states. The parent states in \Eq{eq:parentstates} all minimize the largest energy scale, $U_1$, leaving a large degeneracy due to the local $U(8)$ symmetry \cite{CAL23} rotating among all $f$-modes. This symmetry is preserved by the other density-density term $H_W$, the second-largest energy scale. The next energy scale is $\gamma$, the $f$-$c$ hybridization which ensures that the $\Gamma_1\oplus\Gamma_2$ flavor of $c$-electrons is at low energy near the $\Gamma$ point. Since the $c$-electrons couple to the $f$-mode order, they will split the large $U(8)$ degeneracy. The mean-field term quantifying this effect is \cite{Song20211110MATBGHF},
\bea
h_{\Gamma_1\oplus\Gamma_2}[O_f] &= M \sigma_1 + \frac{J}{2}(\tau_3 \overline{O}_f^T \tau_3+\sigma_3 \overline{O}_f^T \sigma_3),
\eea
where $\overline{O}_f = \sum_n \zeta_n \zeta_n^\dag - \frac{1}{2}\mathbb{1}$ is the $f$-mode order. The larger $J/2 \sim 8$meV flat-$U(4)$-spin coupling term favors a $U(4)$ ferromagnetic order for $O_f$ (Hund's rule), and the smaller $M \sim 4$meV term selects a preferred state, the KIVC, out of this manifold since $M$ breaks the $U(4)$ symmetry \cite{Song20211110MATBGHF}. This hierarchy must be reconsidered when heterostrain is included.

Within the mean field approximation to the heavy fermion theory, we find a phase boundary at a critical value of the heterostrain where the single-particle term $H_f$ (see \Eq{eq:Hfindices}) destabilizes the strong-coupling ferromagnetic states stabilized by $J/2$. We estimate this critical value by comparing the typical magnitude of these terms as $\eps_c \sim J/2M_f \sim 0.2\%$. This small critical heterostrain is in reasonable agreement with the IKS phase diagram in the BM model\cite{KWA21}, where the IVC states are destabilized at $\eps \sim 0.15\%$ for $\nu = -2,0,2$. At $\nu = \pm3$, the critical heterostrain is $< 0.1\%$ in Ref. \cite{KWA21}. Applying the Hartree-Fock approximation to THF model at $\nu=\pm3$ may not be as reliable as at lower fillings since fluctuations are large for our choice of interaction strength corresponding to a dielectric constant of 6 \cite{Hu2023KondoMATBGStrain,HU23i}. Nevertheless, density matrix renormalization group calculations\cite{PhysRevB.108.235128} have confirmed the IKS phase for a range of heterostrain magnitudes at $\nu=-3$  (using dielectric constant of 10). We will now identify the THF parent states describing the heterostrain-induced phases. 
\begin{figure}
\centering
\includegraphics[width=\columnwidth]{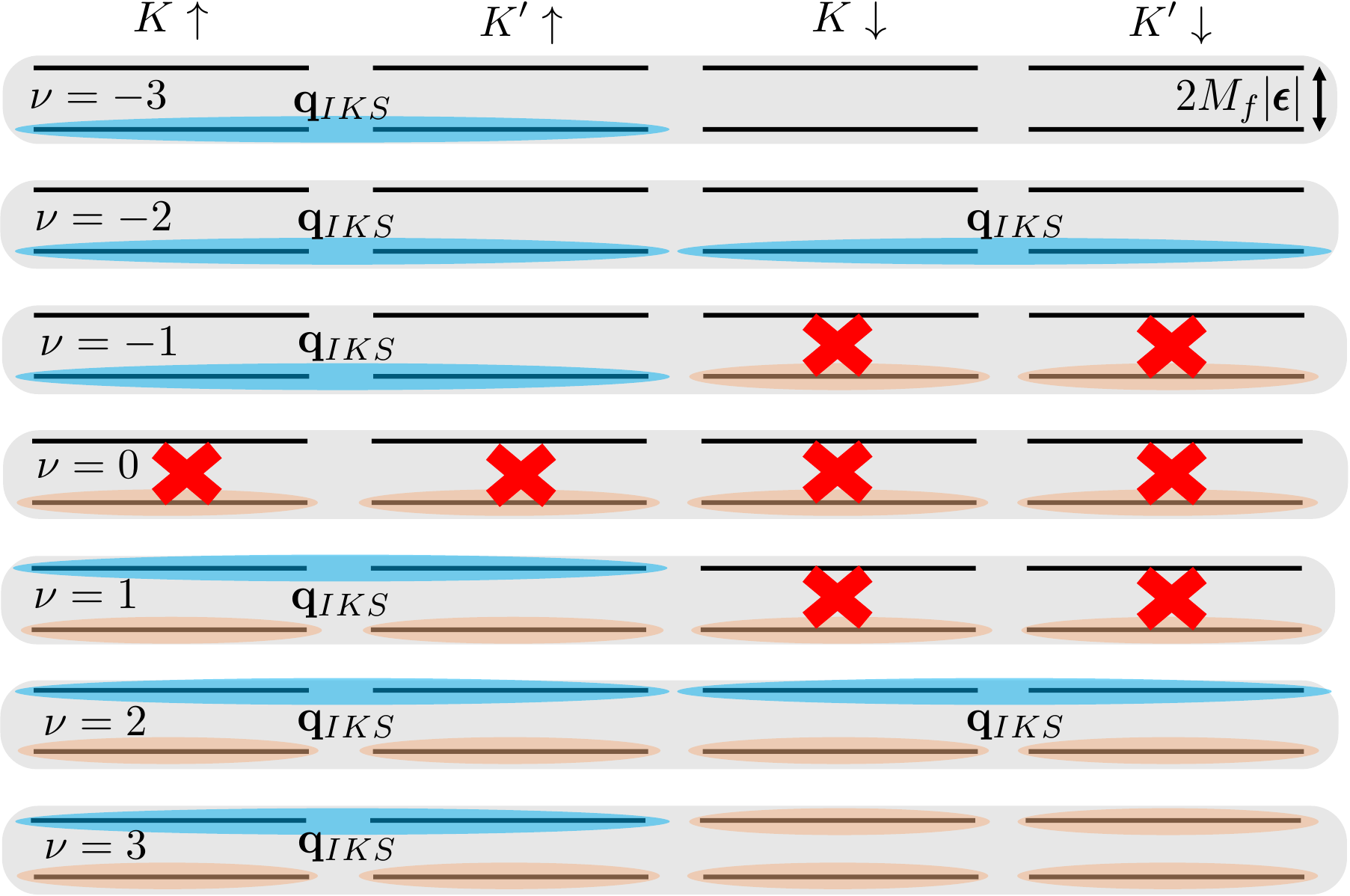}
\caption{Parent state orders in heterostrain-split bands (black). Blue shading denotes IVC $\zeta_{+}$ order between half-filled $K$ and $K'$ $f$-modes, and orange denotes fully filled VP $f$-modes. Red $X$'s denote gapless sectors. At zero relaxation, the parent states are particle-hole symmetric under the exchange of upper/lower $f$-modes and unfilled (black)/ fully filled (orange) $f$-modes. The phase diagram can be understood as a strongly interacting problem after selecting the lowest 4 $f$-modes due to the (non-interacting) heterostrain.
}
\label{figparentstate}
\end{figure}

\begin{figure*}
\centering
\includegraphics[width=17.5cm]{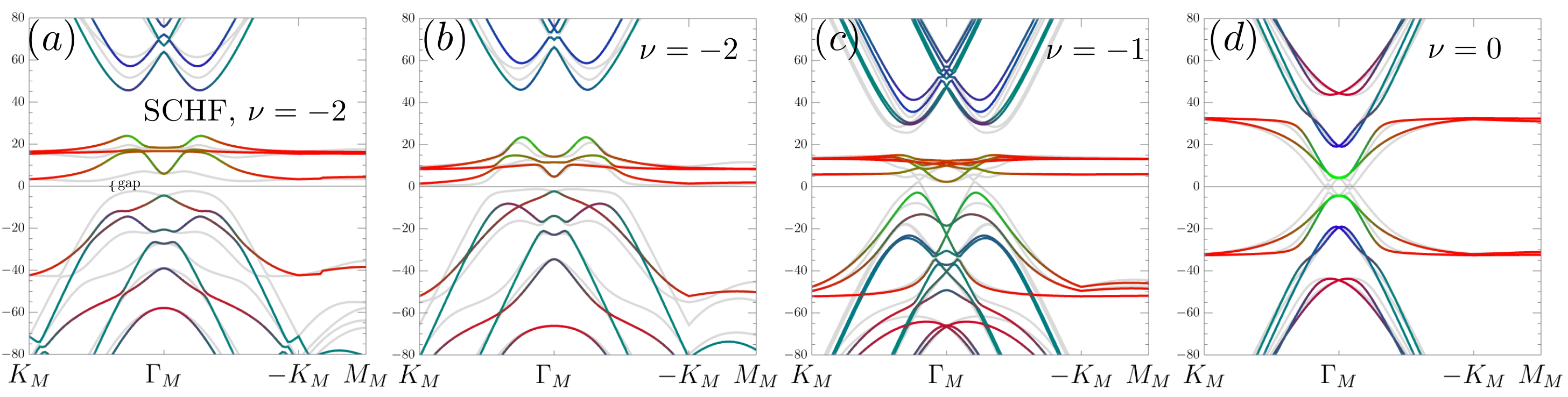}
\caption{Heavy Fermion one-shot band structures in $\eps = 0.15\%$ heterostrain. Hartree-Fock bands are colored red, green, and blue according to their $f$, $c_{\Gamma_1,\Gamma_2}$ and $c_{\Gamma_3}$ weights with the Fermi level set to 0. $(a)$ and $(b)$ show the self-consistent Hartree-Fock and one-shot band structure at $\nu=-2$, respectively. The occupied bands are in quantitative agreement, although the one-shot band structure underestimates the gap. $(c)$ and $(d)$ show the one-shot $\nu = -1$ and $\nu = 0$ bands, respectively, which are also in excellent agreement with the self-consistent bands (see \App{app:oneshotstrain}). Since $\nu = -1,0$ have Dirac nodes off the high symmetry lines, we show rotated paths through the BZ in gray where the Dirac nodes are visible. In $(a) - (c)$, the IKS boost vector is $\mbf{q} = \frac{2}{3} M_M$ and $\mbf{k}$ is the momentum in the valley-boosted frame, and in $(d)$, $\mbf{q}= 0$ and $\mbf{k}$ is the conventional moir\'e momentum. 
}
\label{figIKSHF}
\end{figure*}

\subsection{IKS Parent State Wavefunctions}

Our central assertion is that in the $|\pmb{\eps}|M_f > J/2$ phase, $\zeta_n$ should minimize $H_f$ in \Eq{eq:Hfindices} since it is the largest $f$-mode energy scale after the $U(8)$-symmetric density-density interactions. Thus $\zeta_n$ should be entirely supported by the four eigenvectors of $H_f$ with negative eigenvalues $- M_f |\pmb{\eps}|$, one eigenvector per spin per valley. These eigenvectors define a linear combination of the $p_x \pm i p_y$ orbitals depending on direction of the heterostrain $\pmb{\eps}$. Note $C_{3z} = e^{\frac{2\pi i}{3} \sigma_3}$ is always broken by $M_f \pmb{\eps} \cdot \pmb{\sigma}$, and $C_{2x} = \sigma_1$ is broken if there is compressive heterostrain, $\eps_{-} \neq 0$ (see \App{appreview}) as is considered here. However, $C_{2z}\mathcal{T} = \sigma_1 \hat{K}$ (where $\hat{K}$ is complex conjugation) is unbroken. This will play an important role in our analysis.

For simplicity, we start at $\nu= -3$ where one $f$-mode is filled per unit cell, which we will use as a point of departure for our main study, i.e. the fillings $|\nu| = 2,1,0$. Recall that we chose a compressive heterostrain  $\eps>0$ along the x axis, resulting in $\eps_- < 0$, and assume the spin is polarized in the $\uparrow$ direction. Then, as per the discussion in the paragraph above, the relevant eigenvectors of $H_f$, independent of $\mbf{R}$, are 
\bea
\zeta_K &= (1,i,0,0)^T/\sqrt{2}, \\
\zeta_{K'} &= (0,0,i,1)^T/\sqrt{2},
\eea
in the valley $\otimes$ orbital $ = (K +,K-,K'+,K'-)$ basis. We now consider an IVC state with $\zeta_{\th,\uparrow} = (\cos \th \, \zeta_{K} + \sin \th \zeta_{K'}) \otimes \ket{\uparrow}$, where we have fixed the inter-valley phase factor to 1; this phase is arbitrary due to the spontaneously broken $U(1)$ valley symmetry and because the total energy cannot depend on it, it is effectively not a variational parameter. On the other hand, the angle $\th$ is a variational parameter. For $\theta \neq \pi/4 \mod \pi/2$, this state breaks $C_{2z}$ (represented by $\tau_1 \sigma_1$) and spinless time-reversal (represented by $\tau_1 \hat{K}$) but preserves $C_{2z}\mathcal{T}$. This is important because $C_{2z}\mathcal{T}$ protects the local stability of the gapless Dirac nodes in the band structure \cite{2014ARCMP...5...83V}.

Typically, a state can lower its energy in Hartree-Fock by opening a gap to decrease the energy of the occupied quasi-particles. According to this heuristic, Dirac nodes are unfavorable energetically because they prevent a gap in the Hartree-Fock band structure. Indeed, without strain, gapped states are favored in strong coupling and spontaneously break $C_{2z}\mathcal{T}$ by polarizing to the Chern basis (or preserving $C_{2z}\mathcal{T}$ while breaking translations and annihilating the Dirac nodes to open a gap\cite{Kang2020Feb24DiracNodeMATBG,XIE23a}). These states are at high energy in the presence of strain, however. Attempts to choose $\th$ so as to gap the Dirac nodes in the Hartree-Fock band structure also fail, as we have verified numerically in self-consistent Hartree-Fock. We explain this analytically in \App{app:DiracIVC} by solving the one-shot Hartree-Fock spectrum obtained from $\zeta_{\th,\uparrow}$ in the flat band limit. 

Next, motivated by a tunable order parameter capable of annihilating the Dirac nodes, we follow Ref.\cite{KWA21} and consider the family of ``spiral" states parameterized by the IKS boost vector $\mbf{q}$: 
\bea
\zeta_{+,\uparrow}(\mbf{R}) &= e^{-i \mbf{Q} \cdot \mbf{R}}\zeta_{+,\uparrow}, \quad \text{where}\quad \mbf{Q} = \frac{\tau_3}{2} \mbf{q}. 
\eea
The two component vector $\mbf{q}$ serves as our variational parameter. The valley space Pauli matrix $\tau_3$ acts on states,
\bea
\label{eq:zetaplus}
\zeta_{+,\uparrow} = \frac{1}{\sqrt{2}} (\zeta_{K} +  \zeta_{K'}) \otimes \ket{\uparrow} = (1,i,i,1)/2 \otimes \ket{\uparrow}
\eea
chosen to have maximal IVC i.e. an equal weight in each valley. (Within the model, there remains an exact $SU(2) \times SU(2)$ symmetry corresponding to an independent spin rotation in each valley, and $\zeta_{+,\uparrow}$ is one representative of this manifold generated by this symmetry.) The maximal IVC ensures both spin-less time-reversal and $C_{2z}$ are preserved by $\zeta_{+,\uparrow}(\mbf{R})$. The presence of these symmetries is important to our analytical argument presented later in the text; they are supported by prior numerical calculations in the BM model \cite{KWA21}. We will provide additional justification for maximal IVC, and hence $C_{2z},\mathcal{T}$ symmetries, by showing that the IKS boost becomes energetically favorable through hybridization gap openings \emph{enabled} by IVC. Finally, we have picked the (arbitrary) relative phase such that $\zeta_{+,\uparrow}$ is even under $C_{2z}$. For later use, we will define the orthonormal vector $\zeta_{-,\uparrow} = \frac{1}{\sqrt{2}} (\zeta_{K} - \zeta_{K'})\otimes \ket{\uparrow}$. 

Note that $\zeta_{+,\uparrow}(\mbf{R})$ breaks translations by moir\'e lattice vectors implemented by $T_\mbf{R}$, but preserves the product of moir\'e lattice translation and the valley $U(1)$ transformation: 
\bea
\tilde{T}_\mbf{R} = e^{i \mbf{Q} \cdot \mbf{R}} T_\mbf{R} ,
\eea
referred to as the IKS translation from now on. It will therefore be convenient to define the boosted operators
\bea
\label{eq:boostbasis}
\tilde{f}^\dag_{\mbf{R}, \al \eta s} &\equiv f^\dag_{\mbf{R}, \al \eta s} e^{-i \mbf{R} \cdot \mbf{q} \eta/2}, \quad 
\tilde{c}^\dag_{\mbf{k}, a \eta s} \equiv c^\dag_{\mbf{k} + \mbf{q} \eta /2,a \eta s} \\
\eea
obeying $\tilde{T}_\mbf{a} \tilde{f}^\dag_{\mbf{R}, \al \eta s} \tilde{T}^\dag_\mbf{a} = \tilde{f}^\dag_{\mbf{R}+\mbf{a}, \al \eta s}$ and $\tilde{T}_\mbf{a} \tilde{c}^\dag_{\mbf{k}, a \eta s} \tilde{T}^\dag_\mbf{a} = e^{- i \mbf{k} \cdot \mbf{a}}\tilde{c}^\dag_{\mbf{k}, a \eta s}$. The advantage of this basis is that $\mbf{k}$, the IKS momentum, is a good quantum number even though moir\'e translations are broken. 
The Hamiltonian takes the same form in the boosted operator basis provided we replace $\mbf{k} \to \mbf{k} - \eta\frac{\mbf{q}}{2}$; this only affects the single particle kinetic term (as discussed in the next paragraph and in \App{app:boostH}).

We can now define the IKS parent state
\bea
\ket{\text{IKS}^0_\mbf{q},\nu\!=\!-3} &= \prod_{\mbf{R}} \tilde{f}^\dag_{\mbf{R}}\zeta_{+,\uparrow}\ket{\tilde{\text{FS}}},
\eea
where $\ket{\tilde{\text{FS}}}$  is the  Fermi sea filling the negative energy $\tilde{c}$-electron states of the $H_c$, again in the absence of heterostrain and $\mu$ terms, i.e. up to their quadratic band touching points boosted by $\mbf{q}/2$ in valley $K$ and by $-\mbf{q}/2$ in valley $K'$. Because there is no $f$-$c$ hybridization in the parent states, their kinetic energy $\braket{H_0}$ is actually independent of $\mbf{q}$. This is because the single-particle energy of the $f$-modes is local and thus left unchanged by the valley boost $f^\dag_{\mbf{R}, \al \eta s} \to f^\dag_{\mbf{R}, \al \eta s} e^{-i \mbf{R} \cdot \mbf{q} \eta/2}$ (see $H_f$ in \Eq{eq:Hfindices}). The kinetic energy of the $c$-electrons is also unchanged by the boost since $\delta h_K$ is local and the filling is still up to the quadratic band touching. Moreover, because the parent state contains no $f$-$c$ hybridization, the expectation value of $H_{fc}$ vanishes by Wick's theorem. The interaction energy of the parent states, $\langle H-H_0\rangle$, is also independent of $\mbf{q}$. This is apparent for the $H_U, H_W, H_V$ because they are local density-density terms. To see this for $H_J$ as well, we note that it is local, charge neutral and invariant under valley $U(1)$ transformations, as shown carefully in \App{app:boostH}. Thus the \emph{parent state} energies are equal at all values of $\mbf{q}$. However, upon Hartree-Fock iteration, the descendants of parent states with different $\mbf{q}$ can, and do, have different energies. 
This can be seen by noting that the Hartree-Fock Hamiltonian obtained from the parent states (referred to as ``one-shot'') has the form
$\left(\begin{array}{cc} h_K(\mbf{k}+\frac{\mbf{q}}{2})+\mathcal{E}_K & \Sigma_{KK'}\\
\Sigma^\dagger_{KK'} & h_{K'}(\mbf{k}-\frac{\mbf{q}}{2})+\mathcal{E}_{K'}\end{array}\right)$, where we highlighted the valley space structure as well as the $\mbf{k}$ and $\mbf{q}$ dependence. We also absorbed the strain and $\mu$ terms into the intravalley self-energies $\mathcal{E}_\eta=\delta h_\eta+ \Sigma_{\eta \eta}$, which are  $\mbf{k}$ and $\mbf{q}$ independent when we ignore the small Fock term contribution of $H_V$. Because the $f$-modes are inter-valley coherent in the parent states, the self energy $\Sigma_{KK'}$ is nonzero (it is also $\mbf{k}$ and $\mbf{q}$ independent at this step). Therefore, the eigenstates of the above Hartree-Fock Hamiltonian give rise to the single particle correlation matrix, $\left(\begin{array}{cc}\langle \tilde{c}^\dag_{\mbf{k}, \mu}\tilde{c}_{\mbf{k}, \nu}\rangle & \langle \tilde{c}^\dag_{\mbf{k}, \mu}\tilde{f}_{\mbf{k},\nu}\rangle \\ 
\langle\tilde{f}^\dag_{\mbf{k}, \mu}\tilde{c}_{\mbf{k}, \nu}\rangle
& \langle\tilde{f}^\dag_{\mbf{k}, \mu}\tilde{f}_{\mbf{k}, \nu}
\rangle \end{array}\right)$, whose elements depend on {\em both} $\mbf{k}+\mbf{q}/2$ and $\mbf{k}-\mbf{q}/2$. Such correlation matrix then results in the $\mbf{q}$ dependence of the total energy. Thus we see that $\mbf{q}$ is a variational parameter in the self-consistent energies even though heterostrain has fixed the orbital degree of freedom in $\zeta$. In the following section, we will show how tuning $\mbf{q}$ is able to annihilate the Dirac nodes at $|\nu| = 2$.

At $\nu = -2$, our ansatz for the IKS parent state follows similar reasoning. We first consider the spin degrees of freedom. Any spin polarized ferromagnetic state (two $\uparrow$ $f$-modes per unit cell) fully fills the negative energy $\zeta$ eigenstates in both valleys. But since $\tilde{f}^\dag_{\mbf{R}}\zeta_{+,\downarrow}\tilde{f}^\dag_{\mbf{R}}\zeta_{-,\downarrow}= -\tilde{f}^\dag_{\mbf{R}}\zeta_{K,\downarrow}\tilde{f}^\dag_{\mbf{R}}\zeta_{K',\downarrow}=-f^\dag_{\mbf{R}}\zeta_{K,\downarrow}f^\dag_{\mbf{R}}\zeta_{K',\downarrow}$, this means there is no inter-valley coherence, so all $\mbf{q}$ dependence of the $f$-modes factors out of the wavefunction, removing $\mbf{q}$ as a variational parameter. Thus, the fully spin polarized ferromagnet cannot hope to remove the Dirac nodes and, as we confirm numerically, leads to a higher energy gapless state.
We are led to consider the spin unpolarized sector to access the variational freedom afforded by $\mbf{q}$. We propose
\bea
\label{eq:IKSminus2}
\ket{\text{IKS}^0_\mbf{q},\nu\!=\!-2} &= \prod_{\mbf{R}} \tilde{f}^\dag_{\mbf{R}}\zeta_{+,\uparrow}\tilde{f}^\dag_{\mbf{R}}\zeta_{+,\downarrow}\ket{\tilde{\text{FS}}}, 
\eea
which is simply a copy of the $\nu = -3$ IKS parent state in both spin sectors. Lack of spin polarization is supported by experiment \cite{Cao2018TBGMott,YAN19,2023arXiv230906583H}. Again, one can independently rotate the spins in each valley to generate the full ground state manifold.

The observation that a loss of IVC in the ferromagnetic sector removes $\mbf{q}$-dependence from the wavefunction is important as we reach $|\nu| = 0,1$. At filling $\nu = -1$, adding another negative energy $f$-mode to the $\downarrow$ sector will remove IVC there, leaving a single copy of the IKS in the $\uparrow$ sector. Hence the parent state wavefunction at $\nu = -1$ is given by
\bea
\ket{\text{IKS}^0_\mbf{q},\nu\!=\!-1} &= \prod_{\mbf{R}} \tilde{f}^\dag_{\mbf{R}}\zeta_{+,\uparrow}\tilde{f}^\dag_{\mbf{R}}\zeta_{+,\downarrow}\tilde{f}^\dag_{\mbf{R}}\zeta_{-,\downarrow}\ket{\tilde{\text{FS}}} \\
&\propto \prod_{\mbf{R}} \tilde{f}^\dag_{\mbf{R}}\zeta_{+,\uparrow}f^\dag_{\mbf{R}}\zeta_{K,\downarrow}f^\dag_{\mbf{R}}\zeta_{K',\downarrow}\ket{\tilde{\text{FS}}} \ . \\
\eea
Note that the spin $\downarrow$ sector has no spontaneously broken symmetries. This will have an important effect on the Dirac nodes, which we will elaborate on next. 

Lastly, at $\nu = 0$, we fill all four negative energy eigenvectors (one in each spin and valley). This state has no IVC, and thus there is no configuration of spins where $\mbf{q}$ allows variational freedom. Hence, unlike $\nu = -3,-2,-1$ where the parent states can use IKS order to enlarge the variational space, at $\nu = 0$ they cannot. The parent state is merely a semi-metal with externally broken $C_3$ symmetry and possesses no broken global symmetries:
\bea
\ket{\text{SM}^0,\nu\!=\!0} &= \prod_{\mbf{R}} f^\dag_{\mbf{R}}\zeta_{K,\uparrow}f^\dag_{\mbf{R}}\zeta_{K,\downarrow}f^\dag_{\mbf{R}}\zeta_{K',\uparrow}f^\dag_{\mbf{R}}\zeta_{K',\downarrow}\ket{\text{FS}} \ .
\eea
Because there are no spontaneously broken symmetries at $\nu=0$, i.e. $U(1)$ valley and $C_{2z}\mathcal{T}$ are preserved, the state must be gapless due to Dirac nodes. This is because the Dirac nodes in the two flat bands carry the same winding numbers \cite{PO18a,ZOU18,SON19,AHN19,KAN20a} and cannot be annihilated without breaking other symmetries or involving more bands. We show this concretely in \App{app:nu0SMdirac} by analyzing the one-shot Hartree-Fock Hamiltonian, which we show takes the form of a Bernal bilayer graphene $k \cdot p$ Hamiltonian, and hence has chiral winding number 2. The gaplessness at $\nu=0$, also confirmed beyond mean-field \cite{PAR21a}, is a dramatic difference compared to the strong coupling strain-free phase diagram, where $\nu=0$ has the \emph{largest} correlated gap \cite{BER21b,BUL20a}. 

At positive fillings, the parent state follow from particle-hole transformations on the $f$-modes (see \Fig{figparentstate}). At zero relaxation this particle-hole transformation is exact, and in the presence of relaxation we will find the same parent states are valid, although their excitation gaps will show particle-hole breaking. We address the particle-hole asymmetry in the Sec.\ref{sec:mainPH}. It is also worth remarking that the $|\nu| = 2,3$ insulators, which preserve $C_{2z}$ and $\mathcal{T}$, will have vanishing anomalous Hall signal, consistent with experiments. 

This heavy fermion construction of the phase diagram is in agreement with the ``basis spiral" picture proposed in Ref.\cite{KWA21}, which we now substantiate numerically. 

\section{Hartree-Fock.} Having written out the parent states at all fillings, 
the important question of choosing $\mbf{q}$ remains. To address this question, we first study analytical one-shot and numerical self-consistent Hartree-Fock calculations. As in the unstrained case, we establish that the one-shot results capture the self-consistent Hartree-Fock results very well, hence validating our proposed parent states. 

\Fig{figIKSHF}(a,b) shows the band structure of self-consistent and one-shot HF at $\nu=-2, \eps = 0.15\%$ and $\mbf{q} = \frac{2}{3}M_M$ where $M_M$ is the moir\'e $M$ point, demonstrating remarkable agreement with the self-consistent bands exhibiting a larger gap but with otherwise extremely similar features. The occupied bands contain the two $f$ electrons in the parent state hybridized with $c$-modes, and the electron excitations are heavy (red for $f$-like) while the hole electrons are much lighter (blue-green).  A simple understanding of the band structure can be obtained from the energetics of the $f$-modes, which are decoupled from the $c$-electrons at the BZ edge. The Hartree-Fock Hamiltonian describing these modes is (see \App{app:projected})
\bea
h_{HF,f}[O_f] &= - U_1 (O_f^T - \frac{\mathbb{1}}{2}) + \nu_f (U_1 + 6 U_2) + M_f \eps_- \sigma_2 \tau_3
\eea
for a general $f$-mode order parameter $O_f$ and $\nu_f = \Tr (O_f - \frac{\mathbb{1}}{2})$. The spectrum can be obtained analytically for $O_f = \sum_s \zeta_{+,s} \zeta_{+,s}^\dag$, giving energies (each is spin degenerate)
\bea
\label{eq:fenergies}
E_f = - U_1 - M_f |\eps_-|, \ - M_f |\eps_-|, \ M_f |\eps_-|, \ M_f |\eps_-|
\eea
relative to the Fermi level which is approximately $-\frac{3}{2}U_1 - 12 U_2$. The spin-degenerate occupied levels are $- U_1 - M_f |\eps_-|$ with wavefunctions $\zeta_{+,s}$  at $\nu=-2$, and the remaining three spin-degenerate levels (six in total) are unoccupied. This explains the splittings observed in the numerical spectrum as also illustrated in \Fig{fig2}(a,b). The wavefunctions corresponding to lowest two energies in \Eq{eq:fenergies} are $\zeta_{+,s}$, $\zeta_{-,s}$ which have maximal IVC, whereas the higher two degenerate levels can be chosen to have no IVC in their wavefunctions. 

\Fig{figIKSHF}(c,d) show the one-shot gapless spectra at $\nu=-1,0$ respectively (self-consistent band structures may be found in \App{app:oneshotstrain} and are also markedly similar). Lastly, we can also compare the $f$-order parameter 
\bea
\null [O_f]_{\al \eta s, \be \eta' s'} = \frac{1}{N} \sum_\mbf{k} \braket{\tilde{f}^\dag_{\mbf{k}, \al \eta s} \tilde{f}_{\mbf{k}, \be \eta' s'}}
\eea
computed from one-shot HF applied to the heavy-fermion model to the same order parameter calculated directly in the 6-band self-consistent HF calculations in the BM model (see \App{app:sec:hf_in_bm}) at $\nu=-2$ denoted by $O_{f,\infty}^{BM}$. We find strong agreement quantified by $||O_{f,1}- O_{f,\infty}^{BM}||/||O_{f,\infty}^{BM}|| = 0.07$. Similar accuracy is found for $\nu = -1, 0$ (see \App{app:oneshotstrain}). This constitutes strong numerical support for the validity of the parent states and one-shot HF.

We also observe in \Fig{figIKSHF} that the $\nu=-2$ is gapped while $\nu=-1,0$ are gapless. The difference stems from the parent states having broken valley symmetry in both spin sectors at $\nu=-2$, but not at $\nu =-1$ (which has unbroken valley symmetry in one spin sector) and $\nu=0$. We will explain these features and emphasize their topological origin in what follows.

\section{IKS Mechanism}

As argued, the remaining variational freedom admitted by the parent states in heterostrain is the boost vector $\mbf{q}$. We will show that at $\mbf{q}=0$, the ferromagnetic exchange interaction $J$ ensures a gapless phase. Then we will show that a \emph{nonzero} IKS boost $\mbf{q}$ is able to open a gap, and that the optimal $\mbf{q}$ vector can be estimated from the $f$-$c$ decoupled limit by maximizing the hybridization gap. Throughout, we will operate under the assumption, guided by experiment and numerical calculations, that a gapped state is favored over nearby gapless states within Hartree-Fock. Although it is difficult to obtain the total Hartree-Fock energy analytically, comparing the direct gap between different orders is a good heuristic. This is because opening a quasi-particle gap at the Fermi level is expected to lower the total energy, since the occupied states decrease in kinetic energy while the unoccupied states gain kinetic energy. 
At zero heterostrain for instance, the KIVC opens a larger gap than the valley-polarized state and indeed is favored in self-consistent Hartree-Fock\cite{Song20211110MATBGHF}.

\subsection{Topological Protection of the $\mbf{q}=0$ Gapless Phase.}
\label{sec:q0IKS}

\begin{figure}
\centering
\includegraphics[width=\columnwidth]{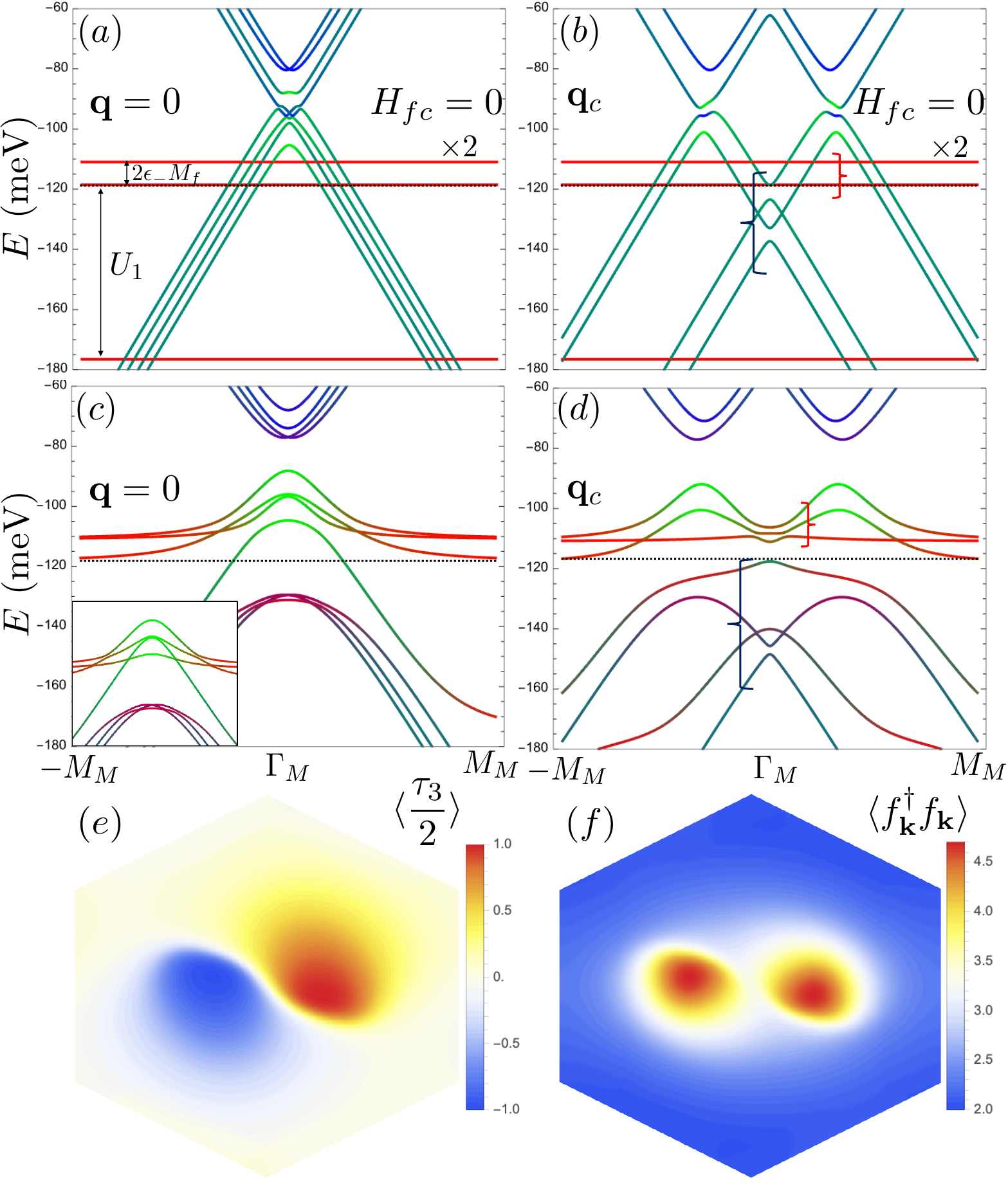}
\caption{One-Shot condition for the boost vector at $\nu  =-2$. (The spectra in the spin $\uparrow$ and $\downarrow$ are identical.) $(a)$ shows the one-shot $\mbf{q}=0$ band structure, which is gapless due to the hole pocket.  The high-energy $f$ modes (2 per spin) are split from the unoccupied low energy $f$-modes (1 per spin) by the heterostrain term $2 \eps_- M_f$. The occupied $f$-modes (1 per spin) are lowered by the large Hubbard $U_1$. $(b)$ shows the unhybridized one-shot band structure with the IKS boost obtained from the analytical estimates in \Eqs{eq:mainEf}{eq:mainEc}. The three conduction $f$-modes are marked with the red bracket, and the four valence $c$-modes are marked with the black bracket. $(c)$ Without the IKS boost, the hybridized band structure cannot develop a gap due to the Dirac nodes. The inset shows the topological semi-metallic band touching points. $(d)$ After hybridization, a full gap opens at the Fermi level (dotted). The boost vector $\mbf{q}_c = .61 M_M$, which is close to the value of $\frac{2M_M}{3}$ used in \Fig{figIKSHF}, where the self-consistent Hartree-Fock spectrum is plotted for comparison. Note that the indirect gap is nonzero, and is enlarged at self-consistency. $(e)$ and $(f)$ show the expectation value of the valley charge $\tau_3/2$ and $f$-mode occupation number in the self-consistent THF ground state at $\nu=-2$. The edge of the BZ is fully IVC, whereas the lobes created by the $c$-electron holes display valley polarization.}
\label{fig2}
\end{figure}

We will first prove that the $\mbf{q}=0$ one-shot spectrum is gapless. \Fig{fig2} shows the one-shot $\nu=-2$ band structure at $\mbf{q}=0$, which reveals a gapless state with a large hole pocket at the $\Gamma$ point where the $\Gamma_1\oplus\Gamma_2$ $c$-electrons are pushed above the Fermi surface (see \Fig{fig2}a). Moreover, the direct gap also vanishes due to Dirac nodes between the valence and conduction band off the high-symmetry lines as seen in the inset of \Fig{fig2}(c).

To explain this analytically, we develop a flat band projection method in a $k \cdot p$ expansion. In \App{app:projected}, we obtain expressions for the 4 bands (per spin) near the Fermi level in a degenerate perturbation theory about the flat band limit. Explicitly, we first find analytical eigenstates for the non-interacting flat bands at $M=0$ in the absence of heterostrain and $\mu$ terms. We then project the one-shot Hartree-Fock Hamiltonian including the heterostrain terms onto these states. Firstly, the projected single-particle Hamiltonian is (in the $K$ valley)
\bea
h_{0,\text{eff}}(\mbf{k}) &= v \eps_- \frac{2 c'' \gamma k_y \sigma_0 + M_f v (-2 k_x k_y \sigma_1 + (k_x^2-k_y^2) \sigma_2)}{v^2 |\mbf{k}|^2 + \gamma^2}  \\
&\quad +  \frac{M \gamma^2}{v^2 |\mbf{k}|^2 + \gamma^2} \sigma_1.
\eea
Evidently, at $M=0$ the above Hamiltonian has a winding number $2$ and a quadratic band touching at ${\bf k}=0$, similar to Bernal bilayer without trigonal warping terms, albeit with a tilt due to the first term proportional to the unit matrix $\sigma_0$. At $M\neq 0$, the last term is akin to a nematic order parameter added to the Bernal bilayer\cite{PhysRevB.81.041401}. It splits the quadratic band touching into two Dirac nodes, but the total winding around a loop that includes both nodes is still $2$. This can be seen graphically by plotting the contours of constant $k_xk_y$ (hyperbolas) and the zero of $k^2_x-k^2_y$ (diagonal lines) and looking for intersections. At $M=0$ there is a double intersection of two ``crosses'' at ${\bf k}=0$ while at $M\neq 0$ there are two intersections corresponding to two Dirac nodes.
Thus, we see that both Dirac cones in the flat bands have the same winding number. This is important because it prevents a gap from opening when the Dirac nodes collide at the $\Gamma$ point, since they have the same winding number: in this two-band Hilbert space with unbroken spin and valley symmetries, it is impossible to annihilate the Dirac nodes.

The potential to open a gap is realized at the interacting level in the IVC parent state \Eq{eq:IKSminus2}, where the Hamiltonian now includes four bands thanks to the breaking of valley $U(1)$ symmetry. Nevertheless, we can prove that the Hartree-Fock bands are still gapless due to the existence of Dirac nodes when the boost vector $\mbf{q} = 0$ by examining the $C_{2z}$ eigenvalue of the valence band at the $\Gamma$ point. In perturbation theory on the Hartree-Fock Hamiltonian, we find the analytical expressions
\bea
E_{-,s}(\mbf{k}=0) &= -2W_3\pm M,\;\;\;\left(C_{2z}\;\text{odd}\right) \\
E_{+,s}(\mbf{k}=0) &= \frac{J}{2}-2 W_3\pm M, \;\;\;\left(C_{2z}\;\text{even}\right) \\
\eea
for the energies of the $C_{2z} = -1$ and $+1$ eigenvalues respectively.  At $\nu = -2$, one band should be occupied in each spin sector. We see that for all $M, W_3$, the $C_{2z} = +1$ energies are higher due to the ferromagnetic coupling $J > 0$. Thus the valence band will always have the $C_{2z} = -1$ irrep at the $\Gamma$ point. Next we recall that at the three $M$ points on the BZ edge, the band is made up of the heterostrain-split $f$ modes with $C_{2z} = +1$ (see \Eq{eq:zetaplus}). Thus the product of the $C_{2z}$ eigenvalues in the valence band is odd, which, along with spin-less $\mathcal{T}$, enforces a gapless point --- it is a symmetry-enforced topological semi-metal\cite{Bradlyn2017TQC}. This is because an odd product of $C_{2z}$ eigenvalues enforces an odd Chern number for an isolated band\cite{2011PhRvB..83x5132H}, whereas $\mathcal{T}$ enforces zero Chern number. This incompatibility requires a gapless point. Thus we have proven that the parent state in \Eq{eq:zetaplus}, selected by heterostrain, yields a gapless phase at $\mbf{q}=0$ which is symmetry-indicated and protected energetically by the ferromagnetic exchange $J$. It remains to show that $\mbf{q}\neq 0$ can open this gap.

\subsection{Gap Opening from Non-abelian Dirac Node Braiding}

\begin{figure} \centering \includegraphics[width=\columnwidth]{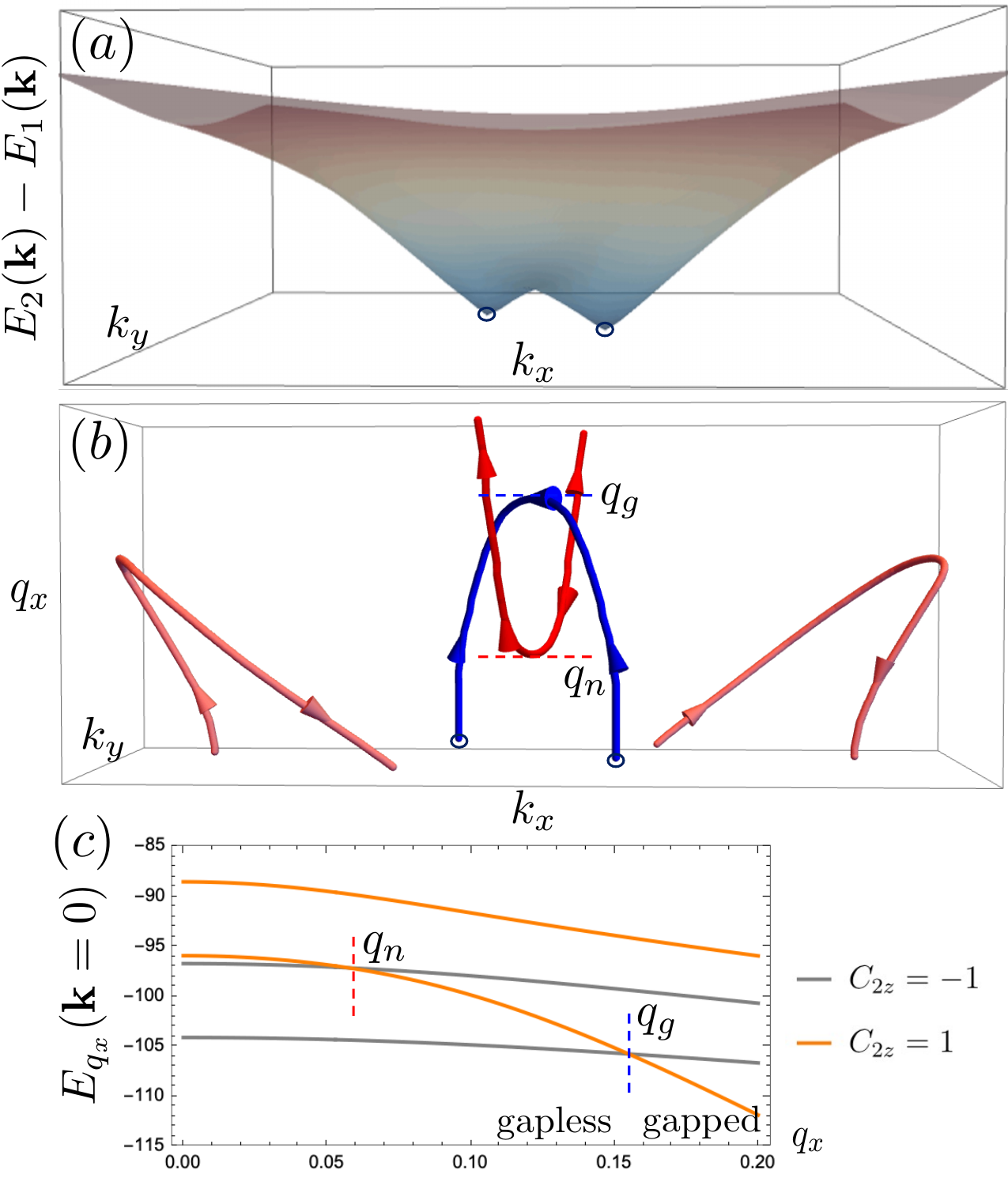} \caption{
Non-abelian Dirac node braiding.
$(a)$ Difference in energy between the valence and conduction one-shot projected Hartree-Fock bands at ${\bf q}=0$, with the two Dirac nodes marked by the small black circles. $(b)$ Braiding of Dirac nodes as a function of $q_x$, showing nucleation of a node/anti-node pair at $q_n$ in the conduction manifold followed by braiding and annihilation of the valence band node at $q_g$, where a direct gap opens at $\Gamma$. Analytical expressions are shown in \Eq{eq:qexpressions}. $(c)$ Level crossing of inversion eigenvalues demonstrates the $C_{2z}$ obstruction being resolved in the Dirac node braiding process. The energy ordering of the $C_{2z}$ eigenvalues at the $\Gamma$-point (${\bf k}=0$) for vanishing boost, ${\bf q}=0$, is robust due to the positive (ferromagnetic) sign of $J$. Because this ordering is inverted relative to the $M_M$-point, the absence of the gap is guaranteed at $q_x=0$.
} 
\label{fig_node}
\end{figure}

To open a gap, the ground state must undergo a level crossing where the $C_{2z} = -1$ eigenvalue of the highest valance band at the $\Gamma$ point is exchanged for a $C_{2z} = +1$ eigenvalue. We can demonstrate analytically that such a level crossing occurs by turning on a boost $\mbf{q} = (q_x,0)$, carrying out the same degenerate perturbation theory as in the prior section at the $\Gamma$ point  (for analytical expressions, see \App{app:projected}). For now, in order to obtain transparent analytical expressions, we restrict consideration to $\mbf{q} = (q_x,0)$ along the axis of the applied heterostrain.) This shows that the $C_{2z}$-protected obstruction can be resolved by the IKS boost. This is a necessary but not sufficient condition for the opening of a gap.

To elucidate the gap opening mechanism, we will use the theory of $C_{2z}\mathcal{T}$-protected non-abelian Dirac node braiding. Recall that in two-band models, Dirac nodes are associated with an integer winding number, which is essentially the (signed) number of Berry curvature monopoles enclosed in a loop taken around the node. Typical examples include graphene, which has Dirac nodes with opposite winding number $\pm1$ at the $K$ and $K'$ points, or Bernal bilayer graphene, with winding number $\pm 2$. However, in multi-band models, the abelian classification of Dirac nodes breaks down. Ref. \cite{Tomas2019NonAbelianNodalLine} demonstrated that in multi-band systems with $C_{2z}\mathcal{T}$, Dirac nodes are described by a generalized quaternionic charge. Succinctly, we can assign a (non-abelian) charge $e_n$ to a Dirac node between band $n$ and $n+1$ such that $e_n^2 = -1$, $\{e_n,e_{n+1}\} = 0$ and $[e_n,e_{m}] = 0$ if $|n-m| > 1$. Two nodes can be annihilated if they have charge $-e_n,e_n$ whose product is 1. Since the charges in neighboring bands anti-commute, the braiding of nodes in adjacent bands can flip a sign. Hence we can understand a non-abelian braiding process 
\begin{align}
e_n e_n &= e_n e_{n+1} (-e_{n+1}) e_n \\
&= e_n e_{n+1} e_n e_{n+1} = e_{n+1} (-e_n ) e_n e_{n+1} = e_{n+1} e_{n+1} \nonumber
\end{align}
where a node-anti-node pair is nucleated in the $n+1$th gap (first line) and undergoes braiding with a pair in the $n$th gap, allowing them to flip sign and annihilate. This process exactly describes the opening of the Hartree-Fock gap driven by the IKS boost. \Fig{fig_node}(a) shows the presence of two Dirac nodes between the valence and conduction bands. In \Fig{fig_node}(b), these nodes ($e_n e_n$) are shown in blue and the nodes between the lowest two conduction bands are shown in red as a function of the IKS boost $q_x$. The sign of $e_n, e_{n+1}$ are denoted by arrows. The arrows reverse sign when their curves passes under a curve of an adjacent gap, capturing their anti-commutation property. We observe a braiding structure where the red nodes nucleate at a finite $q_x=q_n$ (marked by the dashed red line), which allows the blue nodes to subsequently annihilate and open a gap between the valence and conduction bands at $q_x=q_g$ (marked by the dashed blue line). We can obtain simple expressions for the values of $q_g$ and $q_n$ in the projected one-shot model. They are
\bea
\label{eq:qexpressions}
q_n &= 2 \sqrt{\frac{J - 4 M}{U_1}} \frac{\gamma}{v}, \ \ q_g = \frac{J+M}{\sqrt{J U_1/2}}  \frac{\gamma}{v} \ . 
\eea
The scale $q_g$ is an estimate of the minimal $q_g$ such that a direct gap can be opened in Hartree-Fock. In the next section, we will discuss the indirect gap.

In summary, we have shown that the parent state favored by heterostrain cannot open a gap when the boost vector vanishes, $\mbf{q}=0$, due to a symmetry obstruction, but a direct and indirect gap can be opened at some non-zero $q_x$. This explains the appearance of IKS order, but does not yet determine which value of ${\bf q}$ is optimal. For completeness, in \App{app:compare} we use the same symmetry analysis on the $U(4)\times U(4)$ states, which are ground states without strain at $\nu=-2$, to show that they are gapped without a boost.

\subsection{Criterion for the IKS Boost.}

\begin{figure}
\centering
\includegraphics[width=\columnwidth]{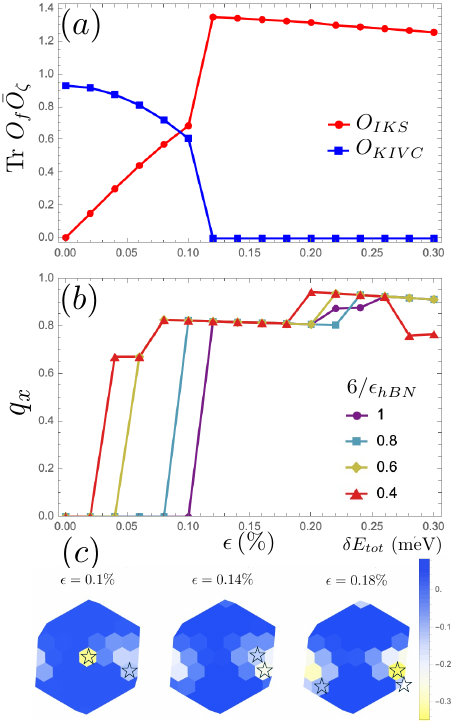}
\caption{6-band BM model simulations. $(a)$ Transition from the KIVC to IKS in the $f$-order parameter $\Tr O_f \bar{O}_\zeta$ where $\bar{O}_\zeta$ is the trace-less part of the parent state order parameters, obeying $\Tr \bar{O}_{IKS} \bar{O}_{KIVC} = 0$. The $f$ mode order parameter is extracted directly from the BM simulations. We use dielectric constant $\eps_{hBN} = 6$. $(b)$ Optimal $\mbf{q}_{IKS}$ as a function of interaction strength $6$, showing a sharp transition to $q_x \sim \frac{2M_M}{3}$ above a critical heterostrain which scales linearly with interaction strength, and then slowly rises with increasing heterostrain. $(c)$ Hartree-Fock total energy as a function of $\mbf{q}$ ($\delta E$ is measured relative to the average at each $\eps$) . At large heterostrain, many different $\mbf{q}$ in the $M_M-K_M$ region compete closely in energy. The stars mark $\mbf{q}$ which have a nonzero indirect gap.
}
\label{fig:BMsim}
\end{figure}

The optimal $\mbf{q}_c$, at the one-shot level, is heuristically expected to be the one which maximizes the quasi-particle energy gained by opening a gap at the Fermi level. From \Fig{fig:BMsim}(c), we indicate which solutions in self-consistent BM simulations have nonzero indirect gaps. Surprisingly, very few IKS boosts $\mbf{q}$ show an indirect gap at all, and their locations are sharply correlated with the lowest energy ground states. Based on this numerical observation, we now devise a method to estimate the  optimal boost $\mbf{q}_c$ which opens the largest indirect gap.

To understand the degrees of freedom available at the Fermi level, we turn to the unhybridized limit where $H_{fc} = 0$. Let us first understand the $\mbf{q}=0$ state from this perspective, whose one-shot band structure is shown in \Fig{fig2}(a). At $\nu = -2$, we must first fill the 2 lowest $f$-modes (1 per spin), which are pushed far down in energy by $U_1$. We will refer to these $f$-modes as the valence $f$-modes, and the higher energy $6$ $f$-modes as conduction $f$-modes. Note that the conduction $f$-modes are split by heterostrain. To achieve overall filling $\nu = -2$, we expect to fill the Dirac-like $c$-electrons up to their charge-neutrality point. However, we see from \Fig{fig2}(a) that there is a significant hole pocket around $\Gamma$ that prevents filling the $c$-electrons up to their charge neutrality point because there are lower energy conduction $f$ modes. Instead, the $f$-modes near the Fermi level will be partially occupied in the first descendant state. A gap cannot be opened even after hybridization as shown in \Fig{fig2}(c) and as explained in \Sec{sec:q0IKS}. 

Let us then consider \Fig{fig2}(b) where a boost is introduced and hole pockets from valley $K, K'$ are shifted relative to each other. Due to the approximately linear dispersion of the $c$-electrons, turning on the boost will lower their energy at the $\Gamma$ point. For a range of boosts, $c$-electron modes near $\Gamma$ are brought down in energy to the vicinity of the conduction $f$-modes, allowing them to be filled. Upon introducing hybridization, the crossings of the $f$- and $c$-modes at generic points in the BZ will turn into avoided crossings and a gap opens since the semi-metal obstruction has been resolved (as explained in \Sec{sec:q0IKS}).

We now address of the question of which boost, $\mbf{q}_c$, is the most favorable. We propose that $\mbf{q}_c$ can be approximated from the unhybridized limit ($H_{fc} = 0$). In this limit (see \Fig{fig2}(b)), the $f$ and $c$ bands cross, and will open a hybridization gap as $H_{fc}$ is turned on. We will restrict our attention to the $\Gamma$ point motivated by our discussion of the $C_{2z}$ eigenvalues which must be exchanged to obtain a gapped phase.

To maximize this gap, we propose that $\mbf{q}_c$ should be picked so that the highest of the 4 low-energy $c$ modes $E_{c,\mbf{q}}(\mbf{k}=0)$ is just at the Fermi level. In this case, the off-diagonal $H_{fc}$ term leads to the greatest hybridization. This is a heuristic principle which allows us to estimate $\mbf{q}_c$ in terms of the analytically accessible unhybridized quasi-particle energies. This  condition for $\mbf{q}_c$ can be seen in \Fig{fig2}(b), where all 4 low-energy $c$-modes are below the Fermi level unlike in \Fig{fig2}(a). The Fermi level is set by the lowest unoccupied $f$-mode at energy 
\bea
\label{eq:mainEf}
E_{f} &= M_f\eps_--\frac{3 U_1}{2}-12 U_2 \\
\eea
denoted by the dashed line in \Fig{fig2}(a). In \App{app:optimalboostfc}, we find that the highest $c$-electron energy at $\mbf{k}=0$ which must be in the valence manifold for a gapped state is 
\bea
\label{eq:mainEc}
E_{c}(q_x) &\approx -\frac{|v q_x| }{2} + \frac{J}{4} - (W_1 + W_3) \\
&\qquad + \frac{1}{2} \sqrt{M^2 + (M' \eps_+ + (2 c'+ c)   \eps_-)^2} 
\eea
up to corrections of order $O(1/|v q_x|)$.   Finally, the criterion $E_f = E_c(\mbf{q})$ gives an estimate of $q_c \sim 0.6 M_M$ at $\eps = 0.2\%$, and varies between $0.52M_M$  and $0.68M_M$ for $\eps \in (0.1\%, 0.3\%)$. Note that $q_c \approx 0.66 M_M$ corresponds to a tripling of the moir\'e unit cell, which is in good agreement with experiment \cite{NUC23}. We can also verify this analytical approximation with numerical calculations within the full 6-band BM model from which the THF Hamiltonian is derived. The results are shown in \Fig{fig:BMsim}. We observe that the optimal $\mbf{q}$ shows a broad minimum which deviates slightly off the $x$-axis at large $\eps$. We leave modeling of these finer details for future studies. 

The unhybridized limit is also a useful way to understand the $\mbf{k}$-dependent valley polarization that appears in the ground state as shown in \Fig{fig2}(e). At the BZ edge where the occupied $f$-modes are polarized along $\zeta_{+}$, no valley polarization appears due to perfect IVC. However, regions of valley polarization, termed ``lobes'' in Ref. \cite{KWA21}, appear in the BZ corresponding to the regions of depopulated $c$-electrons (see \Fig{fig2}(f) for the corresponding increase in $f$-electron occupation). This is plain to see from the unhybridized limit where the $c$-electrons in valley $K'$ are high energy at $\mbf{k} =+\mbf{q}/2$, leading to a greater density of valley $K$ in the ground state, and vice versa at $\mbf{k} = -\mbf{q}/2$. 

\section{Particle-Hole Asymmetry.} 
\label{sec:mainPH}
\begin{figure*}
\centering
\includegraphics[width=17.5cm]{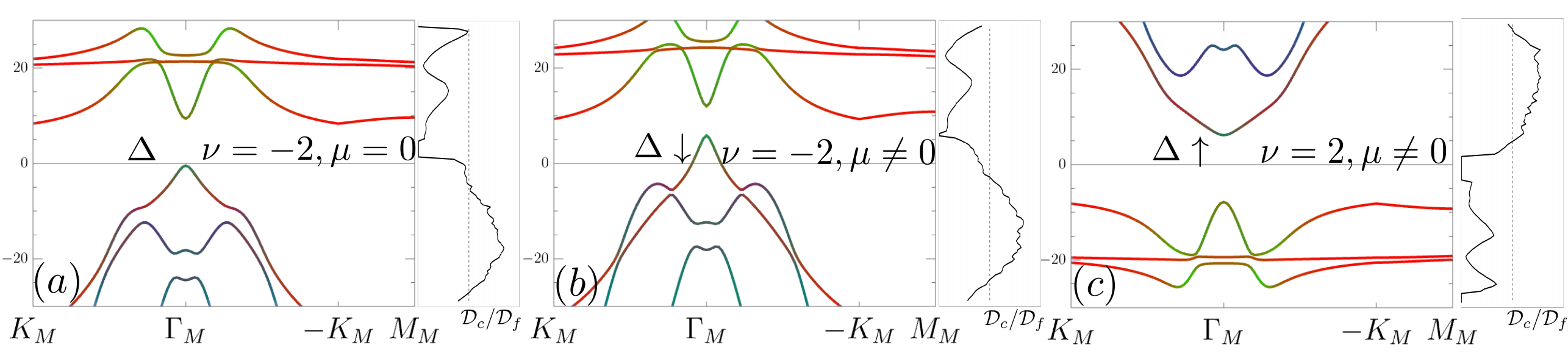}
\caption{$(a)$ Zoom-in of the particle-hole symmetric self-consistent bands at $\nu=-2$. Particle-hole breaking potentials (denoted $\mu \neq 0$ to indicate the inclusion of $\mu_1,\mu_2$) raise the energy of the $c$-electrons (but do not affect the $f$-electrons), leading to a decrease of the gap at $\nu=-2$ in $(b)$ and an increase of the gap at $\nu = +2$ in $(c)$. We show the ratio of the $c$-electron and $f$-electron density of states $\mathcal{D}_c/\mathcal{D}_f$ to explain why raising the chemical potential of the $c$ electrons must decrease the gap at $\nu=-2$ and increase it at $\nu=+2$. The dashed line marks $\mathcal{D}_c/\mathcal{D}_f =1$.
}
\label{figPH}
\end{figure*}
So far we have not incorporated the particle-hole breaking potentials. Both $\mu_1$ and $\mu_2$, the $c$-electron chemical potentials, should be viewed as small corrections since they do not break the $U(4)\times U(4)$ symmetry of the $M=v'=0$ THF model. Thus we can understand their effect in one-shot Hartree-Fock from first order perturbation theory. \Fig{figPH}(a) depicts the IKS band structure at $\nu = -2$ ($\nu =+2$) near the Fermi level, where the $c$-modes dominate the density of states below (above) the gap. This is one of the key physical features of the THF model, explaining the heavy excitations towards charge neutrality and the light excitations away\cite{KAN21,2023NatCo..14.5036D,PhysRevB.103.205415,Song20211110MATBGHF,2023arXiv230908529R,2023PhRvL.131b6502H,Hu2023KondoMATBGStrain,2024arXiv240214057C,2024arXiv240211749L,2024arXiv240212296B}. Hence by adding the positive $\mu_1,\mu_2$ potentials to the Hamiltonian which raise the energy of the $c$ electrons but do not affect the $f$-electrons, the gap will decrease at $\nu = -2$ where the $c$ electrons are occupied, but will \emph{increase} at $\nu = +2$ where the $c$ electrons are \emph{unoccupied}. Indeed,  \Fig{figPH}(b,c) show that the orbital makeup of states near the Fermi level is almost unchanged, and it is only the gap size which is altered. Quantitatively in self-consistent Hartree-Fock, we find that the $\nu=\pm2$ gap is $\sim 6$meV for $\mu_1 = \mu_2=0$. When the potentials $\mu_1=14.5$meV and $\mu_2 = 4.5$meV are turned on, the $\nu=-2$ gap shrinks to $\sim 3$meV while the $\nu = +2$ gap rises to $\sim 12$meV. While these gaps will decrease in dynamical mean field theory or other more accurate treatments, the asymmetry in the gaps at $\pm \nu$ will remain. This clearly shows that finer features of TBG phase diagram can be understood in a simple, analytical fashion and traced to their microscopic origin. Here we have elucidated why relaxation, which increases the dispersion on the electron-side of the single-particle band structure by raising the $c$-electron energy, leads to larger correlated gaps on the electron-side than hole side. This is in contrast to naive expectations that the more dispersive single-particle band is more weakly correlated, and aligns with experiment. 

\section{Conclusion} 

We have used the heavy fermion model of TBG to explain why small heterostrains, on the order of $0.1$\%, transform the phase diagram, while the inclusion of relaxation only causes particle-hole breaking in the charge gaps. Our theory identifies heterostrain as the most relevant perturbation to the emergent $U(4) \times U(4)$ symmetry of the flat bands due to its strong $f$-mode coupling, and we showed that splitting the $f$-modes reduces the possible orders in strong coupling, removing variational degrees of freedom. We found a topological reason that gapped phases are forbidden without an IKS boost, and then estimated its magnitude using the zero-hybridization limit of the $f$- and $c$-modes. All ground states are robust to weak particle-hole breaking, but their gaps change at leading order, explaining the stronger correlated features at positive filling due to the positive signs of the $c$-electron potentials computed directly from the generalized BM model\cite{HER24a}. Our theory can be used to simplify and physically explain key features of the insulators observed most frequently in experiment, and give a unified framework understanding the numerical results of self-consistent Hartree-Fock. One interesting extension of this work is to study the correlated Landau levels \cite{2024arXiv240801599H} within the magnetic Hofstadter bands of the BM model \cite{2024PhRvX..14b1042W,2023arXiv230508171S}, or to study the IKS as a parent state for superconductivity \cite{2024arXiv240200869W}. Additionally, strongly correlated physics can be studied within the heterostrained THF model using dynamical mean-field theory \cite{2023arXiv230908529R,CAL20,DAT23,ZHO24} 
or diagrammatic methods \cite{2024arXiv240816761L}
to obtain more realistic finite temperature phase diagrams. Finally, the framework introduced here can apply to other systems with valley textures due to topological obstructions \cite{2024arXiv240615342W,2024arXiv240615343K}, most notably twisted trilayer graphene \cite{WAN23c,YU23a}, and can be studied with other techniques \cite{PhysRevB.109.125404,2024PhRvB.109f4517W,2024PhRvB.109s5153P,2023arXiv230800748I,2024arXiv240104163H,2024arXiv241214065C}.

\section{Acknowledgments}
The authors are grateful to Yves Kwan, Gautam Rai, Lorenzo Crippa, Sid Parameswaran, Ryan Lee, and Minhao He for helpful conversations, as well as Kevin Nuckolls, Myungchul Oh, and Dillon Wong for collaboration on a related project. B. A. B.'s work was supported by the U.S. Department of Energy (DOE) under Contracts No. DE-SC0016239 and by the Simons Collaboration on New Frontiers in Superconductivity (SFI-MPS-NFS-00006741-01). This work was also supported by a grant from the Simons Foundation (SFI-MPS-NFS-00006741-09, O.V.). J. H.-A. is supported by a Hertz Fellowship, with additional support from DOE Grant No. DE-SC0016239. D.C. acknowledges the hospitality of the Donostia International Physics Center, at which this work was carried out, and acknowledges support by the Simons Investigator Grant No. 404513, Simons collaboration on superconductvity SFI-MPS-NFS-00006741-01, the Gordon and Betty Moore Foundation through Grant No. GBMF8685 towards the Princeton theory program, the Gordon and Betty Moore Foundation’s EPiQS Initiative (Grant No. GBMF11070), Office of Naval Research (ONR Grant No. N00014-20-1-2303), Global Collaborative Network Grant at Princeton University, BSF Israel US foundation No. 2018226. D.C. also gratefully acknowledges the support provided by the Leverhulme Trust. H.H. was supported by the European Research Council (ERC) under the European Union’s Horizon 2020 research and innovation program (Grant Agreement No. 101020833). J. Y.'s work at University of Florida is supported by startup funds at University of Florida.
J. Y.'s work at Princeton University is supported by the Gordon and Betty Moore Foundation through Grant No. GBMF8685 towards the Princeton theory program. J. K. acknowledges the support from the NSFC Grant No.~12074276, the Double First-Class Initiative Fund of ShanghaiTech University, and the start-up grant of ShanghaiTech University. 
O.V. was funded in part by the Gordon
and Betty Moore Foundation’s EPiQS Initiative Grant
GBMF11070 and acknowledges
support from the National High Magnetic Field Laboratory funded by the National Science Foundation (Grant No. DMR-2128556)
and the State of Florida. 

\bibliography{bibfile_references,tbg_ref_DH}
\bibliographystyle{apsrev4-1}

\appendix
\onecolumngrid
\newpage
\section{Review of the Heavy Fermion Model and its Symmetries under Perturbation}
\label{appreview}

In this Appendix, we review the heavy fermion model and its correction terms in heterostrain. In \App{app:HFreview}, we review the form of the original, fully symmetric heavy fermion  model and introduce its global symmetries in the $U(4) \times U(4)$ limit. \App{app:strainapp} introduces the perturbations from heterostrain which break the $U(4) \times U(4)$ symmetry to the minimal $U(2) \times U(2)$. \App{app:relaxapp} then introduces the perturbations from relaxation, which actually preserve the $U(4) \times U(4)$ symmetry when the smallest corrections are neglected. In \App{appintreview}, we review the interaction terms and discuss the boosted basis used for IKS calculations. 

\subsection{Heavy Fermion Hamiltonian and Symmetries}
\label{app:HFreview}

 We briefly review the Heavy Fermion mapping of TBG from Ref.\cite{Song20211110MATBGHF}. The two nearly flat bands of TBG are anomalous in the presence of $C_{2z}\mathcal{T}$ and $P$, meaning that no lattice model can reproduce their wavefunctions. In fact, this anomaly cannot be lifted by including any particle-hole symmetric pair of bands in the lattice model. However, a fully symmetric low energy model can be obtained in a different basis: the heavy fermion basis. It consists of two flavors of Dirac electrons (one carrying the 2D $\Gamma_3$ irrep, one carrying the 2D $\Gamma_1 \oplus \Gamma_2$ rep) and a pair of localized Wannier states forming $p_x$-$p_y$ orbitals at the moir\'e unit cell center (carrying the $E_{1a}$ irrep). The heavy fermion model evades the topological obstruction through the formally infinite Hilbert space of the Dirac fermion.

\begin{table*}[ht]
\centering
\begin{tabular}{c|cccc|cccc|cccccc}
$\theta$ \, $({}^\circ)$  & $\gamma$  & $M$ & $v$ & $v'$ & $\mu_1$ & 
$\mu_2$ & $v_1$ & $v_2$ & $c$ & $c'$ & $c''$ & $M_f$ & $\gamma'$ & $M'$ \\
\hline
\text{BM: }1.05 & -24.8 & 3.7 & -4.3 & 1.6 & 14.4 & 4.5 & 0.2 & -0.4 & -8750 & 2050 & -3362 & 4380 & -3352 & -4580 \\
 \end{tabular}
 \caption{Heavy Fermion parameters for the BM model at $w_0/w_1 = .8$. All valued reported are in meV except for the velocities $v,v',v_1,v_2$ which are in eV\,\AA.}
 \label{tab:relaxedHFparam}
\end{table*}

The BM model can be projected into this basis and the following single-particle Hamiltonian in the $K$ valley is obtained (the basis is ordered $c^\dag_{\Gamma_3}, c^\dag_{\Gamma_1 + \Gamma_2}, f_{E_{1a}}$ each of which is a two-dimensional representation)
\bea
h_K(\mbf{k}) &= \bpm 
 & v(k_x \sigma_0 + i k_y \sigma_3) &  \gamma \sigma_0 + v'(k_x \sigma_1 + k_y \sigma_2 )\\
v(k_x \sigma_0 - i k_y \sigma_3) & M \sigma_1 & \\
 \gamma \sigma_0 + v'(k_x \sigma_1 + k_y \sigma_2 ) & & 0 \\
\epm
\eea
neglecting higher shells of the Dirac electrons for simplicity. The parameters of the model are given in \Tab{tab:relaxedHFparam}. The intra-valley symmetries of this Hamiltonian are
\bea
D[C_{3z}] &= \bpm e^{i \frac{2\pi}{3} \sigma_3} & & \\
& \sigma_0 & \\ & & e^{i \frac{2\pi}{3} \sigma_3} \epm, \qquad D[C_{2z}\mathcal{T}] = \bpm \sigma_1 & & \\
& \sigma_1 & \\ & & \sigma_1 \epm K \\
D[C_{2x}] &= \bpm \sigma_1 & & \\
& \sigma_1 & \\ & &  \sigma_1\epm, \qquad D[P] = -i \bpm \sigma_3 & & \\
& \sigma_3 & \\ & & -\sigma_3 \epm \\
\eea
where $P$ is a unitary particle-hole taking $\mbf{r} \to - \mbf{r}$. The Hamiltonian in the $K'$ valley can again be obtained by time-reversal, and spin $SU(2)$ symmetry is implemented by taking two copies of the $K$ and $K'$ Hamiltonians, leading to a four-fold degeneracy overall. This degeneracy is due to the $U(1)$ charge, $U(1)$ valley, and $SU(2)$ spin symmetries which yield a $U(2) \times U(2)$ symmetry group, the factors corresponding to the two valleys. For clarity, we write the full model
\bea
\label{eq:hHF}
h_{HF}(\mbf{k}) &=  \bpm 
 & v(k_x \tau_3 \sigma_0 + i k_y \tau_0 \sigma_3) &  \gamma \tau_0 \sigma_0 + v'(k_x \tau_3 \sigma_1 + k_y \tau_0 \sigma_2 )\\
v(k_x \tau_3 \sigma_0 - i k_y \tau_0 \sigma_3) & M \tau_0 \sigma_1 & \\
 \gamma \tau_0 \sigma_0 + v'(k_x \tau_3 \sigma_1 + k_y \tau_0 \sigma_2 ) & & 0 \\
\epm \otimes s_0
\eea
where $\tau_i$ and $s_i$ are sets of Pauli matrices representing valley and spin respectively. The inter-valley crystallographic symmetries are spin-less time-reversal and two-fold rotation
\bea
D[\mathcal{T}] &= \tau_1 K, \quad D[C_{2z}] = \tau_1 \sigma_1
\eea
which both take $\mbf{k} \to - \mbf{k}$. The global symmetries are generated by the Hermitian commuting charge $\tau_3$ (valley) and $s_i$ (spin).

The $U(2) \times U(2)$ symmetry group is generated by the algebras $\frac{1}{2}(\tau_0 + \tau_3) \sigma_i$  and $\frac{1}{2}(\tau_0 - \tau_3) \sigma_i$. 
We now consider various limits where the $U(2) \times U(2)$ symmetry is enhanced, ultimately to a $U(4) \times U(4)$ symmetry group. The key to this enhancement is the inter-valley anti-commuting operator 
\bea
D[P C_{2z}] &=  -i \tau_1 \bpm \sigma_2 & & \\
& \sigma_2 & \\ & & -\sigma_2 \epm 
\eea
which is local in $\mbf{k}$. An analogous operator appears in the projected flat band theory of twisted bilayer graphene in the BM model. However, it is worthwhile to recall some key differences between the projected flat band theory (see for instance Ref.\cite{BER21a}) and the heavy fermion theory discussed here. In the projected theory, all symmetry generators are restricted to the flat bands. However, in the heavy fermion theory, the symmetries are preserved by the full $f$ and $c$ Hilbert spaces despite the kinetic energy $c$ electrons. Another key difference concerns the chiral operator which we will now introduce. In the BM model, the chiral operator is the sublattice operator and only anti-commutes with $w_0 = 0$. However, in the heavy-fermion theory, one can define an anti-commuting chiral operator acting on the $f$-$c$ basis for all values of $w_0$ \cite{Song20211110MATBGHF}. We recall the details of the symmetries in the THF model now. 

We exhibit two limits where new anti-commuting local symmetries $C,C'$ emerge. Then $C P C_{2z}$ and $C' P C_{2z}$ become commuting local symmetries that enlarge the symmetry group. 

First we introduce the chiral $U(4)$ symmetry group which arises when $v' = 0$, so that
\bea
D[C] &=  \tau_0 \bpm -\sigma_3 & & \\
& \sigma_3 & \\ & & \sigma_3 \epm 
\eea
anti-commutes with $h_{HF}(\mbf{k})$ in \Eq{eq:hHF}. This now gives a commuting symmetry (multiplying by an overall phase and the $U(1)$ operator $\tau_3$)
\bea
D[C P C_{2z}] &\equiv \tau_2 \bpm \sigma_1 & & \\
& -\sigma_1 & \\ & & \sigma_1 \epm, \qquad [D[C P C_{2z}], h_{HF}(\mbf{k})] = 0 \text{ if } v'= 0 \ . 
\eea
In the chiral limit, $D[C P C_{2z}]$ enhances the $U(2) \times U(2)$ symmetry to a $U(4)$ symmetry. This $U(4)$ symmetry is the analogue of the chiral $U(4)$ in the BM model, but is defined in the THF model by $v'=0$, rather than $w_0 = 0$.

Second we introduce the $U(4)$-flat limit where $M=0$ and 
\bea
D[C'] &= \tau_0 \bpm  -\sigma_0 & & \\
& \sigma_0 & \\ & & \sigma_0 \epm
\eea
anti-commutes with $h_{HF}(\mbf{k})$ in \Eq{eq:hHF}. This now gives a commuting symmetry (multiplying by an overall phase and the $U(1)$ operator $\tau_3$)
\bea
D[C' P C_{2z}] &\equiv \tau_2 \bpm - \sigma_2 & & \\
& \sigma_2 & \\ & & -\sigma_2 \epm , \qquad [D[C' P C_{2z}], h_{HF}(\mbf{k})] = 0 \text{ if } M = 0 \ . 
\eea
In the flat limit, $D[C' P C_{2z}]$ enhances the $U(2) \times U(2)$ symmetry to a $U(4)$ symmetry. This $U(4)$ symmetry has an analogue in the projected BM model when the dispersion is dropped, in which case the entire single-particle term is zero and $P C_{2z}$ commutes with the interaction Hamiltonian. This is a crucial difference with the THF model, where $C' P C_{2z}$ commutes with the single-particle kinetic energy as well. 

It can be checked that if $v'=0$ and $M=0$ (the chiral-flat limit), the total symmetry algebra is $U(4) \times U(4)$, with each factor acting on a separate ``Chern" sector. Details can be found in Ref.\cite{Song20211110MATBGHF}. In the projected BM model, the $U(4) \times U(4)$ can be understood as rotating each of the four $C=1$ and four $C=-1$ flat bands within spin-valley space. In the THF model, the $U(4)\times U(4)$ symmetry group acts on the full $f,c$ Hilbert space.

Our emphasis will be on perturbations away from this $U(4)\times U(4)$ limit. In particular, heterostrain on the order of $.2\%$ is already a much larger symmetry-breaking term than $M$ or $v'$ as we now show. 

\subsection{Strain}
\label{app:strainapp}

We now discuss the topological heavy fermion model characterizing the heterostrained BM model. Ref.\cite{HER24a} showed that heterostrain is captured within the fully-symmetric THF Hilbert space by adding the symmetry-breaking kinetic term $\delta h_\eps$:
\bea
\delta h_\eps = \bpm
c (\eps_{xy} \sigma_1 + \eps_- \sigma_2) & c' (\eps_{xy} \sigma_1 - \eps_- \sigma_2)  & i \gamma' \eps_+ \sigma_3 \\
c' (\eps_{xy} \sigma_1 - \eps_- \sigma_2) &  M' \eps_+  \sigma_2 & c'' (\eps_{xy} \sigma_0 - i \eps_- \sigma_3) \\
-i \gamma' \eps_+ \sigma_3& c'' (\eps_{xy} \sigma_0 + i \eps_- \sigma_3) & M_f (\eps_{xy} \sigma_1 + \eps_- \sigma_2) \\
\epm
\eea
written in valley $K$ for one spin. To obtain the coefficients, we project the heterostrained BM model onto the $f$-mode Wannier functions and $\Gamma$ point conduction wavefunctions of the original THF model. \Tab{tab:relaxedHFparam} contains the results. The large values of the heterostrain coefficients reflects the fact that small heterostrains $\eps \sim .001$ add $\sim 5$meV perturbations to the Hamiltonian. The $M_f$ term is the most significant since it breaks the symmetry of the $p_x$-$p_y$ $f$-orbitals, polarizing them along the $-M_f$ energy eigenmode of $(\eps_{xy} \sigma_1 + \eps_- \sigma_2)$. We showed with perturbation theory in Ref.\cite{HER24a} that a minimal model preserving the key heterostrained features of the band structure consists of only keeping $M_f$ and $c''$. 

We now discuss the symmetry that survive in heterostrain. First, heterostrain breaks $C_{3z}$ and $C_{2x}$ but preserves $C_{2z}$ and $\mathcal{T}$. Moreover, $P$ is also preserved to leading order when strain is introduced via minimal coupling $\mbf{k} \to \mbf{k} \pm \mbf{A}$ in the BM model, where $\mbf{A}$ is the pseudo-gauge field and $\pm$ reflects layer anti-symmetry. Note that this is not true for relaxation, or for more elaborate models of heterostrain \cite{HER24a}. An important result of this is that the anti-commuting local symmetry $D[PC_{2z}]$ is preserved. To check the survival of the $U(4)$ symmetries, we note that $D[C]$ only anti-commutes with $\delta h_\eps$ if $c'=c''=0$, and since $c''$ is relevant to the low energy band structure \cite{HER24a}, the chiral $U(4)$ is broken. Similarly, $D[C']$ anti-commutes with $\delta h_\eps$ if $c=M'=c''=M_f =0$, and since both $c''$ and $M_f$ are relevant to the low energy bands, the flat $U(4)$ is also broken. 

\subsection{Gradient Hopping Terms}
\label{app:relaxapp}

We next consider the effect of gradient hopping terms that break the emergent particle-hole symmetry of the BM model. Ref.\cite{HER24a} derived the kinetic term perturbation to the heavy fermion model, which is (for valley $K$ and spin $\uparrow$)
\bea
\label{eq:HFrelaxedapp}
\delta h_{\Lambda}(\mbf{k}) &= \bpm 
 \mu_1 \sigma_0 + v_2 \mbf{k} \cdot \pmb{\sigma} & v_1 \mbf{k} \cdot \pmb{\sigma}^*  &  0 \\
 v_1 \mbf{k} \cdot \pmb{\sigma}^* & \mu_2 \sigma_0 &  0 \\
0 & 0 & 0 \\
\epm \\
\eea
consisting of the $c$-electron potentials $\mu_1, \mu_2$ and velocities $v_1, v_2$. (Without loss of generality, we have set the $f$-electron chemical potential to zero.) The parameters are given in \Tab{tab:relaxedHFparam}. We showed with perturbation theory in Ref.\cite{HER24a} that a minimal model preserving the key particle-hole breaking features of the band structure consists of only keeping $\mu_1$ and $\mu_2$. 

We now check the symmetries of $\delta h_{\Lambda}$. Although $P$ is broken, we find that $D[CP] = -i\text{ diag}(-\sigma_0,\sigma_0,-\sigma_0)$ is preserved if $v_2=0$. Since $v_2$ is negligible, the chiral $U(4)$ is actually preserved by $\delta h_\Lambda$. Next we check that $D[C'P] = -i\text{ diag}(-\sigma_3,\sigma_3,-\sigma_3)$ is preserved if $v_1=0$. Since $v_1$ is negligible, the flat $U(4)$ symmetry is also preserved. Thus relaxation, despite breaking particle-hole, does not destroy the $U(4)\times U(4)$ limit of the heavy fermion model. This explains the numerical observation taht particle-hole breaking alone is insufficient to destabilize the KIVC \cite{KWA21}.

\subsection{Interaction Terms}
\label{appintreview}

Because the interaction Hamiltonian, first obtained in Ref.\cite{Song20211110MATBGHF} is obtained purely from the overlaps of the Wannier states and conduction wavefunctions on the Coulomb potential, the interaction Hamiltonian in the presence of heterostrain is \textit{unchanged}. This is because we have shown that the THF basis is unchanged in the presence of heterostrain, although symmetry-breaking terms appear in the Hamiltonian. Hence we retain the original fully symmetric interaction of the THF model, with heterostrain appearing only as a perturbation in the kinetic term. 

We now recap the interaction Hamiltonian, which consists of 5 terms:
\bea
H_{int} &= H_U + H_W + H_J + H_V
\eea
where the $f$-$f$ interaction is
\bea
H_U &= \frac{U}{2} \sum_\mbf{R} (:f^\dag_\mbf{R} f_\mbf{R}:)^2  + \frac{U_2}{2} \sum_{\braket{\mbf{R}\mbf{R}'}} :f^\dag_\mbf{R} f_\mbf{R}::f^\dag_{\mbf{R}'} f_{\mbf{R}'}:
\eea
including an on-site term with $U_1 = 58$meV and nearest-neighbor term $U_2 =2.3$meV. We note that these values depend on the choice of dielectric constant, screening length, and twist angle, which are device-dependent, and smaller values of $U_1$ can be easily obtained. The $c$-$c$ repulsion is
\bea
H_V &= \frac{1}{2N} \sum_{\mbf{q},\mbf{k},\mbf{k}} \frac{V(\mbf{q})}{\Omega }  :c^\dag_{\mbf{k}+\mbf{q}} c_\mbf{k}: :c^\dag_{\mbf{k}'-\mbf{q}} c_{\mbf{k}'}:
\eea
where $V(\mbf{q})$ is the screened Coulomb interaction. We only keep the mean-field decoupling of $H_V$ in the Hartree channel, so we can effectively replace $V(\mbf{q})$ by $V(0) = \frac{\pi \xi e^2}{\eps}$ where $e$ is the electric charge, $\xi = 10$nm and the dielectric constant $\eps = 6$.

Next, the $f$-$c$ repulsion is
\bea
H_W &= \frac{1}{N}\sum_{\mbf{R} \mbf{k} \mbf{k}'} e^{- i (\mbf{k}-\mbf{k}')\cdot \mbf{R}}:f^\dag_\mbf{R} f_\mbf{R}: :c^\dag_\mbf{k} W c_{\mbf{k}'}:
\eea
where $W$ is a diagonal matrix acting as $W_1 = 44$meV on the $\Gamma_3$ $c$-electrons and $W_2 = 50$meV on the $\Gamma_1 \oplus \Gamma_2$ electrons. Lastly, the $U(4)$ ferromagnetic exchange interaction is
\bea
H_J &= - \frac{J}{2 N} \sum_{\al s \eta,\al' \eta' s'} \sum_{\mbf{R} \mbf{k} \mbf{k}'} e^{i(\mbf{k}-\mbf{k}')\cdot \mbf{R}}(\eta \eta' + (-1)^{\al + \al'}) :f^\dag_{\mbf{R},\al s \eta}f_{\mbf{R},\al' s' \eta'}::c^\dag_{\mbf{k},\al'+2, s' \eta'}c_{\mbf{k}',\al+2, s \eta}: 
\eea
and $J = 16.38$meV. The interaction preserves the full $U(4) \times U(4)$ symmetry group.

\subsection{Boosted Basis}
\label{app:boostH}

We now discuss the boosted basis relevant for the IKS states which break translation $T_\mbf{R}$ and $U(1)$ valley $\tau_3$ but preserve $\tilde{T}_\mbf{R} = e^{i \frac{\tau_3 \mbf{q}}{2} \cdot \mbf{R} } T_\mbf{R}$. It is convenient to introduce a diagonal basis of this surviving symmetry group as in the main text
\bea
\label{eq:boostbasisapp}
\tilde{f}^\dag_{\mbf{R}, \al \eta s} &= f^\dag_{\mbf{R}, \al \eta s} e^{-i \mbf{R} \cdot \mbf{q} \eta/2}, \quad 
\tilde{c}^\dag_{\mbf{k}, a \eta s} = c^\dag_{\mbf{k} + \mbf{q} \eta /2,a \eta s} \\
\eea
obeying $\tilde{T}_\mbf{a} \tilde{f}^\dag_{\mbf{R}, \al \eta s} \tilde{T}_\mbf{a}^\dag = \tilde{f}^\dag_{\mbf{R}+\mbf{a}, \al \eta s}$ and $\tilde{T}_\mbf{a} \tilde{c}^\dag_{\mbf{k}, a \eta s} \tilde{T}_\mbf{a}^\dag = e^{- i \mbf{k} \cdot \mbf{a}}\tilde{c}^\dag_{\mbf{k}, a \eta s}$. We now write the interaction in this basis.

Firstly, we note that the $H_U$ term contains only the total $f$-mode density $f^\dag_\mbf{R} f_\mbf{R} = \tilde{f}^\dag_\mbf{R} \tilde{f}_\mbf{R} $ since the position-dependent phases cancel. Thus $H_U$ can be rewritten by replacing $f \to \tilde{f}$. Next we consider $H_V$, which reads
\bea
H_V &= \frac{1}{2N} \sum_{a \eta s, a' \eta' s'} \sum_{\mbf{p},\mbf{k},\mbf{k}} \frac{V(\mbf{p})}{\Omega }  :c^\dag_{\mbf{k}+\mbf{p},a \eta s} c_{\mbf{k},a \eta s}: :c^\dag_{\mbf{k}'-\mbf{p},a' \eta' s'} c_{\mbf{k}',a' \eta' s'}: \\
&= \frac{1}{2N} \sum_{a \eta s, a' \eta' s'} \sum_{\mbf{p},\mbf{k},\mbf{k}} \frac{V(\mbf{p})}{\Omega }  :\tilde{c}^\dag_{\mbf{k}+\mbf{p}-\mbf{q} \eta/2,a \eta s} \tilde{c}_{\mbf{k}-\mbf{q} \eta/2,a \eta s}: :\tilde{c}^\dag_{\mbf{k}'-\mbf{p}-\mbf{q} \eta'/2,a' \eta' s'} \tilde{c}_{\mbf{k}'-\mbf{q} \eta'/2,a' \eta' s'}: \\
&= \frac{1}{2N} \sum_{a \eta s, a' \eta' s'} \sum_{\mbf{p},\mbf{k},\mbf{k}} \frac{V(\mbf{p})}{\Omega }  :\tilde{c}^\dag_{\mbf{k}+\mbf{p},a \eta s} \tilde{c}_{\mbf{k},a \eta s}: :\tilde{c}^\dag_{\mbf{k}'-\mbf{p},a' \eta' s'} \tilde{c}_{\mbf{k}',a' \eta' s'}: \\
\eea
 where we shifted the $\mbf{k}$ sum by $\mbf{q} \eta /2$ and $\mbf{k}'$ sum by $\mbf{q} \eta' /2$. Thus $H_V$ can also be rewritten by replacing $c \to \tilde{c}$. 
 
 Next we consider the $f$-$c$ coupling terms. We see that
 \bea
H_W &= \frac{1}{N} \sum_{a \eta s} \sum_{\mbf{R} \mbf{k} \mbf{k}'} e^{- i (\mbf{k}-\mbf{k}')\cdot \mbf{R}}:f^\dag_\mbf{R} f_\mbf{R}: :c^\dag_{\mbf{k}a \eta s} W_a c_{\mbf{k}' a \eta s}: \\
&= \frac{1}{N} \sum_{a \eta s} \sum_{\mbf{R} \mbf{k} \mbf{k}'} e^{- i (\mbf{k}-\mbf{k}')\cdot \mbf{R}}:\tilde{f}^\dag_\mbf{R} \tilde{f}_\mbf{R}: :\tilde{c}^\dag_{\mbf{k}- \mbf{q} \eta/2,a \eta s} W_a \tilde{c}_{\mbf{k}'- \mbf{q} \eta/2, a \eta s}: \\
&= \frac{1}{N} \sum_{a \eta s} \sum_{\mbf{R} \mbf{k} \mbf{k}'} e^{- i (\mbf{k}-\mbf{k}')\cdot \mbf{R}}:\tilde{f}^\dag_\mbf{R} \tilde{f}_\mbf{R}: :\tilde{c}^\dag_{\mbf{k},a \eta s} W_a \tilde{c}_{\mbf{k}', a \eta s}: \\
\eea
after shifting the sums. Lastly we consider
\bea
H_J &= - \frac{J}{2 N} \sum_{\al s \eta,\al' \eta' s'} \sum_{\mbf{R} \mbf{k} \mbf{k}'} e^{i(\mbf{k}-\mbf{k}')\cdot \mbf{R}}(\eta \eta' + (-1)^{\al + \al'}) :f^\dag_{\mbf{R},\al s \eta}f_{\mbf{R},\al' s' \eta'}::c^\dag_{\mbf{k},\al'+2, s' \eta'}c_{\mbf{k}',\al+2, s \eta}:  \\
&= - \frac{J}{2 N} \sum_{\al s \eta,\al' \eta' s'} \sum_{\mbf{R} \mbf{k} \mbf{k}'} e^{i(\mbf{k}-\mbf{k}')\cdot \mbf{R}}(\eta \eta' + (-1)^{\al + \al'}) e^{i \mbf{R} \cdot \mbf{q} \eta/2 - i \mbf{R} \cdot \mbf{q} \eta'/2} :\tilde{f}^\dag_{\mbf{R},\al s \eta}\tilde{f}_{\mbf{R},\al' s' \eta'}::\tilde{c}^\dag_{\mbf{k} - \mbf{q}\eta'/2,\al'+2, s' \eta'}\tilde{c}_{\mbf{k}' - \mbf{q}\eta/2,\al+2, s \eta}:  \\
&= - \frac{J}{2 N} \sum_{\al s \eta,\al' \eta' s'} \sum_{\mbf{R} \mbf{k} \mbf{k}'} e^{i(\mbf{k}+ \mbf{q}\eta'/2-\mbf{k}' -  \mbf{q}\eta/2)\cdot \mbf{R}}(\eta \eta' + (-1)^{\al + \al'}) e^{i \mbf{R} \cdot \mbf{q} \eta/2 - i \mbf{R} \cdot \mbf{q} \eta'/2} :\tilde{f}^\dag_{\mbf{R},\al s \eta}\tilde{f}_{\mbf{R},\al' s' \eta'}::\tilde{c}^\dag_{\mbf{k},\al'+2, s' \eta'}\tilde{c}_{\mbf{k}',\al+2, s \eta}:  \\
&= - \frac{J}{2 N} \sum_{\al s \eta,\al' \eta' s'} \sum_{\mbf{R} \mbf{k} \mbf{k}'} e^{i(\mbf{k}-\mbf{k}')\cdot \mbf{R}}(\eta \eta' + (-1)^{\al + \al'})  :\tilde{f}^\dag_{\mbf{R},\al s \eta}\tilde{f}_{\mbf{R},\al' s' \eta'}::\tilde{c}^\dag_{\mbf{k},\al'+2, s' \eta'}\tilde{c}_{\mbf{k}',\al+2, s \eta}:  \\
\eea
so in all cases, the $c \to \tilde{c}$ and $f \to \tilde{f}$ replacement holds. This can be expected because the interaction arises from projecting the density-density Coulomb interaction, which is locally invariant under $U(4) \times U(4)$ gauge transformations, onto the $f$-$c$ basis.

This means that the expectation value of $H_{int}$ on any state is unchanged upon replacing $c,f$ with $\tilde{c},\tilde{f}$ operators. As such, the one-shot energy dependence on $\mbf{q}$ only comes from the kinetic energy, where breaking translation and $U(1)$ for $\mbf{q}\neq 0$ allows new hybridization gaps to open. This effect will be studied extensively in the following sections.

\section{Hartree-Fock and One-Shot Theory of the Correlated Insulators}
\label{app:oneshotstrain}

In this Appendix, we propose heavy fermion parent states for the integer filling ground states in the presence of heterostrain. These parent states are used for the seeds of self-consistent Hartree-Fock. We find that the overlap of the initial and final Hartree-Fock states is high enough that an analytical one-shot calculation accurately predicts the self-consistent band structure.  We then develop a perturbation theory in the mean-field heavy fermion Hamiltonian which predicts a gap opening at a critical value of the IKS boost vector at $\nu = -2$. This expression is in good agreement with numerical Hartree-Fock calculations. 

\subsection{Parent States}
\label{sec:parentstates}

The crux of a heavy fermion model is a separation between the $c$-electrons, which form a weakly correlated Fermi sea due to their strong kinetic energy, and the local moments (heavy fermions) which possess very weak kinetic energy and are governed by the interactions. We study this system with Hartree-Fock, where the strong interaction effects on the heavy fermions manifests as spontaneous symmetry breaking. We pursue an analytical understanding of the ground state by proposing a model wavefunction or ``parent state." To motivate this approach, let us write the full Hamiltonian
\bea
H &= (H_f + H_{c} + H_{fc}) + (H_U + H_W + H_V + H_J) \ .
\eea
A parent state, by definition, is an eigenstate of 
\bea
H_d &= H_f + H_{c}  + H_U 
\eea
which decouples the $f$-mode and $c$-mode states. The ground state of $H_c$ is the (product state) Fermi sea of the $c$-electrons. Note that we do not consider $H_W$ as part of $H_d$ (like was done in Ref.\cite{Song20211110MATBGHF}), although $H_W$ also has exact eigenstates which are products of decoupled $f$ and $c$ states. This simplifies the discussion since the $c$-Fermi sea is independent of $\nu$, the filling measured from charge neutrality, and the interaction. The effect of $H_W$ will be captured within one-shot Hartree-Fock.

The ground state of $H_d$ depends on $\nu$: there are $\nu + 4$ $f$-modes per unit cell. Since $H$ contains a strong onsite repulsion $H_U$ between all 8 flavors of $f$-modes (valley, spin, and flavor), the ground state requires equal occupation on all sites, i.e. conventional charge density waves are strongly disfavored at integer filling \cite{HU23i}. In the original heavy-fermion model, the single-particle term $H_f$ is zero, and there is a massive degeneracy of possible ground states which is resolved primarily by $H_{fc}$ and $H_J$. In the case of heterostrain, instead we have
\bea
\label{eq:Hf}
H_f = M_f \sum_\mbf{R} f^\dag_{\mbf{R}}(\eps_{xy} \tau_0 \sigma_1 + \eps_- \tau_3 \sigma_2) s_0 f_\mbf{R}
\eea
where, as a reminder, $\tau, \sigma,s$ are the Pauli matrices on valley, sublattice, and spin respectively. The eigenvalues of $H_f$ are $\pm M_f \sqrt{\eps_{xy}^2 + \eps_{-}^2}$ which split the eight $f$ modes into two branches with these energies. We see that the essential effect of heterostrain is to impose external symmetry breaking at the level of the parent states. Since the single-particle energy splitting is quite large, $\sim 10meV$ for strain $\sim \eps = .002$, we find it is energetically favorable for the parent state to fill the lower energy $f$-modes and induce symmetry-breaking orders within this subspace. 

For simplicity, we set $\eps_{xy} = 0$ which amounts to choosing the uniaxial heterostrain along the $x$-axis. We now discuss each filling in turn by identifying the $f$-mode order at the appropriate filling in the low energy sector:
\bea
\ket{GS_0,\nu} = \prod_{\mbf{R},n}^{\nu+4} f^\dag_\mbf{R} \zeta^{f,\nu}_n \ket{\text{FS}} \ .
\eea
The states $\ket{GS,\nu}$ capture the heterostrain-induced symmetry-breaking of the $f$-modes. Their associated one-particle reduced density matrix, commonly called the order parameter, is
\bea
\label{eq:parentstateorder}
O_f = \braket{GS_0,\nu|f^\dag_{\mbf{R},A} f_{\mbf{R},B}|GS_0,\nu} = [\sum_n \zeta^{f,\nu}_n {\zeta^{f,\nu}_n}^*]_{BA}
\eea
where $A,B$ label the tensor product basis $\al, \eta,s$. (Note that $O_f$ can be a nonzero matrix even if no symmetries are spontaneously broken.) The resulting IKS orders are immediately obtained by oppositely boosting the valleys in opposite directions:
\bea
\label{eq:parentIKS}
\ket{GS_0,\mbf{q},\nu} = \prod_{\mbf{R},n}^{\nu+4} f^\dag_\mbf{R} [\exp \lp i \frac{\tau_3}{2}\mbf{q}\cdot \mbf{R} \rp \zeta^{f,\nu}_n] \ket{\tilde{\text{FS}}} =  \prod_{\mbf{R},n}^{\nu+4} \tilde{f}^\dag_\mbf{R}  \zeta^{f,\nu}_n \ket{\tilde{\text{FS}}} \ .
\eea
Since the IKS boost commutes with $H_U$ and the parent states are still decoupled in $f$ and $c$, $\ket{GS,\mbf{q},\nu}$ are also good parent states. 

We will now write out the parent states at each integer filling and compute their one-shot Hartree-Fock spectrum and order parameter within the THF model. We also perform numerical calculations in the 6-band BM model and numerically compute the $f$-mode order parameter 
\bea
\null [O_f]_{\al \eta s, \be \eta' s'} = \frac{1}{N} \sum_\mbf{k} \braket{\tilde{f}^\dag_{\mbf{k}, \al \eta s} \tilde{f}_{\mbf{k}, \be \eta' s'}}
\eea
using the explicit $f$-mode wavefunctions obtained from the BM model. Here the boost in $\tilde{f}_{\mbf{k},\al \eta s}$ is chosen to be that of the global ground state, which $\mbf{q} \neq 0$ for $\nu \neq 0$. We choose a $12\times 12$ mesh in the BM model simulations (see \App{app:sec:hf_in_bm}) so that $\mbf{q} = \frac{2}{3}M_M$ is a point on the mesh, and is typically the IKS ground state for a range of heterostrain values. 

\subsubsection{$\nu = -3$}

\begin{figure*}[h]
\centering
\includegraphics[width=.99\textwidth]{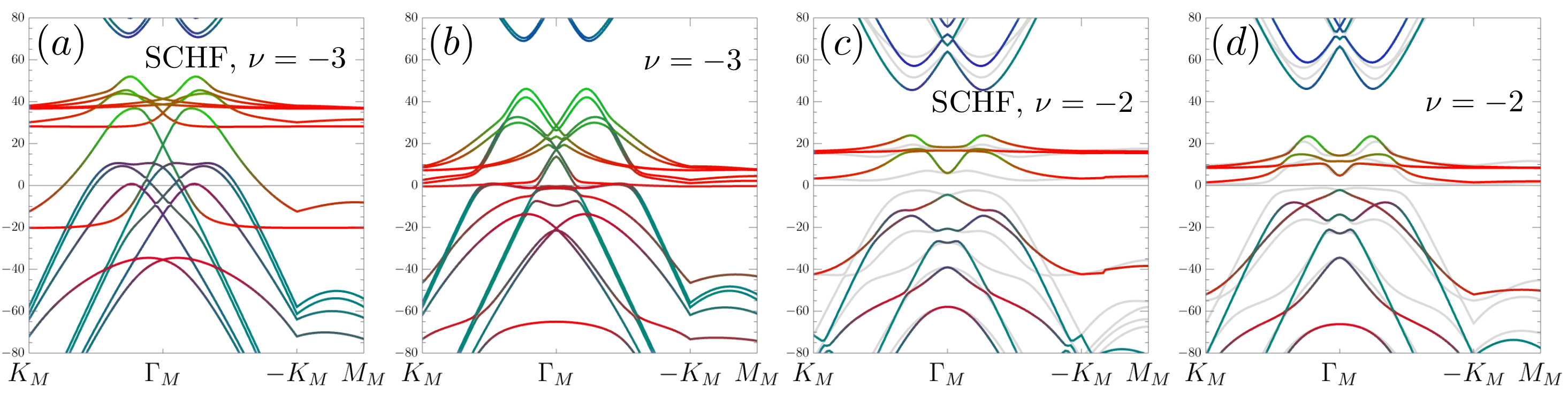}
\caption{Hartree-Fock band structures at $\nu = -3,-2$ for $\mbf{q} = \frac{2}{3}M_M, \eps = .15\%$. Bands are colored to indicate their $f$-mode character (see Fig. 1. of the Main Text). $(a)$ and $(b)$ compare the self-consistent and one-shot bands at $\nu=-3$, whereas $(c)$ and $(d)$ compare them at $\nu=-2$. The gray lines show the same band structure along a rotated path where Dirac nodes appear at $\nu = -1,0$. Note that $\nu=-3$ suffers from strong charge fluctuations due to the large value of the interactions. To achieve convergence, we have scaled down the interactions in $(a)-(b)$ by $20\%$. Even so, we observe large differences between one-shot and self-consistent Hartree-Fock (SCHF) compared to the strong qualitative agreement at $\nu=-2$.
}
\label{fig:HFstrainparentnu3}
\end{figure*}

We begin with $\nu = -3$ where the parent state consists of a single $f$-mode per unit cell. We note that fluctuations are strong at $\nu = -3$ for our choice of dielectric constant and restricted Hartree-Fock may not be appropriate \cite{2023PhRvL.131b6502H}. In fact, exact diagonalization in the projected BM model shows the presence of charge density waves at $\nu = -3$ for some parameters \cite{Xie20220929TBGPhaseDia,XIE23a}. 
Nevertheless, we include $\nu = -3$ in our study for completeness, and because it offers a simple starting point for higher fillings (where Hartree-Fock is expected to accurately describe the physics). We consider the low energy spin-polarized, spin-less $\mathcal{T}$ preserving eigenstates of the $f$-block \Eq{eq:Hf}. Working in the spin $\uparrow$ sector without loss of generality, and recalling that $\eps_- \leq 0$ in our conventions, we find that the lowest energy eigenvectors of \Eq{eq:Hf} are
\bea
\label{eq:Uminus3}
\zeta = (1,i,0,0)^T/\sqrt{2}, \quad (0,0,i,1)^T/\sqrt{2}
\eea
in the ordered basis $K+, K-, K'+, K'-$ where $K,K'$ are the two valleys $\pm$ labels the $p_x\pm i p_y$ orbitals. Any linear combination of these eigenvectors will have the same energy. To select a state of out this manifold, we impose spin-less time-reversal symmetry $D[\mathcal{T}] = \tau_1 K$ on the ground state, so that $D[\mathcal{T}]\zeta = e^{i \al} \zeta$. The overall phase $e^{i \al}$ is a gauge choice because $D[\mathcal{T}]$ is anti-unitary. Since $D[\mathcal{T}]$ exchanges valleys, the amplitudes of the two states in \Eq{eq:Uminus3} must be equal, imposing maximal inter-valley coherence. Their relative phase can be chosen arbitrarily because of the $U(1)$ valley symmetry. For concreteness, we fix the phase by imposing $D[C_{2z}] \zeta = + \zeta$. (Choosing the opposite phase is equivalent to a valley $U(1)$ gauge transformation. In later sections, we will compare the $C_{2z}$ eigenvalues at the $\Gamma$ point and BZ edge. Their relative sign is invariant under the valley $U(1)$ gauge freedom.) The resulting eigenvector and many-body parent state at $\mbf{q} = 0$ are
\bea
\zeta^{f,\nu=-3} = (1,i,i,1)/2, \quad \ket{GS_{0},-3} = \prod_\mbf{R} \frac{f^\dag_{\mbf{R},+,K,\uparrow}+if^\dag_{\mbf{R},-,K,\uparrow}+if^\dag_{\mbf{R},+,K',\uparrow}+f^\dag_{\mbf{R},-,K',\uparrow}}{2}\ket{\text{FS}}
\eea
and $\ket{\text{FS}}$ is the Fermi sea of negative energy $c$-electrons. The parent state $\ket{GS_{0},-3}$ contains maximal $C_3$-breaking and IVC due to the equal amplitude superposition of $f$-modes. This can be seen explicitly from the order parameter (\Eq{eq:parentstateorder})
\bea
\label{eq:Ofminus2app}
O_{f,0} = \frac{1}{4} (\tau_0 \sigma_0 + \tau_1 \sigma_1 - \tau_2 \sigma_3 - \tau_3 \sigma_2) \otimes \bpm 1 & \\ & 0\epm
\eea
where the final Kronecker product is a projector on the spin $\uparrow$ states. One can see explicitly that the $C_{2z}$ and $\mathcal{T}$ symmetries commute with $O_f$. We note that, since $U(1)$ valley is spontaneously broken, numerical Hartree-Fock can converge to any $U(1)$ valley rotation of this state. 

\Fig{fig:HFstrainparentnu3} contains Hartree-Fock spectra in an IKS state with boost vector $\mbf{q} = \frac{2}{3}M_M$. We emphasize again that Hartree-Fock might not be trustworthy at $\nu = -3$ due to strong fluctuations of the $f$ modes beyond the Slater approximation. We include \Fig{fig:HFstrainparentnu3} only for completeness. We reserve a quantitative comparison of the one-shot heavy fermion ansatz for higher fillings below. 

\subsubsection{$\nu = -2$}

The parent state at $\nu = -2$ is obtained by stacking two spin-polarized $\nu = -3$ state of opposite spins to obtain a spin singlet, again with maximal $C_3$-breaking and inter-valley coherence. The parent state is 
\bea
\ket{GS_{0},-2} = \prod_{\mbf{R},s = \uparrow,\d} \frac{f^\dag_{\mbf{R},+,K,s}+if^\dag_{\mbf{R},-,K,s}+if^\dag_{\mbf{R},+,K',s}+f^\dag_{\mbf{R},-,K',s}}{2}\ket{\text{FS}} 
\eea
with the associated $f$-order parameter
\bea
\label{eq:Ofnu2}
O_{f,0} = \frac{1}{4} (\tau_0 \sigma_0 + \tau_1 \sigma_1 - \tau_2 \sigma_3 - \tau_3 \sigma_2) s_0 \ . 
\eea
Both spin sectors have inter-valley coherence and will respond to the IKS boost. However, we note that even without a boost, this state displays the $C_{3z}$-breaking and Kekul\'e distortion at the graphene lattice scale observed in low heterostrain samples in experiment. It is clear that the zero-strain strong coupling ferromagnetic states like VP, KIVC, and TIVC incur a strain energy penalty $\sim |2 M_f \eps_-|$ per particle, since they have support on the high-energy modes split by heterostrain. 

Moreover, we find in \Fig{fig:HFstrainparentnu3} that the one-shot band structure shows a strong similarity with the fully self-consistent band structure. This underscores the predictive power of the analytical one-shot Hamiltonian obtained from the order parameter in \Eq{eq:Ofnu2}. Furthermore, we can quantitatively assess the similarity of the one-shot and self-consistent order parameters from the absolute error
\bea
||O_{f,1} - O_{f,\infty}||_F = .022
\eea
remarkably indicating that the one-shot calculation is nearly 98\% accurate. The error in the parent state is $||O_{f,0} - O_{f,\infty}|| = .26$. This larger value can be easily understood by comparing $\Tr O_{f,0} = 2$ to $\Tr O_{f,\infty} = 2.57$, which shows that the self-consistent order parameter has increased its $f$-mode occupation and slightly depleted the Fermi sea. This effect is not captured in the parent state (the ``zero-shot" Hartree-Fock state). Nevertheless, we find that 
\bea
||O_{f,0} O_{f,\infty}||^2_F = 2
\eea
to within machine precision, showing that $O_{f,\infty}$ is completely contained the eigenspace of the parent state $O_{f,0}$. 

Lastly, we can compare the one-shot order parameter obtained from the Heavy Fermion model to the $f$-mode order parameter extracted from a 6-band BM model Hartree-Fock calculation with the same boost vector. We again find good agreement
\bea
||O_{f,1} - O^{BM}_{f,\infty}||_F = .16 
\eea
showing that, despite the numerous simplifications in the heavy fermion Hamiltonian, both in the interaction and strain coupling, it (even at the one-shot Hartree-Fock level) still produces a quantitatively reliable comparison to the 6-band BM model. 

\subsubsection{$\nu = -1$}

\begin{figure*}[h]
\centering
\includegraphics[width=.99\textwidth]{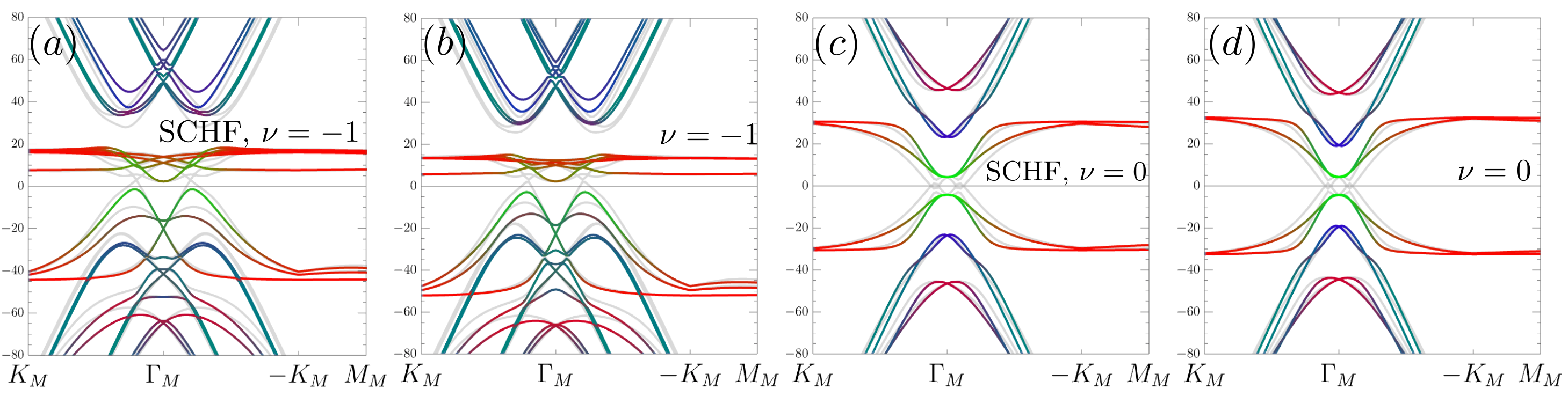}
\caption{Hartree-Fock band structures at $\nu = -1$ for $\mbf{q} = \frac{2}{3}M_M, \eps = .15\%$ and $\nu = 0$ for $\mbf{q} = 0, \eps = .15\%$. Bands are colored to indicate their $f$-mode character (see Fig. 1. of the Main Text). $(a)$ and $(b)$ compare the self-consistent and one-shot bands at $\nu=-1$, whereas $(c)$ and $(d)$ compare them at $\nu=0$. The gray lines show the same band structure along a rotated path where Dirac nodes are visible. Both fillings show excellent agreement between the one-shot and self-consistent calculations.}
\label{fig:HFstrainparentnu1}
\end{figure*}

Since $\nu = -1$ is a half-integer spin state, all the low energy $f$-modes in one spin sector ($\d$) are occupied. The other spin sector ($\uparrow$) contains a copy of the $\nu = -3$ parent state. Note that all the inter-valley coherence is restricted to the $\uparrow$ sector, since the Slater determinant of the two states in \Eq{eq:Uminus3} which span the $\d$ sector is diagonal in valley. Explicitly, the parent state is
\bea
\ket{GS_{0},-1} = \prod_{\mbf{R}}\frac{-if^\dag_{\mbf{R},+,K,\d} + f^\dag_{\mbf{R},-,K,\d} }{\sqrt{2}} \frac{if^\dag_{\mbf{R},+,K',\d} + f^\dag_{\mbf{R},-,K',\d} }{\sqrt{2}}  \frac{f^\dag_{\mbf{R},+,K,\uparrow}+if^\dag_{\mbf{R},-,K,\uparrow}+if^\dag_{\mbf{R},+,K',\uparrow}+f^\dag_{\mbf{R},-,K',\uparrow}}{2}\ket{\text{FS}} 
\eea
with the associated $f$-order parameter
\bea
\label{eq:Ofnu1}
O_{f,0} = \frac{1}{2}(\tau_0 \sigma_0 + \tau_3 \sigma_2) \otimes \bpm 0 & \\ & 1\epm+ \frac{1}{4} (\tau_0 \sigma_0 + \tau_1 \sigma_1 - \tau_2 \sigma_3 - \tau_3 \sigma_2) \otimes \bpm 1 & \\ & 0\epm \ .
\eea
We see explicitly that only the $\uparrow$ sector has inter-valley coherence, e.g. $\tau_1\sigma_1,\tau_2 \sigma_3$ which breaks $e^{i \th \tau_3}$. Thus we can interpret the parent state as a $\nu = -3$ state stacked with a valley-symmetric $C_3$-breaking background. 

We find in \Fig{fig:HFstrainparentnu1} that the one-shot band structure is in quantitative agreement with the fully self-consistent band structure.  Furthermore, we can quantitatively assess the similarity of the one-shot and self-consistent order parameters from the absolute error
\bea
||O_{f,1} - O_{f,\infty}||_F = .023
\eea
indicating the one-shot calculation is nearly 97\% accurate. The error in the parent state is $||O_{f,0} - O_{f,\infty}|| = .214$. This larger value can be easily understood by comparing $\Tr O_{f,0} = 3$ to $\Tr O_{f,\infty} = 3.22$, which shows that the self-consistent order parameter has increased its $f$-mode occupation and slightly depleted the Fermi sea. This effect is not captured in the parent state. Nevertheless, we find that 
\bea
||O_{f,0} O_{f,\infty}||^2_F = 3
\eea
to within machine precision, showing that $O_{f,\infty}$ completely contained the eigenspace of the parent state $O_{f,0}$. 

Lastly, we can compare the one-shot order parameter obtained from the Heavy Fermion model to the $f$-mode order parameter extracted from a 6-band BM model Hartree-Fock calculation with the same boost vector. We again find good agreement
\bea
||O_{f,1} - O^{BM}_{f,\infty}||_F = .20 
\eea
which is similar to the small value at $\nu = -2$.

\subsubsection{$\nu = 0$}

The last state we consider is $\nu = 0$ (all other states can be obtained by particle-hole symmetry since the parent state $f$ mode order is not altered by the presence of $\mu_1,\mu_2$ when we relaxation is included, which do not break any global symmetries of the parent  Hamiltonian). At $\nu = 0$, the parent state fills four $f$-modes. Since heterostrain splits the $f$-modes into four low energy and four high energy states, there is only one possible parent state:
\bea
\ket{GS_{0},0} = \prod_{\mbf{R},s=\uparrow,\d}\frac{-if^\dag_{\mbf{R},+,K,s} + f^\dag_{\mbf{R},-,K,\sigma} }{\sqrt{2}} \frac{if^\dag_{\mbf{R},+,K',s} + f^\dag_{\mbf{R},-,K',s} }{\sqrt{2}}  \ket{\text{FS}} 
\eea
which is diagonal in valley and hence has no inter-valley coherence. This means that $\ket{GS_{0},0}$ is invariant under boosts and will not lower its energy by forming an IKS state at the level of the parent state. The associated order parameter is
\bea
O_{f,0} = \frac{1}{2}(\tau_0 \sigma_0 + \tau_3 \sigma_2) s_0
\eea
and will not exhibit a Kekul\'e distortion due to lack of valley breaking. Note that no symmetries are spontaneously broken in this state, and the nontrivially order parameter $O_{f,0}$ merely reflects $C_{3z}$ breaking due to heterostrain.

\Fig{fig:HFstrainparentnu1} confirms that the one-shot calculation is an extremely good match to the fully self-consistent calculation, which is a valley-preserving semi-metal. (Later in \App{app:projected}, we will obtain analytical expressions for these bands.) To quantitatively compare the order parameters, we compute the absolute errors
\bea
||O_{f,0} - O_{f,\infty}||_F &= .076 \\
||O_{f,1} - O_{f,\infty}||_F &= .018 \\
\eea
showing that the one-shot, and even the parent state itself, reproduce the self-consistent state to high accuracy. This can be explained in part by the fact that $\Tr O_{f,\infty} =4$ so the parent state is has the same $f$ occupation of the self-consistent state. 

Lastly, we can compare the one-shot order parameter obtained from the Heavy Fermion model to the $f$-mode order parameter extracted from a 6-band BM model Hartree-Fock calculation with the same boost vector. We again find remarkably good agreement
\bea
||O_{f,1} - O^{BM}_{f,\infty}||_F = .11 
\eea
which is even smaller than the value compared to $\nu = -2$ and $\nu = -1$.

In sum, we find that at fillings $0,\pm1,\pm2$, the one-shot theory of the IKS states is quantitatively predictive, yielding analytical model wavefunctions that capture the $C_3$-breaking, inter-valley coherence (or lack thereof at $\nu=0$), and IKS order of the self-consistent Hartree-Fock ground states. We have compared the one-shot Heavy Fermion $f$ order parameters to those of the self-consistent 6-band BM, and found absolute errors $\lesssim .2$.

\subsection{Flat Band Projection within Heavy Fermion Theory}
\label{app:SPperturbation}

Having established the qualitative and nearly quantitative accuracy of a one-shot theory based on heavy fermion parent states, we now develop a perturbation theory to obtain analytical expressions for the band structures of the one-shot Hamiltonians, which is equivalent to projecting the six-band heavy fermion model onto the two flat/active bands. We will use this expression at $\nu = -2$ to show how valley boosts open up a gap in the one-shot spectrum and thereby stabilize IKS order. 

We begin our derivation by studying the unstrained, single-particle heavy fermion model. This model contains a few key parameters, the Dirac velocity $v$, the $fc$ coupling $\gamma$ and $v'$, and the $c$-electron mass $M$. When $M \to 0$, the active bands are perfectly flat and the model possesses an emergent $SO(2)$ rotation symmetry, which will allow us to obtain simple expressions for the energies and eigenstates. This will be the starting point of our perturbation theory. 
 
Explicitly, the Hamiltonian can be written in terms of $\mbf{k} = k(\cos \th, \sin \th)$ for $M \to 0$ as
\bea
\label{eq:hKO2}
h_K(\mbf{k}) &\to  U(\th) \bpm 
0 & v k  \sigma_0  &  \gamma \sigma_0 + v' k \sigma_1 \\
v k  \sigma_0  & 0 & \\
\gamma \sigma_0 + v' k \sigma_1 & & 0 \\
\epm U^\dag(\th), \qquad U(\th) = \text{diag }(e^{i \th}, e^{2 i \th}, 1, e^{3i \th}, e^{i \th}, e^{2i \th} ) \\
\eea
where the $e^{i \th}, e^{2 i \th}$ factors represent the angular momenta of complex $\Gamma_3$ irrep, and $1, e^{i 3 \th}$ represent the two real irreps $\Gamma_1\oplus \Gamma_2$ (note that the lattice breaks $SO(2)$ to $C_3$, such that angular momentum is only defined mod 3). After the basis transformation by $U(\th)$, no $\th$ dependence appears in the matrix in \Eq{eq:hKO2}, so the spectrum is $SO(2)$ invariant. Note that the $M$ term breaks this symmetry by reintroducing $\th$ dependence since it couples the $1, e^{i 3 \th}$ irreps. Throughout this section for notational brevity, we neglect the $e^{-k^2\la^2/2}$ damping factor that multiples $\gamma$ and $v'$. However, it may be trivially restored by taking $\gamma \to \gamma e^{-k^2\la^2/2}$ and $v' \to v' e^{-k^2\la^2/2}$ at the end of any algebraic calculation.  

The eigenvectors of the 2 flat bands (which have exactly $E=0$ in the flat band limit $M\to 0$) are 
\bea
\label{eq:appUK}
U_{+,K}(\mbf{k}) &= \frac{1}{\sqrt{2(v^2k^2 + (v'k + \gamma)^2)}}U(\th)\bpm
0, & 0, & v'k + \gamma, & v' k + \gamma, & -v k, & -v k \epm^T \\
U_{-,K}(\mbf{k}) &= \frac{1}{\sqrt{2( v^2 k^2 + (v' k - \gamma)^2)}}U(\th)\bpm
0, & 0, & -v'k + \gamma, & v' k - \gamma, & -v k, & v k \epm^T \ . \\
\eea
In the case $v' = 0$, which is exclusively the limit in which we will perform analytical calculations in this work and captures the essential features of the strained bands quite well (see Ref. \cite{HER24a}), it is simpler to work with the alternative orthonormal basis
\bea
U_{1,K} &= (U_+(\mbf{k}) + U_-(\mbf{k}))/\sqrt{2}, \quad U_{2,K} = (U_+(\mbf{k}) - U_-(\mbf{k}))/\sqrt{2}, \qquad (v' = 0) \\
U_{K}(\mbf{k})  &= \bpm U_{1,K}(\mbf{k}) & U_{2,K}(\mbf{k})  \epm
\eea
The eigenvectors in the other valley $U_{i, K'}(\mbf{k})$ are obtained by time-reversal. 

We can now re-introduce $M$ and $\delta h_\eps$ with degenerate perturbation theory within projection onto the flat bands. We find
\bea
U^\dag_K(\mbf{k}) (\delta h_\eps + h_M) U(\mbf{k}) &= \frac{\eps_-}{v^2 |\mbf{k}|^2 + \gamma^2} \bpm
 2 \gamma  c''  v k_y &  -i M_f v^2 (k_x-i k_y)^2 \\
 i M_f v^2 (k_x+i k_y)^2 & 2 \gamma c'' v k_y \\
\epm + \frac{\gamma^2 M}{v^2 |\mbf{k}|^2 + \gamma^2} \bpm
 0 & 1 \\ 1 & 0 \\
\epm
\eea
where we have neglected the $M'$ term for simplicity, since the strain Hamiltonian is dominated by the $c''$ and $M_f$ terms (see Ref.\cite{HER24a}).

\begin{figure}
\centering
\includegraphics[width=\columnwidth]{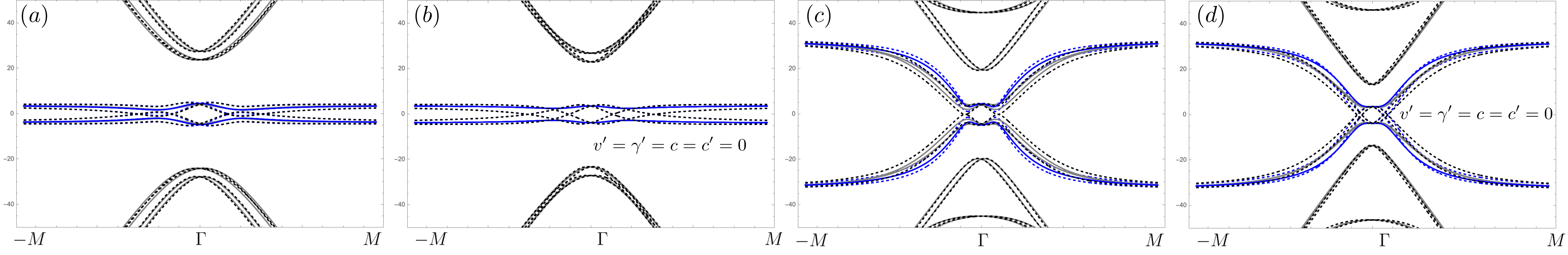}
\caption{Comparison of the numerical (gray) and analytical (blue) projected bands for the non-interacting $(a,b)$ and  one-shot Hartree-Fock $(c,d)$ bands at $\nu=0$ (see \Eq{eq:hintapp0}). Panels $(b)$ and $(d)$ simplfy the model by setting subdominant terms $v',\gamma',c,c'$ to zero, which still capture the essential features of the bands. All plots are shown at $\mbf{q}=0$, $\eps = 0.15\%$ and include both valleys. The dashed lines are obtained on a rotated path through the BZ to show Dirac nodes.}
\label{fig:cnpproj}
\end{figure}

We have shown that projecting the model onto the $\eps = 0, M=0, v'=0$ flat bands provides a quantitatively accurate perturbation theory for the non-interacting bands (see \Fig{fig:cnpproj}(a)). We will now perform this projection on the mean-field Hamiltonian.

\subsection{One-Shot Mean-Field Hamiltonian}
\label{app:projected}

The mean-field heavy fermion Hamiltonian takes into account the full spin and valley degrees of freedom, leading to a $24\times24$ matrix $h_0(\mbf{k}) + h_{int}$ where $h_{int}$ is the mean-field form of the interaction obtained by contracting with the order parameter. Note that $h_0(\mbf{k})$ is the non-interacting Hamiltonian containing all spins and valleys. We consider one-shot states where only the $f$ order parameter is nonzero. Then mean-field interaction Hamiltonian $h_{int}$ can be written using App. 4 of Ref.\cite{Song20211110MATBGHF}. The result (using $P = O_f^T$ and $\nu_f = \Tr O_f - 4$) is
\bea
\label{eq:hintapp}
h_{int} &= \bpm 
\nu_f W_1 \mathbb{1}  & & \\
& \nu_f W_3 \mathbb{1} - \frac{J}{2}(\tau_3 (P - \frac{1}{2}) \tau_3 + \sigma_3 (P - \frac{1}{2}) \sigma_3 ) & \\
& & - U_1 (P - \frac{1}{2}) + \nu_f (U_1 + 6 U_2) \mathbb{1}
\epm  \ . \\
\eea
We will consider $\nu = -2$ where the ground state is a spin singlet, all bands are doubly degenerate. In principle the same calculation can be performed for $\nu = -3$, but since Hartree-Fock shows large charge fluctuations for the interaction strength we use, in this work we focus on $\nu = -2$. As such we only need to study a single spin sector. Using the parent states in \App{sec:parentstates} for now at $\mbf{q}=0$, we compute
\bea
\tau_3 (P - \frac{1}{2}) \tau_3 + \sigma_3 (P - \frac{1}{2}) \sigma_3 = - \frac{1}{2} - \frac{1}{2} \tau_1 \sigma_1 \\
\eea
in the spin $\uparrow$ sector. We now project the one-shot Hamiltonian in the spin $\uparrow$ sector onto the flat bands, obtaining
\bea
\label{eq:projectedhapp}
\bar{h}_{HF, \uparrow}(\mbf{k}) &= \mathcal{U}^\dag(\mbf{k}) \lp h_0(\mbf{k}) + h_{int} \rp \mathcal{U}(\mbf{k}) \\
\eea
where the columns of $\mathcal{U}(\mbf{k}) $ are $U_{1,K}(\mbf{k}) , U_{2,K}(\mbf{k}) , U_{1,K'}(\mbf{k}) , U_{2,K'}(\mbf{k})$. Along the $k_x$ direction, we obtain simple expressions for the bands (each is spin-degenerate):
\bea
\label{eq:appspinhamproj}
\text{spectrum }\bar{h}_{HF, \uparrow}(k_x,0) &= \frac{(J-4 W_3) \gamma^2 - 3 v^2 k_x^2 (U_1+8U_2) \pm 2\sqrt{v^4 k_x^4 (M_f \eps_-)^2 + M^2 \gamma^4 } }{2(v^2 k_x^2 +\gamma^2)}, \\
& \frac{- 2 W_3 \gamma^2 - 2 v^2 k_x^2 (U_1+6U_2) \pm \sqrt{v^4 k_x^4 (U_1/2 -M_f \eps_-)^2 + M^2 \gamma^4 } }{v^2 k_x^2 +\gamma^2} \\
\eea
which are compared with the full band structure in \Fig{fig:appnu2q0}. We see that there is no indirect gap, since filling $\nu = -2$ is achieved by occupying the bottom projected band (the $-$ sign in the second line of \Eq{eq:appspinhamproj}) in each spin sector. We can show this analytically by comparing the energies of the lowest occupied band at $\Gamma$ and the lowest unoccupied band at the BZ edge assuming the bottom band is filled at each $\mbf{k}$. The maximum of the bottom band has energy $-M - 2W_3 = -104.13$meV, and the minimum of the next highest band has energy $- \frac{3}{2}U_1 - 12 U_2 - |M_f \eps_-| < - \frac{3}{2}U_1 - 12 U_2 = -114.8$meV. Since $|M_f \eps_-| < 5$meV is smaller than $U_1$, we see that the unboosted spectrum must have no indirect gap based on the hierarchy of interactions, $W_3$ vs $U$. As discussed in the Main Text, a direct gap is also forbidden by symmetry due to the presence of protected Dirac nodes. Next, we generalize this calculation to a general IVC angle and show that Dirac nodes remain protected.

\subsubsection{Presence of Dirac Nodes for General IVC Angle without IKS Boost}
\label{app:DiracIVC}

We now show that generalizing the above calculation to a generic IVC angle does not gap the Dirac nodes. This is motivation for the necessity of introducing the IKS boost. To do this, we study the projected one-shot mean-field Hamiltonian obtained from the generic IVC order (without IKS boost) parameter obtained from $(\cos \th \zeta_{K} + \sin \th \zeta_{K'})$ as discussed in the Main Text. The order parameter is
\bea
O = P^T &= \frac{1}{4} \lp 1 + \tau_3 \sigma_2 + \sin 2\th (\tau_1 \sigma_1 - \tau_2 \sigma_3) + \cos 2\th (\tau_3 \sigma_0 - \tau_0 \sigma_2) \rp \otimes s_0 \ .
\eea
The spin Pauli matrix $s_0$ corresponds to filling $\nu = -2$. At the end of this section, we will explain how our finding also holds for $\nu = -3$. Using \Eq{eq:hintapp}, we obtain the one-shot matrix $h_{int}(\th)$ and project the Hamiltonian using the $\mathcal{U}(\mbf{k})$ eigenvectors as before. In the spin $\uparrow$ sector, we obtain a $4\times4$ matrix that can be decomposed into generalized Pauli matrices $\Gamma_{ab} = \sigma_a \otimes \sigma_b$. Although we use the notation $\sigma$, these Pauli matrices do not refer to the heavy-fermion orbital, and $\Gamma$ acts instead on the four spin-$\uparrow$ projected bands. The decomposition is
\bea
\label{eq:decompapp}
4 (v^2 k^2 + \gamma^2)\mathcal{U}^\dag(\mbf{k}) (h_0(\mbf{k}) + h_{int}(\th)) \mathcal{U}(\mbf{k}) &= (-v^2 k^2(7 U_1 + 48 U_2) + (J-8W_3) \gamma^2 )\Gamma_{00} \\
&\quad+ (4 \gamma^2 M + 2 v^2 k_x k_y(U_1 - 4 M_f \eps_-)) \Gamma_{01}\\
&\quad -4 v^2 k_x k_y \cos \th \sin \th \Gamma_{10} \\
&\quad + (v^2 U_1 k^2 + J \gamma^2) \sin 2\th \Gamma_{11} \\
&\quad - 2 v^2 k_x k_y U_1 \cos 2\th \Gamma_{31} \\
&\quad +( (v^2 U_1 k^2 + J \gamma^2)\cos 2\th +8 c'' \eps_- v k_y \gamma)\Gamma_{30} \\
&\quad + v^2(k_x^2-k_y^2) (U_1 \cos \th \Gamma_{02} + U_1 \sin \th \Gamma_{23} - (U_1 + M_f \eps_-)\Gamma_{32}) \ .
\eea
To show the existence of Dirac nodes, it turns out to be sufficient to examine the $k_x = -k_y$ line, where the last line in the equation above vanishes. We see from the remaining matrices that $\Gamma_{01}$ is a symmetry of the Hamiltonian along this line. Thanks to this symmetry, the Hamiltonian splits into two $2\times2$ blocks with opposite $\Gamma_{01}$ eigenvalues. Thanks to this emergent symmetry, the exact spectrum can be found and the Dirac nodes can be identified as protected crossings in the spectrum between different $\Gamma_{01}$ symmetry sectors. Let us denote the lowest energy band in the $\Gamma_{01}=\pm$ sector by $E_{\pm}(k_x = - k_y)$. We will now prove that these bands undergo a crossing. We compute that
\bea
E_+(0) = 2W_3 + M, \quad E_-(0) = 2W_3 - M
\eea
so that $E_+(0) > E_-(0)$ since $M > 0$, but that at the BZ edge (which can be accessed in our $k \cdot p$ approximation by taking $|k| \to \infty$), we obtain
\bea
E_+(k \to \infty) = - \frac{5}{2} U_1 - 12 U_2+ M_f \eps_-, \quad E_-(k \to \infty) = - \frac{3}{2}U_1 - 12 U_2 - M_f \eps_-
\eea
so that $E_+(k \to \infty) < E_-(k \to \infty)$, recalling that $U_1 > 0$ and $M_f \eps_-<0$. Thus a level crossing must occur along the $k_x = -k_y$ at  $\mbf{k} = \pm (k^*,-k^*)$. Our results hold for all IVC angles $\th$. Lastly, we point out that if $M < 0$, as can be achieved for instance by increasing the twist angle \cite{CAL23}, no Dirac nodes appear along this line. However, it can be easily checked that they occur instead on the line $k_x = + k_y$ projected by the same $\Gamma_{01}$ symmetry. At $M= 0$, the two Dirac nodes come together at $\mbf{k}=0$ and form a double vortex band touching.

Lastly, we discuss the generalization to $\nu = -3$, which has the same valley-orbital order parameter but is spin polarized. From \Eq{eq:hintapp}, we observe that the only change at $\nu = -3$ within the spin $\uparrow$ sector is to change $\nu_f \to -3$. This is equivalent to a redefinition of $W_1, W_3$ and $U_2$. But from \Eq{eq:decompapp}, we see that these parameters only enter the constant term $\Gamma_{00}$ in the projected Hamiltonian. Hence changing this term will not change the Dirac nodes. 
 
\subsubsection{Charge Neutrality semi-metal spectrum and Dirac nodes}
\label{app:nu0SMdirac}

For completeness, we also study $\nu=0$ where the predicted order parameter in nonzero heterostrain is full filling of all lower-band $f$-modes with no global symmetry breaking, and $O_f = \frac{1}{2}(\tau_0 \sigma_0 - \tau_3 \sigma_2) s_0$ for heterostrain along the $x$ axis. The one-shot mean-field Hamiltonian is
\bea
\label{eq:hintapp0}
h_{int,\nu=0} &= \bpm 
0  & & \\
&  - \frac{J}{2}(\tau_3 (P - \frac{1}{2}) \tau_3 + \sigma_3 (P - \frac{1}{2}) \sigma_3 ) & \\
& & - U_1 (P - \frac{1}{2})
\epm  \\
\eea
where $P = O_f^T$. Projecting to $U(\mbf{k})$ yields a $2\times2$ matrix which can be solved to yield the eigenvalues (written in polar coordinates $k,\th$)
\bea
\label{eq:radical}
E^{\nu=0}_\pm(\mbf{k}) = \frac{4 \gamma  c'' v k \epsilon _- \sin \th\pm \sqrt{4 v^4k^4 M_f^2  \epsilon _-^2-4 v^4 k^4 M_f U_1 \epsilon _-+v^4k^4 U_1^2 +4 \gamma ^2 v^2k^2 M \sin (2 \theta ) \left(U_1-2 M_f \epsilon _-\right)+4 \gamma ^4 M^2}}{2 \left(\gamma ^2+v^2k^2 \right)}
\eea
which are shown in \Fig{fig:cnpproj}(b) and compare favorably with the numerical spectrum. With these expressions, we can solve for the Dirac points exactly, which must occur between the conduction and valence bands since $C_{2z}\mathcal{T}$ is unbroken. We find that the Dirac points occur where the argument of the radical in \Eq{eq:radical} vanishes, leading to 
\bea
\th = - \frac{\pi}{4}, \quad v^2 k^2 = \frac{M\gamma^2}{U_1/2 -  M_f \eps_-} = \frac{M\gamma^2}{U_1/2 + M_f |\eps_-|} \ .
\eea
Thus we have obtained an analytical understanding of the semi-metallic phase at charge neutrality. 

\begin{figure}
\centering
\includegraphics[width=\columnwidth]{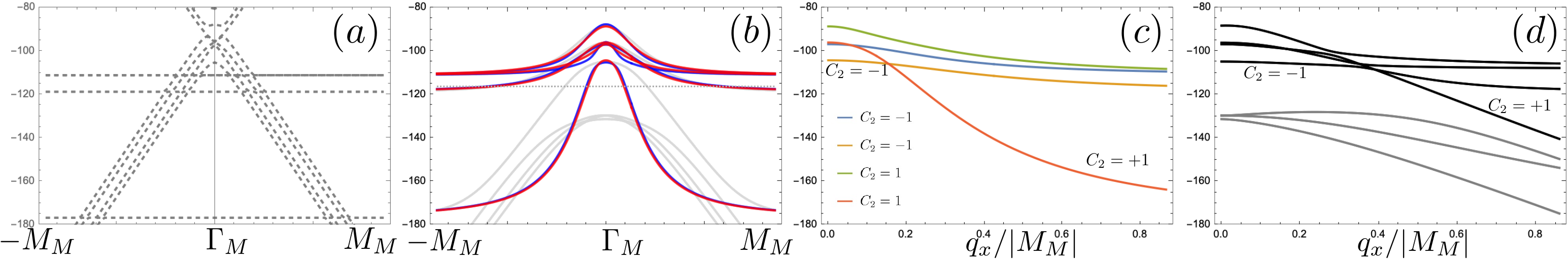}
\caption{$(a)$ Decoupled ($\gamma = \gamma' = c'' = 0$) mean-field band structure (dashed) at $\nu=-2$ with a hole pocket visible at $\mbf{q}=0$. $(b)$ Full mean-field band structure  at $\mbf{q}=0$ (gray) compared with the projected calculation (blue) and the projected calculation with $M=0$ (red). The Fermi level (dotted) is slightly above the flat band energy at the edge of the BZ due to the hole pocket at $\Gamma$. A very good match is obtained for the 3 lowest conduction bands per spin, and the $\mbf{k}=0$ energy of the highest valence band is also accurate. At lower energies, mixing with the $\Gamma_3$ $c$-electrons invalidates the flat band projection. $(c)$ Perturbation theory expressions for the $\mbf{k}=0$ energies as a function of $\mbf{q}$, the IKS boost. We see a level crossing at $|\mbf{q}| \sim .2 M$ after which the lowest $C_{2z}$ eigenvalue switches from $-1$ to $+1$. $(d)$ The same trend is reproduced in numerical one-shot calculations. 
}
\label{fig:appnu2q0}
\end{figure}

\subsection{Boosted Mean-Field Spectrum}
\label{app:projectedMF}

The projected IKS one-shot Hamiltonian has the same order parameter in the boosted $f$ basis, but now differs in the kinetic term:
\bea
\label{eq:boostedf}
\bar{h}_{HF, \uparrow,\mbf{q}}(\mbf{k}) &= \mathcal{U}_\mbf{q}^\dag(\mbf{k}) \lp h_{0,\mbf{q}}(\mbf{k}) + h_{int} \rp \mathcal{U}_\mbf{q}(\mbf{k}) \\
\eea
where the boosted kinetic term and its boosted flat band eigenvectors are
\bea
\mathcal{U}_\mbf{q}(\mbf{k}) = \lp U_{1,K}(\mbf{k}+\frac{\mbf{q}}{2}) , U_{2,K}(\mbf{k}+\frac{\mbf{q}}{2}) , U_{1,K'}(\mbf{k}-\frac{\mbf{q}}{2}) , U_{2,K'}(\mbf{k}-\frac{\mbf{q}}{2})  \rp, \quad h_{0,\mbf{q}}(\mbf{k}) = h_{0,K}(\mbf{k}+\frac{\mbf{q}}{2}) + h_{0,K'}(\mbf{k}-\frac{\mbf{q}}{2}) \ .
\eea
To show that the IKS boost opens up an indirect gap, we compute the eigenvalues at the $\mbf{k}= 0$ point in the boosted BZ. This can be accomplished analytically, yielding
\bea
\label{eq:Eqkequal0app}
E_\mbf{q}(\mbf{k}=0) &= \frac{-16 \gamma ^2 W_3-3 q^2 v^2 (U_1+8 U_2) \pm  2\sqrt{16 \gamma ^4 M^2+M_f^2 q^4 v^4 \eps_-^2}}{8 \gamma ^2+2 q^2 v^2}, \\
& \qquad \frac{4 \gamma ^2 (J-4 W_3)\pm\sqrt{64 \gamma ^4 M^2+q^4 v^4 (-2 M_f \eps_-+U_1)^2}-4 q^2 v^2 (U_1+6 U_2)}{8 \gamma ^2+2 q^2 v^2} \ . \\
\eea
Note that it is also possible to obtain the bands $\mbf{k}\neq 0$ for all $\mbf{q}$, but these expressions contain quartic roots and are too cumbersome for analytics. We plot these boosted $\mbf{k}=0$ energies as a function of $\mbf{q}$ in \Fig{fig:appnu2q0}(c) and compare with the corresponding one-shot energies of the full mean-field Hamiltonian in \Fig{fig:appnu2q0}(d). We find good agreement, with both showing that $E_\mbf{q}(\mbf{k}=0)$ is decreasing with $\mbf{q}$, and that a level crossing occurs near $|\mbf{q}| = .2 M_M$ (visible as the crossing of the red and orange lines in \Fig{fig:appnu2q0}(c)). Both facts are quite important. To understand the level crossings, we compute the eigenvectors. They are written $\mathcal{U}_\mbf{q}(\mbf{k}=0) v_\mbf{q}(\mbf{k}=0)$ where $v_\mbf{q}(\mbf{k}=0)$ are the 4-component vectors
\bea
v_\mbf{q}(\mbf{k}=0) &= (-1,  \pm \arg \lp  - 4 M \gamma^2 + i M_f q^2 v^2 \eps_-\rp,  \pm \arg \lp 4 M \gamma^2 - i M_f q^2 v^2 \eps_-\rp, 1)^T /2 , \\
 &= (1, \pm \arg \lp 8 M \gamma^2 + i q^2 v^2(U_1 - 2M_f \eps_-)\rp, \pm  \arg \lp 4 M \gamma^2 - i M_f q^2 v^2 \eps_-\rp, 1)^T /2 \\
\eea
which we find have $C_{2z}$ eigenvalues $-1,-1$ and $+1,+1$ respectively. Since the bands have $C_{2z}$ eigenvalues $+1$ at the 3 $M$ points at the edge of the BZ (where they are of $f$-character and the $C_{2z} = +1$ gauge has been fixed in our conventions of the order parameter in the Main Text), an odd $C_{2z}$ eigenvalue at $\Gamma$ enforces a topological semi-metal.  This is because the product of $C_{2z}$ eigenvalues is odd implying a nonzero Chern number if the band is gapped (i.e. if it were isolated), but the presence of spinless time-reversal symmetry requires the Chern number to vanish if gapped (i.e. if isolated). Recall that the Fermi level at $\nu = -2$ in the projected Hilbert space is chosen to occupy the lowest two bands, or the lowest band in each spin sector. It is this lowest band (in each spin sector) which has $C_{2z} = -1$ at $\Gamma$ and $C_{2z} = +1$ at the $M$ points. Lastly, we would like to emphasize a technical point, which is the non-periodicity of the projected model in the moir\'e BZ. This arises because of the $k \cdot p$ approximation made in obtaining the spectrum around the $\Gamma$ point, and we only use this model to study the energy spectrum in this regime.  The full model, with higher $c$-electron shells, has moir\'e BZ periodicity and has well-defined $C_{2z}$ eigenvalue fixed by the $f$-mode wavefunctions. The wavefunctions can be found \cite{Song20211110MATBGHF} by summing over plane-waves, but the expressions become more cumbersome.

Thus there can be no direct gap until the $q_x$ is increased beyond the level crossing. Second, the decrease in the $\Gamma$ point energy below the conduction band minimum shows that an indirect gap can also open. The flat band projection approximation breaks down at larger $\mbf{q}$ because of mixing with the $\Gamma_3$ $c$-electrons that is not accounted for in the projected theory. Thus we will develop an understanding of the IKS boost $\mbf{q}$ through an alternative limit, the decoupled limit where hybridization is set to 0, in \App{app:optimalboostfc}.

\subsubsection{Braiding Thresholds}

We briefly discuss the approximations made in obtaining the thresholds $q_n$, where the Dirac node pair in the lowest two conduction bands appears, and $q_g$, where the Dirac nodes annihilate in between the valence and conduction bands. 

Inspecting \Fig{fig:appnu2q0}(c) and \Eq{eq:Eqkequal0app}, we observe that $q_n$, where the blue and red lines cross and a Dirac node pair is nucleated, is obtained from the solution to
\bea
& \frac{-16 \gamma ^2 W_3-3 q^2 v^2 (U_1+8 U_2) +  2\sqrt{16 \gamma ^4 M^2+M_f^2 q^4 v^4 \eps_-^2}}{8 \gamma ^2+2 q^2 v^2} \\
& \qquad \qquad  \qquad \qquad  \qquad \qquad =  \frac{4 \gamma ^2 (J-4 W_3)-\sqrt{64 \gamma ^4 M^2+q^4 v^4 (-2 M_f \eps_-+U_1)^2}-4 q^2 v^2 (U_1+6 U_2)}{8 \gamma ^2+2 q^2 v^2} \ .
\eea
Since the $q^4$ terms in the square root are negigble at small $\mbf{q}$, we drop them and obtain the approximate solution
\bea
q_n = 2 \sqrt{\frac{J - 4 M}{U_1}} \frac{\gamma}{v},
\eea
as in \Eq{eq:qexpressions} of the Main Text. Numerically this approximation is better than $95\%$ accurate. 

Next we examine $q_g$, which is the solution to
\bea
& \frac{-16 \gamma ^2 W_3-3 q^2 v^2 (U_1+8 U_2) -  2\sqrt{16 \gamma ^4 M^2+M_f^2 q^4 v^4 \eps_-^2}}{8 \gamma ^2+2 q^2 v^2} \\
& \qquad \qquad  \qquad \qquad  \qquad \qquad =  \frac{4 \gamma ^2 (J-4 W_3)-\sqrt{64 \gamma ^4 M^2+q^4 v^4 (-2 M_f \eps_-+U_1)^2}-4 q^2 v^2 (U_1+6 U_2)}{8 \gamma ^2+2 q^2 v^2} \ .
\eea
where the orange and red lines cross in \Fig{fig:appnu2q0}(c). Since $q_g$ is larger, it is not justified to drop the $O(q^4)$ term in the radical. Instead, we proceed by approximating $\eps_- = 0$, which removes the $q^4$ term on the lefthand side but not on the righthand side, where the $U_1$ prefactor is still large. The solution is then approximated by
\bea
q_g = \sqrt{\frac{J + M}{JU_1/2}} \frac{\gamma}{v}
\eea
which is better than $90\%$ accurate. It is obvious that higher order corrections can be kept to improve the accuracy. 

\subsubsection{Comparison with other Orders}
\label{app:compare}

\begin{figure}
\centering
\includegraphics[width=\columnwidth]{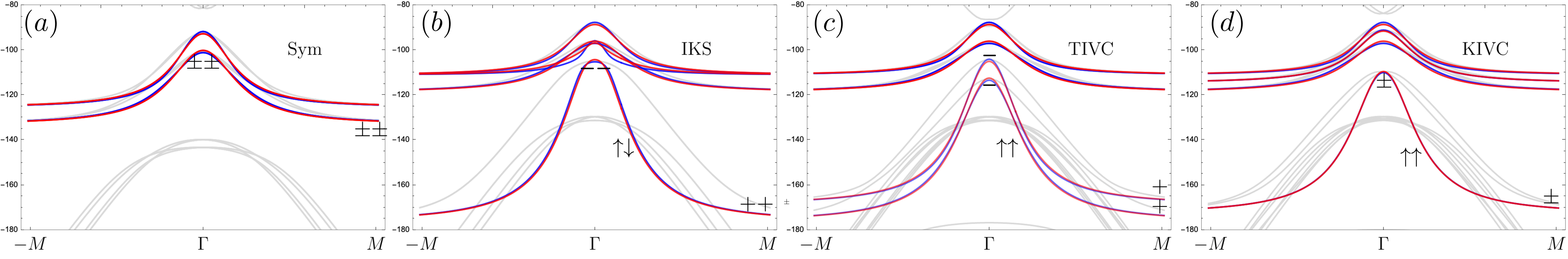}
\caption{One-shot $\nu=-2$ band structures (gray) compared with the flat band projected one-shot band structure using all parameters (blue) vs $M=0$ (red). ($a$)-($d$) show the bands using order parameters corresponding to the symmetric state, IKS at $\mbf{q}=0$, TIVC, and KIVC respectively. The $C_{2z}$ eigenvalues at the high-symmetry points are marked (multiple marks if the band is multiply degenerate at such points), and the spin character of the two valence bands are marked. One can see that the projected bands are a good approximation in the conduction bands and near the $\Gamma$ and $M$ points in the valence bands, but are worse in the middle of the BZ where the other $c$ electrons hybridze strongly and deform the bands. Note that the symmetry-breaking orders IKS, TIVC, and KIVC all have 2 dispersive bands that are lowered by $U$ (in the IKS and KIVC, they are degenerate) which are supposed to be fully occupied at $\nu = -2$, although Dirac points or negative indirect gaps prevent the Fermi level from lying in a gap. The symmetric state has no bands pushed by $U$ away from the Fermi level. 
}
\label{fig:appnu2qother}
\end{figure}

For completeness, we compare the one-shot IKS results discussed above with other $f$-orders at $\nu = -2$ in the presence of heterostrain. In particular, we consider the KIVC, which is favored in the absence of heterostrain, as well as the TIVC and symmetric state. The explicit $f$-mode order parameters we consider are (see Ref.\cite{Song20211110MATBGHF})
\bea
O_f &= \frac{1}{2}(\tau_0 \sigma_0 + \tau_2 \sigma_2) \otimes \ket{\uparrow}\bra{\uparrow}, \qquad \text{(KIVC}) \\
O_f &= \frac{1}{2}(\tau_0 \sigma_0 + \tau_1 \sigma_1) \otimes \ket{\uparrow}\bra{\uparrow}, \qquad \text{(TIVC)} \\
O_f &= \frac{1}{4}\tau_0 \sigma_0 s_0, \qquad \text{(Symmetric)} \\
\eea
which, with \Eq{eq:hintapp}, fully determine the HF Hamiltonian at $\nu = -2$. Here we have made use of the gauge-freedom of the inter-valley phase factor to choose $O_f$ such that it commutes with the $C_{2z}$ representation $\tau_1 \sigma_1$. The IKS preserves spin-ful time-reversal $\mathcal{T}^s = i \tau_1 s_2 K$, the TIVC preserves spin-less time-reversal $\mathcal{T} = \tau_1 K$, the KIVC preserves the modified Kramers' time-reversal $\mathcal{K} = i \tau_2 K$, and the symmetric state preserves all symmetries. The presence of $C_{2z}$ and an anti-unitary symmetry allows us to determine the presence of Dirac nodes, which we now discuss. 

For each $O_f$, we diagonalize the projected one-shot Hamiltonian in \Eq{eq:projectedhapp} and obtain the $C_{2z}$ eigenvalues at the $\Gamma$ and $M$ point in each spin sector at $\nu = -2$. Note that the TIVC and KIVC are spin-polarized, while the IKS is a spin-singlet. The results are shown in \Fig{fig:appnu2qother}. Let us recall the proof of Dirac nodes for the IKS: in each spin sector, the lowest valence band has an odd number of $C_{2z}=-1$ eigenvalues, and hence there must be a Dirac node connecting the valence and conduction band in each spin-sector (as we have verified numerically). In the TIVC, the two valence bands are both in the spin $\uparrow$ sector and each has an odd number of $C_{2z}=-1$ eigenvalues. This means that they must have a Dirac node between them (as we have verified numerically), but there is not a protected Dirac node between the valence and conduction bands since the there is an even number of total $C_{2z} = -1$ eigenvalues in the valence spin $\uparrow$ sector. We have also verified numerically that there is a global direct gap in agreement with this result. In the KIVC, $\mathcal{K}^2 = -1$ ensures that the two valence bands are degenerate at the time-reversal invariant points, and we find that the total number of $C_{2z} = -1$ eigenvalues in the valence spin $\uparrow$ sector is even so there are no protected Dirac nodes between the valence and conduction bands.  We have also verified numerically that there is a global direct gap in agreement with this result. Lastly, in the symmetric state, $C_{2z}\mathcal{T}$ is preserved in each spin-valley sector, so there are Dirac nodes between the interaction-renormalized flat bands. 

In summary, we have shown that the IKS state, but not the TIVC or KIVC, have symmetry-projected Dirac nodes preventing a direct gap at $\mbf{q}=0$. By turning on $\mbf{q}$ to a finite value, it is possible for the Dirac nodes to annihilate and open a direct gap, as we now show.

\subsection{Criterion for the optimal IKS Boost}
\label{app:optimalboostfc}

\begin{figure}
\centering
\includegraphics[width=\columnwidth]{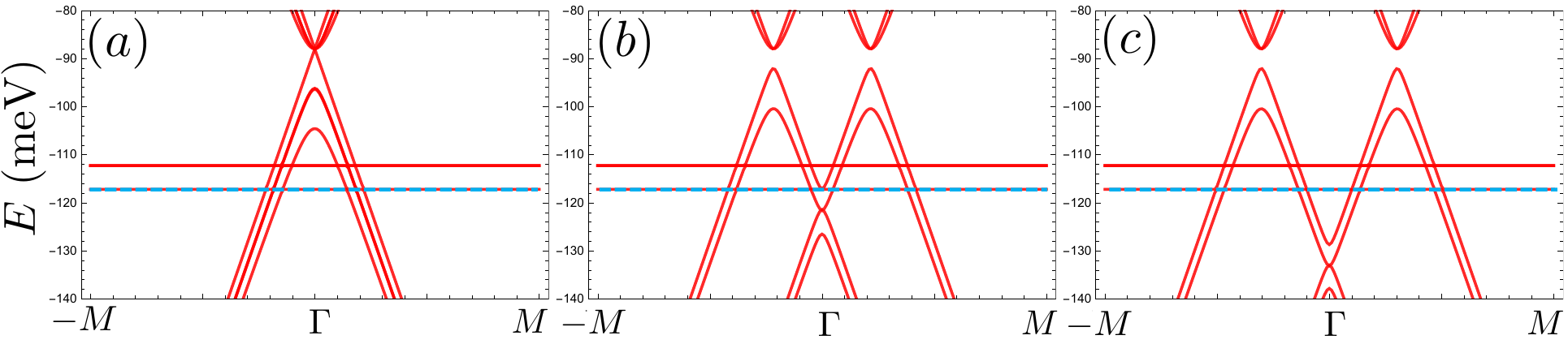}
\caption{Decoupled ($\gamma = \gamma' = c'' = 0$) mean-field band structures at $q = 0$, $q = .5|M_M|$ and $q = .6|M_M|$ in $(a), (b)$, and $(c)$ respectively, at $\eps = .001$. Our proposed optimality condition is satisfied by $(b)$ where the highest of the four valence $c$-electron states touches the lowest conduction $f$ mode energy. When hybridization is re-introduced, this will maximize the mixing between $f$ and $c$ in order to open the IKS gap. The small splittings that appear are due to the the interaction $J$ which couples the $\Gamma_1,\Gamma_2$ modes, as well as the single-particle strain terms $c,c',M'$, discussed in \App{app:strainapp}, which couple all  $\Gamma_1,\Gamma_2, \Gamma_3$ irreps due to the breaking of $C_{3z}$ and $C_{2x}$. 
}
\label{fig:decoupledqexamples}
\end{figure}

We now propose a simple criterion for the optimal IKS boost. To motivate this criterion, we first compute the one-shot $f$-block eigenvalues of $h_0(\mbf{k}) + h_{int}$ in the presence of strain and IKS order, which are
\bea
\label{eq:appflevelsdecoupled}
E_{f\text{-only}} &= -M_f\eps_--\frac{3 U_1}{2}-12 U_2 \quad (4), \\
& \quad M_f\eps_--\frac{3 U_1}{2}-12 U_2 \quad (2), \\
& \quad M_f\eps_--\frac{5 U_1}{2}-12 U_2 \quad (2), \\
\eea
in decreasing order, where the degeneracy is marked in parentheses. These energies describe the $f$-modes at the edge of the BZ where hybridization with the $c$-electrons is very weak due to their large kinetic energy. The lowest two eigenvalues $M_f\eps_--\frac{5 U_1}{2}-12 U_2$ are pulled far below the Fermi level by the large $U$ interaction (they are outside the plotting range in \Fig{fig:decoupledqexamples}). The parent state occupies these two $f$ modes and the Fermi sea of $c$-electrons. As discussed in the last section, the one-shot spectrum can gap at the Fermi level through hybridization with the $c$-electrons thanks to the IKS boost, and this gapping decreases the energy of the quasi-particle Hartree-Fock bands. Note that such hybridization occurs with the $f$-modes near the Fermi level, the lowest of which is $M_f\eps_--\frac{3 U_1}{2}-12 U_2$ (see \Fig{fig:decoupledqexamples} marked in blue) which is split from the higher ones by heterostrain. We can treat this $f$ mode and the nearby $c$ mode as a two-level system at $\mbf{k}=0$, in which case the maximum hybridization occurs when the $f$ and $c$ modes are at equal energy. We will use this criterion to approximate the optimal value of $q$ analytically. 

\begin{figure}
\centering
\includegraphics[width=\columnwidth]{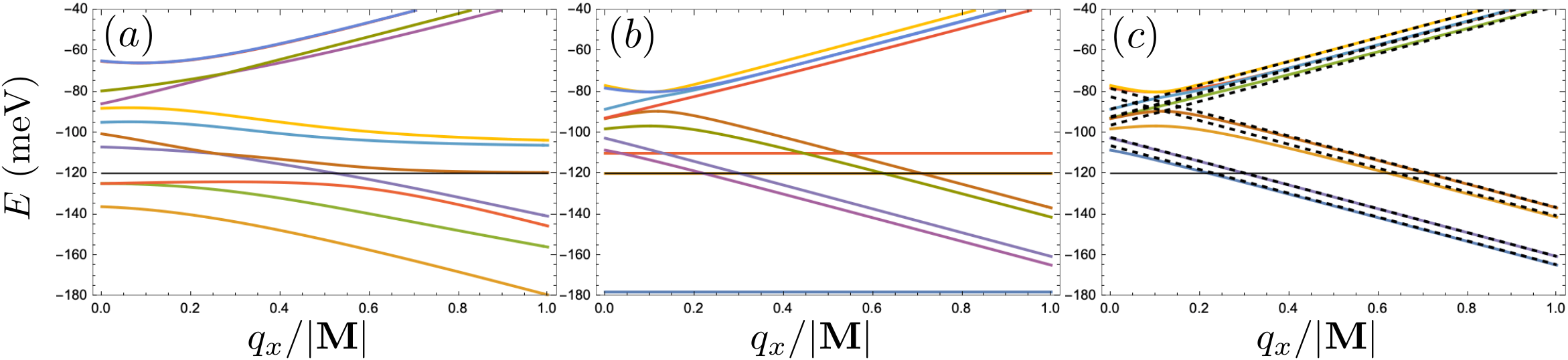}
\caption{$(a)$ Full numerical one-shot eigenvalues at $\mbf{k}=0$ as a function of $q$. $(b)$ Decoupled (where $H_{cf} = 0$) numerical one-shot eigenvalues at $\mbf{k}=0$ as a function of $q$. The flat levels are those of the $f$-modes, given in \Eq{eq:appflevelsdecoupled}. $(c)$ Decoupled analytical one-shot eigenvalues for the conduction bands (no $f$-modes included) at $\mbf{k}=0$ as a function of $q$ (dashed) with numerical $c$-electron eigenvalues in color for comparison. In all figures, the horizontal thin line shows the lowest conduction $f$ mode energy, which is a proxy for the Fermi energy.}
\label{fig:optimalq_app}
\end{figure}

While we cannot obtain a closed form expression for the $c$-mode energies including all terms, we can obtain very good approximations in the large $q$ limit where $|v_F q_x| \gg J,W,M$. Recall that although $\mbf{q}$ is periodic on the BZ, the truncation of the THF model to 1 shell of conduction electrons breaks BZ periodicity. While we find that this artificial periodicity breaking is numerically small for $\mbf{q}$ in the first BZ, our analytical expressions are not periodic in $\mbf{q}$ which allows us to formally expand in large $|\mbf{q}|$ to obtain simplified expressions. Surprisingly, we find this approximation is numerically accurate over a large range of $|\mbf{q}|$. A comparison of the numerical one-shot eigenvalues of $h_0(\mbf{k}) + h_{int}$, the numerical decoupled eigenvalues (setting $\gamma = \gamma' = c'' = 0)$, and our approximation can be found in \Fig{fig:optimalq_app}. We find that the characteristic polynomial of the $8\times8$ $c$-mode block of the Hamiltonian (per spin) splits into two $4$th order factors. Expanding the cumbersome but closed form of the quartic roots, we find the eigenvalues
\bea
\label{eq:appclevelsk0}
E_c(q_x) &\approx \{ -\frac{|v q_x| }{2} + \frac{J}{4} - (W_1 + W_3) \pm \frac{1}{2} \sqrt{M^2 + (M' \eps_+ + (c+2c') \eps_- )^2}, \\
&-\frac{|v q_x| }{2} - (W_1 + W_3) \pm \frac{1}{2} \sqrt{M^2 + (M' \eps_+ + (c+2c') \eps_- )^2} , \\
& \frac{|v q_x| }{2} + \frac{J}{4} - (W_1 + W_3) \pm \frac{1}{2} \sqrt{M^2 + (M' \eps_+ + (c+2c') \eps_- )^2}, \\
&\frac{|v q_x| }{2} - (W_1 + W_3) \pm \frac{1}{2} \sqrt{M^2 + (M' \eps_+ + (c+2c') \eps_- )^2} \} . \\
\eea
This expansion is verified numerically to be very accurate, with less than $1\%$ error at $q_x = \frac{2}{3}M_M$ and $\eps = .0015$. We emphasize that all $c$-electron strain parameters ($c,c',M'$) have been included in the determination of the eigenvalues $E_c(q_x)$. 

To determine the optimal $q_x$, we propose the criterion that the highest occupied $c$-electron energy at $\mbf{k}=0$ equal the lowest unoccupied $f$-electron energy. When hybridization is turned on, these levels will couple and open a gap at the Fermi level. In each spin sector, this is because the 4 $f$ levels are split into 3 conduction bands and 1 valence band (given in \Eq{eq:appflevelsdecoupled}), while the 8 $c$ electrons are split into 4 valence and 4 conduction bands (given in \Eq{eq:appclevelsk0}). This gives the matching criterion for $q_c$:
\bea
M_f\eps_--\frac{3 U_1}{2}-12 U_2  &= -\frac{|v q_c| }{2} + \frac{J}{4} - (W_1 + W_3) + \frac{1}{2} \sqrt{M^2 + (M' \eps_+ + (c+2c') \eps_- )^2} \ .
\eea
A detailed discussion of the determined optimal $q_x$ value is given in the Main Text. 

\newcommand{\tMBZ}{\text{MBZ}}
\newcommand{\tIKS}{\text{IKS}}
\newcommand{\cre}[2]{\hat{#1}^\dagger_{#2}}
\newcommand{\des}[2]{\hat{#1}_{#2}}
\newcommand{\one}{\mathbb{1}}
\newcommand{\expec}[1]{\left\langle #1 \right\rangle}
\newcommand{\vk}{\mbf{k}}
\newcommand{\vQ}{\mbf{Q}}
\newcommand{\vq}{\mbf{q}}
\newcommand{\vG}{\mbf{G}}
\newcommand{\tHF}{\text{HF}}

\section{Bistritzer-MacDonald Hartree-Fock Calculations}
\label{app:sec:hf_in_bm}
In this appendix, we outline the methodology used in our Hartree-Fock simulation of the continuum Bistritzer-MacDonald model. After formalizing the notation, we describe the extraction of the $f$-electron order parameter from the Hartree-Fock self-consistent solution in the band-projected continuum Hamiltonian.

\subsection{Notation}\label{app:sec:hf_in_bm:notation}

Written in the plane wave basis, the Bistrizer-MacDonald Hamiltonian~\cite{BIS11} in the absence~\cite{SON19,BER21,SON21,BER21a,LIA21,BER21b,XIE21,SON22} or presence~\cite{KWA21,PhysRevB.105.245408,VAF23,KAN23b,HER24a} of heterostrain and relaxation effects can be generically written as
\begin{equation}
	\label{app:eqn:spHamiltonian}
	\hat{H}_{0} = \sum_{\substack{ \vk \in \tMBZ \\ \eta, \alpha, \beta, s}}  \sum_{\vQ,\vQ' \in \mathcal{Q}_{\pm}} \left[h^{\left(\eta\right)}_{\vQ,\vQ'} \left( \vk \right) \right]_{\alpha \beta} \cre{c}{\vk,\vQ,\eta,\alpha,s} \des{c}{\vk,\vQ',\eta,\beta,s},
\end{equation}
where the Hamiltonian matrix's explicit form under different conditions is detailed in the cited references. The TBG Hamiltonian in \Eq{app:eqn:spHamiltonian} can be diagonalized as
\begin{equation}
	\label{app:eqn:diag_sp_Hamiltonian}
	\hat{H}_{0} = \sum_{\vk \in \text{MBZ}} \sum_{\eta, n, s} \epsilon_{n,\eta} \left( \vk \right) \cre{\gamma}{\vk,n,\eta,s} \des{\gamma}{\vk,n,\eta,s},
\end{equation}
where 
\begin{equation}
	\label{app:eqn:en_band}
	\cre{\gamma}{\vk,n,\eta,s} \equiv \sum_{\vQ\in \mathcal{Q}_{\pm},\alpha} u_{\vQ \alpha; n \eta} \left( \vk \right) \cre{c}{\vk,\vQ,\eta,\alpha,s},
\end{equation}
are the energy band operators with energies $\epsilon_{n,\eta} \left( \vk \right)$. Here, $u_{\vQ \alpha; n \eta} \left( \vk \right)$ are the eigenstate wave functions of energy band $n$ for the first-quantized single-particle TBG Hamiltonian~\cite{BER21a},
\begin{equation}
	\sum_{\beta} \sum_{\vQ' \in \mathcal{Q}_{\pm}} \left[ h^{\left(\eta\right)}_{\vQ,\vQ'} \right]_{\alpha \beta} u_{\vQ' \beta; n \eta} \left( \vk \right) = \epsilon_{n,\eta} u_{\vQ \alpha; n \eta} \left( \vk \right). 
\end{equation} 
For each valley and spin, the integer $n>0$ represents the $n$-th conduction band, while $n<0$ denotes the $|n|$-th valence band. Additionally, the $f$-electrons are expressed in terms of the plane wave operators as~\cite{SON22}
\begin{equation}
	\label{app:eqn:f_fermions_mom_def}
	\cre{f}{\vk,\alpha,\eta,s} = \sum_{\vQ,\beta} v^{\eta}_{\vQ\beta;\alpha} \left( \vk \right) \cre{c}{\vk,\vQ,\beta,\eta,s},
\end{equation}
where $v^{\eta}_{\vQ\beta;\alpha}$ are the $f$-electron momentum-space wave functions.

The interaction Hamiltonian of TBG is given by~\cite{KAN19,BUL20a,BER21a}
\begin{equation}
	\label{app:eqn:BM_interaction_ham_TBG}
	\hat{H}_I = \frac{1}{2} \frac{1}{N_0 \Omega_0} \sum_{\vq} \sum_{\vG \in \mathcal{Q}_0} V \left( \vq + \vG \right) \delta \rho \left( - \vq - \vG \right) \delta \rho \left( \vq + \vG \right),
\end{equation}
where the density operator is
\begin{equation}
	\label{app:eqn:BM_offset_density_op}
	\delta \rho \left( \vq + \vG \right) = \rho \left( \vq + \vG \right) - \frac{1}{2} \delta_{\vq,\mbf{0}} \delta_{\vG,\mbf{0}} = \sum_{\eta,\alpha,s} \sum_{\vk} \sum_{\vQ \in \mathcal{Q}_\pm} \left( \cre{c}{\vk + \vq, \vQ - \vG, \alpha, \eta, s} \des{c}{\vk, \vQ, \alpha, \eta, s} - \frac{1}{2} \delta_{\vq,\mbf{0}} \delta_{\vG,\mbf{0}} \right),
\end{equation}
and the Fourier transform of the double-gate screened Coulomb interaction potential is
\begin{equation}
	\label{app:eqn:double_gate_interaction}
	V \left( \mbf{q} \right) = \left( \pi U_{\xi} \xi^2 \right) \frac{\tanh \left( |\mbf{q}| \xi/2 \right)}{|\vq|\xi/2},
\end{equation}
In \Eq{app:eqn:double_gate_interaction}, $U_{\xi} = \frac{e^2}{4 \pi \epsilon_0 \epsilon \xi}$ is the interaction energy scale, with $\epsilon$ being the dielectric constant and $\xi$ the distance between the two screening gates. We employ $\xi = 10\, \text{nm}$ (see Refs. \cite{STE20,LIU21c})and $U_{\xi} = 24 \, \text{meV}$, which corresponds to a dielectric constant $\epsilon \approx 6$.

To explore the low energy physics of TBG near charge neutrality, we project into the lowest $N_b$ conduction and highest $N_b$ valence bands. The projected single-particle Hamiltonian $H_0$ is obtained by restricting the summation in \Eq{app:eqn:diag_sp_Hamiltonian},
\begin{equation}
	\label{app:eqn:diag_sp_Hamiltonian_proj}
	H_{0} = \sum_{\vk \in \text{MBZ}} \sum_{\substack{\eta, n, s \\ |n| \leq N_b}} \epsilon_{n,\eta} \left( \vk \right) \cre{\gamma}{\vk,n,\eta,s} \des{\gamma}{\vk,n,\eta,s}.
\end{equation}
The projected interaction Hamiltonian $H_I$ is expressed using form factors~\cite{BER21a},
\begin{equation}
	\label{app:eqn:ff_def}
	M^{\eta}_{mn} \left( \vk, \vq + \vG \right) = \sum_{\alpha} \sum_{\vQ \in \mathcal{Q}_{\pm}} u^*_{\vQ - \vG \alpha; m \eta} \left( \vk + \vq \right) u_{\vQ \alpha; n\eta} \left( \vk \right)
\end{equation}
as
\begin{align}
	H_I = &  \frac{1}{2 N_0 }\sum_{\vk,\vk',\vq} \sum_{\substack{m,n,m',n' \\ |m|,|n|,|m'|,|n'| \leq N_b}} \sum_{s,s',\eta,\eta'} V^{\eta \eta'}_{mn; m'n'} \left( \vk,\vk'; \vq \right) \left( \cre{\gamma}{\vk - \vq,m,\eta,s} \des{\gamma}{\vk,n,\eta,s} - \frac{1}{2} \delta_{\vq, \mbf{0}}\right) \nonumber \\ 
	&\times \left( \cre{\gamma}{\vk' +  \vq',m',\eta',s'} \des{\gamma}{\vk',n',\eta',s'} - \frac{1}{2} \delta_{\vq', \mbf{0}}\right),
\end{align}
where the interaction tensor is
\begin{equation}
	V^{\eta \eta'}_{mn; m'n'} \left( \vk,\vk'; \vq \right) = \frac{1}{\Omega_0} \sum_{\vG} V \left( \vq + \vG \right) M^{\eta}_{mn} \left( \vk, - \vq - \vG \right) M^{\eta'}_{m'n'} \left( \vk', \vq + \vG \right).
\end{equation}
In practice, when computing the interaction tensor, we employ the $C_{2z}T$ gauge fixing condition of Ref.~\cite{BER21a}, which renders the form factors real, thus halving the memory requirements associated with its storage.

The band-basis fermions $\cre{\gamma}{\vk,n,\eta,s}$ preserve the moir\'e translation symmetry. To study phases where the moir\'e translation symmetry is broken, but a modified IKS translation symmetry $e^{-i \mbf{q} \cdot \mbf{R} \tau_3/2} T_\mbf{R}$ with momentum $\vq_{\tIKS}$ is preserved, we introduce modified fermions (see \Eq{eq:boostbasisapp})
\begin{equation}
	\cre{\tilde{\gamma}}{\vk,n,\eta,s} \equiv \cre{\gamma}{\vk + \frac{\eta}{2} \vq_{\tIKS},n,\eta,s}.
\end{equation}
In the tilde basis, $\mbf{k}$ is a good quantum number corresponding to a conserved IKS momentum, but broken valley and translation symmetries. In this basis, the projected single-particle and interaction Hamiltonians are
\begin{align}
	H_{0} = & \sum_{\vk \in \text{MBZ}} \sum_{\substack{\eta, n, s \\ |n| \leq N_b}} \tilde{\epsilon}_{n,\eta} \left( \vk \right) \cre{\tilde{\gamma}}{\vk,n,\eta,s} \des{\tilde{\gamma}}{\vk,n,\eta,s}, \\
	H_I = &  \frac{1}{2 N_0 }\sum_{\vk,\vk',\vq} \sum_{\substack{m,n,m',n' \\ |m|,|n|,|m'|,|n'| \leq N_b}} \sum_{s,s',\eta,\eta'} \tilde{V}^{\eta \eta'}_{mn; m'n'} \left( \vk,\vk'; \vq \right) \left( \cre{\tilde{\gamma}}{\vk - \vq,m,\eta,s} \des{\tilde{\gamma}}{\vk,n,\eta,s} - \frac{1}{2} \delta_{\vq, \mbf{0}}\right) \nonumber \\ 
	&\times \left( \cre{\tilde{\gamma}}{\vk' +  \vq',m',\eta',s'} \des{\tilde{\gamma}}{\vk',n',\eta',s'} - \frac{1}{2} \delta_{\vq', \mbf{0}}\right), 
\end{align}
where we define the shifted energy band eigenvalues and the shifted interaction tensor
\begin{align}
	\tilde{\epsilon}_{n,\eta} \left( \vk \right) &= \epsilon_{n,\eta} \left( \vk + \frac{\eta}{2} \vq_{\tIKS} \right), \\
	\tilde{V}^{\eta \eta'}_{mn; m'n'} \left( \vk,\vk'; \vq \right) &= V^{\eta \eta'}_{mn; m'n'} \left( \vk + \frac{\eta}{2} \vq_{\tIKS}, \vk' + \frac{\eta'}{2} \vq_{\tIKS}; \vq \right).
\end{align}

\subsection{Hartree-Fock Hamiltonian and self-consistent solution}

From now on, the restriction $|n| \leq N_b$ on the band index summations will be implicit (recall that we take $N_b = 3$ keeping the 6 lowest energy bands. We aim to solve the projected many-body TBG Hamiltonian $H_0 + H_I$ at the Hartree-Fock level. We focus on solutions that preserve IKS translation symmetry for a given IKS momentum $\vq_{\tIKS}$, such that  
\begin{equation}
	\expec{ \cre{\tilde{\gamma}}{\vk,n,\eta,s} \des{\tilde{\gamma}}{\vk',n',\eta',s'}} = 0, \quad \text{for} \quad \vk \neq \vk',
\end{equation}
where $\expec{\dots}$ denotes the expectation value in the grand canonical ensemble at the Hartree-Fock level. Each solution is characterized by a density matrix
\begin{equation}
	\rho_{n \eta s; n' \eta' s'} \left( \vk \right) = \expec{ \cre{\tilde{\gamma}}{\vk,n,\eta,s} \des{\tilde{\gamma}}{\vk,n',\eta',s'}} - \frac{1}{2} \delta_{nn'} \delta_{\eta \eta'} \delta_{ss'}.	
\end{equation}
The Hartree-Fock Hamiltonian for a given density matrix is
\begin{equation}
	H_{\tHF} = \sum_{\vk} \sum_{\substack{n \eta s \\ n' \eta' s'}} h^{\tHF}_{n \eta s; n' \eta' s'} \left( \vk \right) \cre{\tilde{\gamma}}{\vk,n,\eta,s} \des{\tilde{\gamma}}{\vk,n',\eta',s'},
\end{equation}
where the Hartree-Fock Hamiltonian matrix is given by
\begin{align}
	h^{\tHF}_{n \eta s; n' \eta' s'} \left( \vk \right) =& \tilde{\epsilon}_{n,\eta} \left( \vk \right) \delta_{n n'} \delta_{\eta \eta'} \delta_{s s'} \nonumber \\
	&+ \frac{1}{N_0} \sum_{\vk'} \sum_{m,m'} \sum_{\eta'',s''} \tilde{V}^{\eta'' \eta}_{mm'; nn'} \left( \vk',\vk; \mbf{0} \right) \rho_{m \eta'' s''; m' \eta'' s''} \left( \vk' \right) \delta_{ss'} \delta_{\eta \eta'} \nonumber \\
	&- \frac{1}{N_0} \sum_{\vk'} \sum_{m,m'} \sum_{\eta'',s''} \tilde{V}^{\eta' \eta}_{m n'; n m'} \left( \vk,\vk'; \vk - \vk' \right) \rho_{m \eta' s'; m' \eta s} \left( \vk' \right).
\end{align}
At self-consistency, the density matrix and the Hartree-Fock Hamiltonian are related by 
\begin{equation}
	\label{app:eqn:hf_scf_fin_temp}
	\rho^{T} \left( \vk \right) = \left\lbrace \exp \left[  \beta \left( h^{\tHF} \left( \vk \right) - \mu \one \right) \right] + \one \right\rbrace ^{-1} - \frac{1}{2} \one,
\end{equation}
where $\one$ is the identity matrix. In \Eq{app:eqn:hf_scf_fin_temp}, $\beta = \frac{1}{T}$ is the inverse temperature, and the chemical potential $\mu$ is fixed by requiring that the total filling is  
\begin{equation}
	\nu = \frac{1}{N_0} \sum_{\vk} \rho \left( \vk \right).
\end{equation}

To obtain integer-filled solutions under heterostrain, we start with a random initial density matrix $\rho \left( \vk \right)$ and use the EDIIS~\cite{KUD02,GAR12} and DIIS~\cite{PUL80,PUL82} methods to reach self-consistency. We assume different IKS momenta $\vq_{\tIKS}$ and select the solution with the lowest energy. A small finite temperature $T = 10 \, \text{K}$ (or $1$meV) is used to speed up convergence in any eventual gapless phases. The Hamiltonian is projected into the two active bands and the four closest remote bands for each spin and valley ($N_b = 3$).

After achieving self-consistency, the $f$-electron order parameters are extracted by projecting the density matrix in the $f$-electron basis as
\begin{equation}
	O^{BM}_{f,\infty} = \frac{1}{N_0} \sum_{\vk} P \left( \vk \right) \rho \left( \vk \right) P^{\dagger} \left( \vk \right),
\end{equation}
where the projection matrix is
\begin{equation}
	\label{app:eqn:proj_mat_to_f}
	P_{\alpha \eta s; n \eta' s'} \left( \vk \right) = \sum_{\vQ \in \mathcal{Q}_{\pm},\beta} \left( v^{\eta}_{\vQ,\beta;\alpha}  \left( \vk + \frac{\eta}{2} \vq_{\tIKS} \right) \right)^* u_{\vQ \beta; n \eta} \left( \vk + \frac{\eta}{2} \vq_{\tIKS} \right)  \delta_{\eta \eta'} \delta_{s s'}.
\end{equation}
In \Eq{app:eqn:proj_mat_to_f}, $v^{\eta}_{\vQ,\beta;\alpha}  \left( \vk \right)$ represents the wave function of the $\cre{\tilde{f}}{\vk,\alpha,\eta s}$ fermion from \Eq{app:eqn:f_fermions_mom_def}, while $u_{\vQ \beta; n \eta} \left( \vk \right)$ are the energy-band wave functions defined in \Eq{app:eqn:en_band}.

\end{document}